\documentclass[preprint,flushrt]{aastex}
\usepackage{natbib}
\usepackage{lineno}
\newcommand\pflux{\mbox{${\rm \, ph \,\, cm^{-2} \, s^{-1}}$}}
\linenumbers
\slugcomment{To be submitted to Astrophysical Journal - Not Yet Refereed }

\shorttitle{Fermi LAT detected blazars}
\shortauthors{Abdo et al.}

\title{Bright AGN Source List from the First Three Months of the {\it Fermi} Large Area Telescope All-Sky Survey}

\author{A.~A.~Abdo$^{1,2}$, 
M.~Ackermann$^{3}$, 
M.~Ajello$^{3}$, 
W.~B.~Atwood$^{4}$, 
M.~Axelsson$^{5,6}$, 
L.~Baldini$^{7}$, 
J.~Ballet$^{8}$, 
G.~Barbiellini$^{9,10}$, 
D.~Bastieri$^{11,12}$, 
B.~M.~Baughman$^{13}$
% ...
}

\altaffiltext{1.} {National Research Council Research Associate}
\altaffiltext{2.} {Space Science Division, Naval Research Laboratory, Washington, DC 20375}
\altaffiltext{3.} {W. W. Hansen Experimental Physics Laboratory, Kavli Institute for Particle Astrophysics and Cosmology, Department of Physics and Stanford Linear Accelerator Center, Stanford University, Stanford, CA 94305}
\altaffiltext{4.} {Santa Cruz Institute for Particle Physics, Department of Physics and Department of Astronomy and Astrophysics, University of California at Santa Cruz, Santa Cruz, CA 95064}
\altaffiltext{5.} {The Oskar Klein Centre for Cosmo Particle Physics, AlbaNova, SE-106 91 Stockholm, Sweden}
\altaffiltext{6.} {Stockholm Observatory, Albanova, SE-106 91 Stockholm, Sweden}
\altaffiltext{7.} {Istituto Nazionale di Fisica Nucleare, Sezione di Pisa, I-56127 Pisa, Italy}
\altaffiltext{8.} {Laboratoire AIM, CEA-IRFU/CNRS/Universit\'e Paris Diderot, Service d'Astrophysique, CEA Saclay, 91191 Gif sur Yvette, France}
\altaffiltext{9.} {Istituto Nazionale di Fisica Nucleare, Sezione di Trieste, I-34127 Trieste, Italy}
\altaffiltext{10.} {Dipartimento di Fisica, Universit\`a di Trieste, I-34127 Trieste, Italy}
\altaffiltext{11.} {Istituto Nazionale di Fisica Nucleare, Sezione di Padova, I-35131 Padova, Italy}
\altaffiltext{12.} {Dipartimento di Fisica ``G. Galilei", Universit\`a di Padova, I-35131 Padova, Italy}
\altaffiltext{13.} {Department of Physics, Center for Cosmology and Astro-Particle Physics, The Ohio State University, Columbus, OH 43210}
% ...

\begin{abstract}
The first three months of sky-survey operation with the {\it Fermi Gamma Ray Space Telescope (Fermi)} Large Area Telescope (LAT) reveals 132 bright sources at $|$b$|>$10$^\circ$ with test statistic greater than 100 (corresponding to about 10$\sigma$). Two methods, based on the CGRaBS, CRATES and BZCat catalogs, indicate high-confidence associations of 106 of these sources with known AGNs. This sample is referred to as the LAT Bright AGN Sample (LBAS). It contains two radio galaxies, namely Centaurus A and NGC 1275, and 104 blazars consisting of 57 flat spectrum radio quasars (FSRQs), 42 BL Lac objects, and 5 blazars with uncertain classification. Four new blazars were discovered on the basis of the LAT detections. Remarkably, the LBAS includes 10 high-energy peaked BL Lacs (HBLs), sources which were so far hard to detect in the GeV range.  Another 10 lower-confidence associations are found.  Only thirty three of the sources, plus two at $|$b$|<$10$^\circ$, were previously detected with {\it EGRET}, probably due to the variable nature of these sources. The analysis of the gamma-ray properties of the LBAS sources reveals that the average GeV
spectra of BL Lac objects are significantly harder than the spectra of
FSRQs. No significant correlation between radio and peak gamma-ray fluxes is observed. Blazar log N - log S and luminosity functions are constructed to investigate the evolution of the different blazar classes, with positive evolution indicated for FSRQs but none for BLLacs.  The contribution of  LAT-blazars to the total extragalactic $\gamma$-ray intensity is estimated. 

\end{abstract}


\begin{document}


\section{Introduction}
The {\it Gamma ray Large Area Space Telescope} ({\it GLAST }) 
was launched on 11 June 2008, and renamed the {\it Fermi Gamma Ray Space Telescope} 
shortly after entering its scientific operating mission, which began on 11 August, 2008. 
The Large Area Telescope (LAT) on {\it Fermi} provides an increase in sensitivity
by more than an order-of-magnitude over its predecessor {\it EGRET}, 
the Energetic Gamma Ray Experiment Telescope on the 
{\it Compton Gamma Ray Observatory} \citep{thompson93}, and the Italian Space Agency Satellite
{\it AGILE} \citep[Astro-rivelatore Gamma a Immagini Leggero;][]{tavani08}. In sky survey
mode, the LAT observes all parts of the sky every 3 hours, providing effectively uniform 
exposure on longer timescales. 
%\citep{michelsonLAT08}

%The {\it Fermi} LAT is providing the first clear view of the universe in $\approx 1$ -- 100 GeV $\gamma$ rays. 
One of the major scientific goals of the 
{\it Fermi Gamma Ray Space Telescope} is to provide new data about $\gamma$-ray activity
of AGNs.  Rapidly varying fluxes and large luminosities of extragalactic $\gamma$-ray sources 
are best explained if the $\gamma$ rays are emitted from collimated jets of 
charged particles moving at relativistic speeds \citep{br78,mgc92}.   
{\it Fermi}-LAT observations will help determine how these particles are accelerated, where the gamma rays are emitted, 
what the energy and power budgets of the supermassive black-hole engines are,  
what this says for the fueling and growth of black holes, and the reasons for the differences between
 radio-loud and radio-quiet AGNs, and FSRQs and BL Lac objects. These are just a few of 
the questions that $\gamma$-ray AGN studies with the {\it Fermi}-LAT are helping to answer
\citep[see][for more discussion of these goals]{LATpaper}.

In a companion publication \citep{SourceList}, 132 bright sources at $|b|>10^\circ$ with test statistic (TS) $>100$ are found in the preliminary three month
{\it Fermi} all-sky survey. As expected from the EGRET legacy, a large fraction of these sources 
are AGNs. Detailed results of the subset of the {\it Fermi} bright source  list that are associated with AGNs are presented here. 

Sixty-six high-confidence blazars are listed in the Third {\it EGRET} catalog of high-energy gamma-ray sources 
\citep[3EG catalog;][]{3EGcatalog}, with the majority of them, $\approx 77\%$, 
identified as flat spectrum radio quasars (FSRQs), and the remaining $\approx 23\%$ identified as belonging to the BL Lac 
class.\footnote{In contrast to the prominent optical emission lines found in FSRQs, BL Lac objects are 
radio-loud, rapidly variable sources displaying nearly featureless continua with emission-line 
equivalent widths $< 5~ \AA$ \citep[for review, see][]{up95}.} The recently released catalog of high-confidence {\it AGILE} gamma-ray sources\footnote{http://www.asdc.asi.it/agilebrightcat/} \citep{AGILEcatalog} shows a somewhat higher percentage of BL Lacs.
Unlike AGN surveys at optical or X-ray energies,  
in which the majority of AGNs are radio quiet \citep[e.g.,][]{dellaCeca94,ivezic02}, 
all AGNs detected at $\gtrsim 100$ MeV energies are also significant radio sources.
This includes the 3EG and {\it AGILE} blazars, which are so far identified 
with flat spectrum (radio spectral index $\alpha_r>-0.5$ at GHz frequencies) radio-loud AGNs, 
and most show superluminal motion \citep{jorstad01,kellermann04}. Moreover, the redshift distribution is
broad, with the largest redshift AGN known in the 3EG catalog 
%is 3EG J0500$-$0159 
at  $z = 2.286$.

Here we present a source list of bright AGNs found in the set of the 132 bright LAT sources at $|$b$|>$10$^\circ$. Identification of variable $\gamma$-ray sources with blazars depends on the statistical likelihood
of positional association and correlated variability of the $\gamma$-ray emissions 
with lower-frequency radiations \citep[e.g.][]{srm03}. 
The 106  sources having high-confidence associations with known blazars and radio-galaxies 
constitute the LAT Bright AGN Sample (LBAS). Included in this list are mean fluxes, weekly peak fluxes, spectral indices,  
locations, and variability information. Only sources with confidence levels greater than $10\sigma$ are retained 
in the LBAS. This list is not, however, complete, 
as we already know of many more sources at lower significance.
The limiting flux depends on both the source sky location and the spectral hardness.    

In Section 2, observations with the LAT, analysis methods, and the source detection procedure are presented. 
Section 3 describes the association method and gives the list of bright {\it Fermi}-LAT detected blazars.
Key properties of the LBAS, including flux and spectral index, are presented in Section 4. 
The LBAS is compared with {\it EGRET} blazars in Section 5.  
Section 6 considers the radio/gamma-ray connection. 
Population studies, including source types and redshifts, are presented in Section 7, 
where the  $\log\,$N - $\log\,$S flux distributions and luminosity functions of the LBAS are constructed. 
The results are discussed in Section 8, including implications of the results for 
blazar evolution. We summarize in Section 9.

In the following we use a $\Lambda$CDM cosmology with values given within
1$\sigma$ of the WMAP results \citep{komatsu08}, namely  $h = 0.71$, $\Omega_m = 0.27$ and $\Omega_\Lambda = 0.73$.
Here the Hubble constant $H_0=100h$ km s$^{-1}$ Mpc$^{-1}$ is used. 

\section{Observations with the Large Area Telescope}

The {\it Fermi}-LAT is a pair-conversion gamma-ray telescope sensitive to photon energies greater than 20 MeV. It is made of a tracker (composed of two sections, front and back, with different capabilities), a calorimeter and an anticoincidence system to reject the charged-particle background. The LAT has a large peak  effective area  ($\sim 8000$ cm$^2$ for 1 GeV photons in the event class considered here), viewing $\approx 2.4$ sr of the full sky with excellent angular resolution (68\% containment radius $ \approx 1^\circ$ at $E = 1$ GeV for the front section of the tracker and about a factor of 2 larger for the back section). A full description of the LAT instrument and its predicted performance are reported in \cite{LATpaper}.  During the first year, the telescope operates in sky-survey mode observing the whole sky every 3 hours. The overall coverage of the sky is fairly uniform, with variations of around $\simeq$15\% around the mean value.

The LAT data used here were collected during the first 3-month all-sky survey, from August 4 to October 30 2008. We refer to the companion paper \citep{SourceList}  for a full description of the data selection and analysis. In order to avoid background contamination from the bright Earth limb, time intervals where the Earth entered the LAT Field-of-View (FoV) were excluded from this study (corresponding to a rocking angle $<$\,47 deg). In addition, events that were reconstructed within 8$^\circ$ of the Earth limb were excluded from the analysis (corresponding to a zenith angle cut of 105$^\circ$). Due to uncertainties in the current calibration,  only photons belonging to the "Diffuse" class with energies above 100 MeV were retained. These  photons provide the purest gamma-ray dataset. The energy range was even more restricted in the source detection and spectral fitting analyses described below, where only photons with E $>$200 MeV were selected.  The list of sources reported in Tables 1 and 2 was obtained as the result of the  source detection, localization and significance estimate analyses described in detail in \cite{SourceList}. 

The source detection step made use of two wavelet algorithms, ({\it mr\_filter}) \citep{sp98} and ({\it PGWAVE}) \citep{ciprini07}. The algorithms were run  independently for different energy bands associated with different localization power and the results were cross-checked. The positions of the sources for which the detection significance was above threshold (4$\sigma$) were  then refined using  ({\it pointfit}), a simplified likelihood method (see \cite{SourceList}). This algorithm uses photons with E$>$500 MeV and returns the optimized sky position as well as an estimate of the error radius for most detected sources. As discussed in \cite{SourceList}, the final error in the source position was estimated by multiplying the error radius returned by the algorithm by a factor close to 1.4  and adding 0.04$^\circ$ in quadrature (estimated from the residuals between the estimated and expected position of Vela). The 95\% confidence error radius was then evaluated assuming a 2-D normal distribution.

To better estimate the source significance, we used the maximum likelihood algorithm implemented in ({\it gtlike}) a tool that is part of the  standard {\it Fermi}-LAT {\it ScienceTools} software package\footnote{http://fermi.gsfc.nasa.gov/ssc/data/analysis/documentation/Cicerone/}.  The flux, photon index and test statistic (TS) of each source in the energy range 0.2-100 GeV were determined by analyzing regions of interest (ROI) typically 15$^\circ$ in radius. The model of the ROI used to fit the data was built taking into account of all the sources detected within a given ROI. The  isotropic background and Galactic Diffuse background models used in the fit are discussed in \cite{SourceList}.  Each source was modeled with a simple power law (k$\mathrm{E^{-\Gamma}}$) for photons E$>$ 200 MeV. The flux [E$>$100 MeV] (F$_{100}$), which is conventionally reported, was then calculated with the fitted parameters. This flux will be used throughout this paper. The  spectral energy distributions of some bright sources show clear evidence for a break or curvature. A fit with a single power law function is certainly not the most appropriate choice for these sources but the resulting photon index does reflect the spectral hardness. A more detailed spectral analysis of the LBAS sources is beyond the scope of this paper. The source fluxes were also estimated  by fitting independent power law functions in two energy bands (0.1-1 GeV) and (1-100 GeV) and summing up the two obtained fluxes.  These fluxes (F$_25$ in Table 3) are the same as those reported in the {\it Fermi} bright source list paper \citep{SourceList}. For most sources, the fluxes obtained by the two methods are consistent within 30\%.

The same procedure was applied to generate weekly light curves (spanning a 12-week period). From those, the weekly peak flux as well as a  variability index (corresponding to a simple $\chi^{2}$ criterion)  were derived. The variability tag reported in this paper is set for sources  associated with a probability of being constant lower than 1\%.  A few representative light curves are displayed in Fig. \ref{fig:lcurves}.

This analysis was performed with the preflight instrument response functions (P6\_V1). In flight, the presence of pile-up signals in the LAT tracker and calorimeter left by earlier particles was revealed in periodic-trigger events. This feature leads to a reduction of the real acceptance as compared to the predicted one as fewer events pass the rejection cuts, most notably for low-energy photons. The magnitude of this reduction is still under investigation, but the fluxes reported here may be lower than the true ones by as much as 30\%  and the photon indices greater than the true ones by as much as  0.1 (true spectra could be softer by 0.1 unit in the photon index). Because of the current uncertainty, no correction has been applied to the results. This uncertainty applies uniformly to all sources. Our relative errors  are much smaller \citep[about 3\% on the flux,][]{SourceList}. With the acceptance used in this analysis, the measured fluxes of the 3 bright pulsars, Vela, Geminga and Crab  \citep{SourceList} are found to be compatible within 11\% with those reported in the 3EG catalog.  
 
Fig. \ref{fig:coverage} shows the 3-month flux sensitivity for TS=100 and a photon index=2.2 as a function of the sky position, calculated by a semi-analytical, maximum likelihood estimate of the significance. This estimate takes the actual exposure, the PSF and the different backgrounds (galactic diffuse, extragalactic diffuse and instrumental) into account. The limiting flux  is higher at low galactic latitude due to a higher galactic diffuse background and close to the celestial south pole (l $\simeq 302^\circ$, b $\simeq -27^\circ$) where the exposure is lower.  

The final result of the detection analysis is a list of 205 sources with a ($\mathrm{TS\,>100}$, $\sim\,10\sigma$), composing the LAT Bright Source (0FGL) list \citep[see Table 6. in][]{SourceList}. For comparison, 31 sources detected by EGRET have a significance greater than $10\sigma$ in the 3EG \citep{3EGcatalog} and EGR \citep{EGRcatalog}  catalogs. Of these, only 13 were detected at $|b|>$10$^\circ$.  In the 0FGL, a total of 132 sources, including 7 pulsars, are present at $|b|>$10$^\circ$. We have explored the possibility of associating AGNs with the 125 remaining sources.

\section{Source association}

Any source association procedure primarily relies on spatial coincidence. 
Fig. \ref{fig:r95_ts} shows the 95\% error radius vs ($\mathrm{TS}$) for the sources considered here. This radius  depends on both the flux and the photon index, with a mean of 0.14$^\circ$. For comparison, the average corresponding radius for the blazars in the 18 month {\it EGRET} sky survey is 0.62$^\circ$. Of the 186 $|b|>10\arcdeg$ 3EG sources, 66 (35\%) had ``high'' (but unspecified) confidence positional associations with blazars in the 3EG catalog.  Another 27 positional coincidences were noted at lower significance.  Although subsequent work \citep[e.g.][]{mhr01,srm03} did find additional associations, $\sim$40\% of the high-latitude 3EG sources remained unidentified.

Although the LAT localization accuracy is much better than those of previous gamma-ray telescopes, it is not good enough to enable a firm identification of a LAT source based solely on spatial coincidence.  For the LAT, a firm identification is assumed only if correlated variability is observed at different wavelengths. % in the case of a blazar, or pulsation is detected  for a pulsar or binary In all other cases,  positional coincidence as a simple high probability association of a LAT source with an object belonging to a plausible class of gamma-ray emitters. 
 In order to find associations between LAT sources and AGNs, two different approaches were pursued. The first method is based on a procedure similar to that developed by \cite{srm03} for associating {\it EGRET} blazars with radio counterparts using an observational figure of merit (FoM). The second one is based on the calculation of source association probabilities following a Bayesian approach \citep{deRuiter77,ss92}, similar to that used by \cite{mhr01} to associate {\it EGRET} sources with radio sources.  This method is  described in \cite{SourceList}.

Several catalogs were used by the two association methods, the most important ones being the Combined Radio All-Sky Targeted Eight GHz Survey \citep[CRATES;][]{crates} catalog and the {\it Roma-BZCAT}\footnote{http://www.asdc.asi.it/bzcat} \citep{BZcatalog}. The CRATES catalog contains precise positions, 8.4 GHz flux densities, and radio spectral indices for more than 11,000 flat-spectrum sources over the entire $|b| > 10\arcdeg$ sky.
The  {\it Roma-BZCAT} is a master list of blazars based on an accurate examination of literature data and  presently includes about 2700 sources, all observed at radio and
optical frequencies and showing proper characteristics of blazars.
Sources are classified as BL Lacertae objects (BZB), flat spectrum radio 
quasars (BZQ) or as blazars of uncertain type (BZU).

\subsection{The Figure-of-Merit Method}

The figure of merit (FoM) approach requires a large, uniform all-sky sample of radio sources from which to draw; for this purpose, we use the Combined Radio All-Sky Targeted
Eight GHz Survey \citep[CRATES;][]{crates} catalog.
%, which contains precise positions, 8.4 GHz flux densities, and radio spectral indices for more than 11,000 flat-spectrum sources over the entire $|b| > 10\arcdeg$ sky.  
In order to quantify the correlation between CRATES sources and LAT detections, we
compare the average number of positional coincidences between LAT sources and CRATES
sources to the number of positional coincidences between LAT sources and
sources drawn from 1,000 randomized simulations of the radio sky.  We count as
a positional coincidence any occurrence of a radio source (real or simulated)
within twice the 95\% error radius of a LAT source, and we generate the
simulated radio skies by scrambling the Galactic coordinates of the CRATES
sources while keeping their radio flux densities, spectral indices, and counterpart RASS fluxes intact.

We define the excess fractional source density of radio/$\gamma$-ray matches as $n = 1-(N_{\mathrm{rand}}/N_{\mathrm{CRATES}})$ and we compute this
quantity in bins of radio flux density $S_{8.4}$ at 8.4 GHz, radio
spectral index $\alpha$, and X-ray flux $F_X$ from the ROSAT All-Sky Survey
\citep[RASS;][]{RASSbright}.  These functions---$n(S)$, $n(\alpha)$, and $n(F_X)$---
constitute the counterpart spectral energy distribution (SED) components of the FoM.  The final component is the dependence on the offset between the radio position and the LAT position,
which we model simply as $n_\mathrm{pos} = 1 - \mathrm{CL}$, where CL is the
confidence limit of the LAT localization contour passing through the radio
position.  The FoM is then given by $100 \times n(S) \times n(\alpha) \times
n(F_X) \times n_\mathrm{pos}$.  To evaluate the significance of the FoM, we
again generated, in the manner described above, 1,000 random simulations of the
radio sky and computed the average distribution of FoM.  We compared this to
the distribution of FoM for the real CRATES sky by again computing the
excess fractional source density as a function of FoM.  This fractional excess
can be directly interpreted as a probability $P_i$ of radio/$\gamma$-ray
association for source $i$, giving an immediate mapping from FoM to
association probability for each individual source (i.e., $1-P_i$ is the
probability of a false positive association). We find that 1,000 simulated skies result in sufficient statistics in each FoM bin to ensure that the mapping is robust. Very similar results are obtained with 10,000 simulations.

The results of this association procedure are shown in Table 1 and Table 2.
Most of the associated radio sources are in the Candidate Gamma-Ray Blazar
Survey \citep[CGRaBS;][]{cgrabs}, an optical survey of the 1,625 CRATES
sources that were most similar in their radio and X-ray properties to the 3EG
blazars.  Optical spectroscopy of the sources with unknown redshifts is
ongoing.  We also considered the possibility of an association with a
non-CRATES radio source when no CRATES association was found. Indeed, a FoM
can be computed for any object for which the necessary radio data are
available.  Thus, for those LAT sources without CRATES associations, we drew
candidate counterparts from the 1.4 GHz NRAO VLA Sky Survey
\citep[NVSS;][]{NVSScatalog} or the 843 MHz Sydney University Molonglo Sky Survey
\citep[SUMSS;][]{SUMSScatalog}, searched NED for archival 8.4 GHz data, and
calculated the FoM for each candidate.  These procedures find high-confidence ($P > 0.90$) associations for 101 of the 125 non-pulsar sources in the 0FGL list with $|b| > 10\arcdeg$ for an association rate of 81\%.  We also find low-confidence FoM associations ($0.40 < P < 0.90$) for 14 more sources, bringing the total association rate to 92\%.  Thus, the radio-bright
blazar population continues to dominate the extragalactic sky.

The individual association probabilities can be used to estimate the number of
false positives in a given sample:\ if the probabilities $P_i$ are sorted from
highest to lowest, then the number of false positives in a sample of $k$
sources is $N_\mathrm{false} \approx \sum_{i=1}^k{(1-P_i)}$. Among the high-confidence associations, there are $\sim$3 false positives, and less than one of the 74 most probable associations should be false. 

We also studied the power of the FoM analysis to reject a blazar association
for a LAT source.  We considered NVSS/SUMSS sources in the direction of the
unassociated LAT sources and computed the FoM that each source would have if
(A) it were as bright as the 4.85 GHz flux density upper limit from the Green
Bank 6 cm survey \citep[GB6;][]{GB6catalog} or the Parkes-MIT-NRAO survey
\citep[PMN;][]{PMNcatalog} (unless the source had an actual GB6/PMN detection, in
which case we used the measured flux density) or (B) its radio spectrum were
as severely inverted as $\alpha = +0.75$ between 1.4 GHz and 4.85 GHz,
whichever constraint was tighter.  From the low-frequency radio spectrum (or
upper limits), we extrapolated the implied 8.4 GHz flux density.  If the
resulting FoM indicated that the source could conceivably be a flat-spectrum
blazar, then we drew no conclusion, but if we found that the ``best-case''
association probability were 0\%, then we concluded that the LAT source was
not associated with any typical member of the population of flat-spectrum
blazars, and we refer to such cases as ``anti-associations.''  Note that the
spectral index $\alpha = +0.75$ is an extremely conservative cutoff. The most
inverted radio spectrum for any actual association has $\alpha < 0.65$.  We
are able to secure anti-associations for 10 sources.  In fact, five of these
turn out to be high-latitude LAT pulsars and pulsar candidates. This shows that, given a reliable LAT error circle, 
the FoM analysis is capable of indicating definitively that a source is not
a blazar

\subsection{Summary of association results}

The combination of the FoM (described above)  and positional association  methods yields a number of 106 high-confidence ($P \ge$ 0.90)  associations (constituting the LBAS) and 11 low-confidence (0.40$<P <$0.90) associations listed in Table 1 and 2 respectively. Simple extrapolation of these numbers implies that the LAT should be detecting some 20-25 blazars through the Galactic plane at $|b|<10^\circ$. Indeed, several have already been located, e.g. 0FGL J0036.7+5951 (1ES 0033+595), 0FGL J0730.4$-$1142 (PKS 0727$-$11), 0FGL J0826.0$-$2228 (PKS 0823$-$223), 0FGL J1802.6$-$3939 (PMN J1802$-$3940), 0FGL J1833.4$-$2106 (PKS 1830$-$211). A more complete search for Galactic background blazars, incorporating spectrum, variability and multiwavelength properties is in progress.

Tables 1 and 2 report, for each source, the LAT name, the name of the associated source  based on the FoM method, the value of the FoM parameter and its probability, the name of the positionally associated source  and its probability,  the redshift and the AGN class.  Fig. \ref{fig:sky_map} shows the sky location of the LBAS AGNs. 
%A  relative deficit of sources is observed in the region of  the celestial south pole (l$\simeq$ 330$^\circ$, b$\simeq$ -30$^\circ$) where the exposure is lower and thus the flux limit higher (Fig. \ref{fig:coverage}).    

One source, 0FGL J10340+6051 reported in Table 1, merits special comment. Two radio associations were found by the FoM method for this $\gamma$-ray source, one with very high probability and one with lower, but still significant, probability reported in Table 2. Although the high-probability source likely dominates the $\gamma$-ray emission, it is entirely plausible that the low-probability source contributes non-negligibly
to the total $\gamma$-ray flux.  We believe that as the LAT detects more
sources and confusion of the $\gamma$-ray sky increases, the power of the FoM
formalism will become increasingly important to the identification of
multiple lower-energy counterparts of complex $\gamma$-ray sources.  

Fig. \ref{fig:ang_sep} shows the overall, normalized angular separation distributions for both sets of sources  (i.e. high- and low-confidence associations).  The solid curve corresponds to the expected distribution ($\chi^2$ distribution with 2 d.o.f. ) for real associations, the dashed one for accidental associations. This figure provides confidence that most associations are real. From this figure, it appears that the 1.4 correction factor applied to the error radius is somewhat overestimated. This overly conservative factor will be significantly reduced with additional analysis updates.

Four new blazars were discovered. Two of these, CRATES J1012+2439, and CRATES J1032+6051 were classified as FSRQ blazars while  CRATES J0144+2705 is a BL Lac. The classification of these three sources was made on the basis of the broad lines observed in their optical spectrum obtained after the LAT detection (Shaw et al., in preparation). The forth new LAT detected blazar is CLASS J1054+2210. Its classification as a BL Lac object was made possible by the analysis of its optical spectrum available at the  SDSS on-line archive. As discussed above, CRATES J1032+6051 is the source which has a low probability to be associated with 0FGL J10340+6051.

The  other sources listed in Table 1 and 2 were classified as FSRQ or BLLac following the {\it Roma}-BZCat and CRATES/CGRaBS catalogs. Some sources, which cannot be properly classified because of the scarcity of available data or which show optical spectra intermediate between those of BL Lacs, FSRQs or radio galaxies, were assigned to the  ``uncertain class'' ( ``Unc" label in the tables).

Based on this classification, the LBAS comprises  57 FSRQs, 42  BL Lac objects,  5 blazars of uncertain type, and 2 radio-galaxies (RGs). The relevant {\it EGRET} sample of reference corresponds to that of  the 18 month {\it EGRET} all-sky survey during Phase 1 of the CGRO mission \citep{fichtel94,dermer07}. This survey had relatively uniform exposure, and contained 60 sources, 46 FSRQs, 14 BL Lacs.  BL Lacs make up 40\% of the LBAS blazars, a fraction significantly higher than found with {\it EGRET} (23\%).  The detection of hard sources (BL Lac objects, see below) by the LAT  is intrinsically favored over soft ones (FSRQs). This is partly due to the strongly energy-dependent PSF. The larger bandpass and higher energy for the peak sensitivity (in the $\sim$ 1-5 GeV range)  of the LAT as compared to {\it EGRET} adds to this effect.

Eleven LBAS sources are associated with blazars already detected in the TeV energy range by the ground based imaging air Cherenkov telescopes. Among these,  7 are classified as high-frequency peaked BL Lacs (HBLs): 1ES 1011+496, Mrk 421, PG 1553+11, Mrk 501, 1ES 1959+650, PKS 2005$-$489 and PKS 2155$-$304; 3 are low-frequency peaked BL Lacs (LBLs): 3C 66A, W Com and BL Lac and  one is a FSRQ: 3C 279.
These 11 sources represent more than 50\% of the TeV blazars detected so far (21). The results of simultaneous observations that cover the optical, X-ray, and high energy gamma-ray bands (LAT and H.E.S.S.)  of PKS 2155-304 are reported in \cite{ahar09}. Another three HBLs in the LBAS are not yet detected in the TeV range: KUV00311$-$1938, 1ES0502+675, B3 0133+388. A total of 10 HBLs  are thus present in the LBAS, a remarkable feature given that sources in this class were difficult to detect in the GeV range. Many of these sources were not particularly flaring at other wavelengths during the period of observation. 

We compared  the broad-band (radio, optical, X-ray) properties of our sample of {\it Fermi}-LAT detected blazars with those of the known blazars listed in the {\it Roma}-BZCat catalog and found that the broadband properties of the {\it Fermi}-LAT detected BL Lacs and FSRQs are consistent with the parent population of FSRQs and BL Lacs. This is illustrated in Fig. \ref{fig:ps_rx} displaying the soft X-ray flux vs radio flux density (at 1.4 GHz) diagram for the Fermi-LAT blazars and the full blazar catalog.

The LBAS includes 13 sources (10 FSRQs and 3 BL Lacs) that were detected in a flaring state promptly announced to the community through Astronomical Telegrams. Among these, 0FGL J2254.0+1609, associated with 3C 454.3, is the brightest gamma-ray extra-galactic source observed in the 3-month {\it Fermi}-LAT survey and is studied in detail in \cite{abdo454.3}. 

The {\it Fermi}-LAT has discovered  gamma-ray emission from a source having an high-confidence association with NGC 1275, the supergiant elliptical galaxy at the center of the Perseus galaxy cluster. EGRET observations yielded only an upper limit to the NGC 1275 gamma-ray emission.  All the details about the gamma-ray properties of this source will be reported in \cite{abdoNGC1275}.

Cen A is the nearest radio galaxy to us and it was one of the few radio galaxies associated with a 3EG source \citep[J1324$-$4314;][]{sreekumar99}. It is included in the LBAS and the position of its nucleus is well inside the 95\% confidence error radius of the source 0FGL~J1310.6$-$4301. The measured {\it Fermi} flux is F$_{100}\simeq 2.3\times 10^{-7}$ ph cm$^{-2}$ s$^{-1}$, about a factor of 2 greater than that measured by {\it EGRET} \citep{sreekumar99}. 

Recently, two more sources reported in the 3EG catalog were tentatively associated with radio galaxies, 3C~111 (Hartman et al. 2008), and possibly NGC~6251 \citep{mukherjee02,foschini05}. These objects are not LBAS sources but the number of radio-galaxies detected at high-energy is expected to increase in the near future as more data accumulate.

Table 4 lists the 33 sources associated with 3EG sources (two more located at $|b|<$10$^\circ$ were also incorporated). Three bright EGRET blazars associated with 0827+243, PKS 1622$-$297 and 1730$-$130 (NRAO 530), whose average EGRET fluxes are in the range of (25 - 47) $\times 10^{-8}$ ph$(E >100$ MeV) cm$^{-2}$ s$^{-1}$  do not appear in the  LBAS. Presumably, these blazars are simply in a lower flux state than when EGRET was in operation. These 3 sources are also among the 22 sources in the pre-launch LAT monitored list\footnote{http://fermi.gsfc.nasa.gov/ssc/data/policy/LAT\_Monitored\_Sources.html}. Of these 22 sources, 17 have high-significance LAT detections in the first 3-months of data. The remaining two monitored sources (H 1426+428, 1ES 2344+514) did not have previous 3EG detections and thus were not expected to be very bright GeV sources.

We note that the LBAS object B2 0218+35 is a well-known gravitational lens. The source PMN J0948+0022, associated with 0FGL J0948.3+0019, has a flat radio spectrum, but shows an optical spectrum with only narrow emission lines, making it an  ``uncertain"-type object in the {\it Roma}-BZCat.

\section{Gamma-ray properties of the LBAS}

\subsection{Introduction}

Table 3 lists the key properties of the 116 sources associated with AGNs (sources with low-confidence associations are in italics): the name, equatorial and galactic coordinates, the $TS$ parameter measuring the significance of the detection, the photon index ($\Gamma$), the photon flux F$_{100}$,  the weekly peak flux,  the photon flux F$_{25}$ and the variability flag. The uncertainties are statistical only. From Table 3 (last column), 40 FSRQs (70\%), 12 BL Lacs (29\%) and 1 Uncertain blazar (0FGL J0714.2+1934) present in the LBAS show evidence for variability.  The observed variability for FSRQs is thus higher than for BL Lacs.  One must be careful in interpreting this result as the flux distributions are different for the two classes (see Fig. \ref{fig:flux_index_lim}), making the detection of variability easier for FSRQs. An in-depth variability analysis of the LBAS is beyond the scope of this paper.     

Table 4 gives similar parameters for the subset of 35 sources (including both high-confidence and low-confidence associations, plus two at $|$b$|<10^\circ$) corresponding to 3EG sources. This subset will be discussed in more detail in section 5.

The source photon index is plotted as a function of the flux in Fig. \ref{fig:flux_index_lim}. It is already visible in this figure that the photon indices of BL Lac objects (open circles) and FSRQs (closed circles) are quite distinct.  The flux sensitivity (calculated in the same way as for the map shown in Fig. \ref{fig:coverage} and depicted as solid lines for two different galactic latitudes)  is fairly strongly dependent on the photon index. The upper envelope in the spectral index - flux ($>$100 MeV) plot reflects that the peak sensitivity of the LAT is at energies much higher than 100 MeV.  These ranges of spectral index and apparent flux limits translate to approximately constant limits above 1 GeV. For a photon index of 2.2, the 10 $\sigma$ flux sensitivity F$_{100} \simeq 5 \times$ 10$^{-8}$ ph cm$^{-2}$ s$^{-1}$, about 3 times lower than that of the Third {\it EGRET} catalog.

\subsection{Flux}

It makes sense to compare the LBAS fluxes with those reported in the Third EGRET Catalog for the EGRET sample. As several analyses \citep[e.g.][]{mp00,dermer07} used the peak flux (maximum flux in all EGRET viewing periods), instead of the mean flux because of the fairly non-uniform coverage in the EGRET Catalog, comparisons will be performed both for the mean and peak flux distributions. For {\it EGRET}, both distributions are biased as observations were preferentially made of sources known to be highly variable in the gamma-ray band, and some of the observations were triggered by ToO requests when an object was brightly flaring in other wavebands. No such bias exists for the LAT.     
      
Fig. \ref{fig:flux_dis}a compares the mean flux distribution measured in the LBAS with that measured in the  {\it EGRET} sample. The high-flux ends of these distributions look similar.  This observation points to a nearly constant global gamma-ray luminosity of detectable blazars at a given time, as can naively be expected. In stark contrast, the weekly peak flux distributions (Fig. {\ref{fig:flux_dis}b) look different, the peak fluxes being significantly higher in the {\it EGRET} sample.  This feature probably arises from the shorter sampling period for the {\it Fermi}-LAT as compared to {\it EGRET}. In the 3-month period considered here, a given source had much less opportunity to explore very different states than in the 4.5 years over which the {\it EGRET} observations were conducted.  Another illustration of this effect  is given in Fig. {\ref{fig:flux_dis}c,d where the peak flux vs the mean flux and the peak flux/mean flux ratio distributions are shown respectively. The inference that the gamma-ray blazars have characteristic variability timescales of months to years is well confirmed by the observation that only $\simeq$ 30\% of the {\it EGRET} blazars are still detected by the LAT at a comparable flux.

\subsection{Photon index}

The photon index distribution  gives insight into the emission and acceleration processes acting within the AGN jets, as it enables some of the physical parameters involved in these processes to be constrained. Moreover, it can be used to test whether  the BL Lac and FSRQ populations have different $\gamma-$ray emission properties.

Fig. \ref{fig:sp_dist} top  displays the photon index distribution for all the LBAS sources. This distribution looks fairly similar to that observed for the {\it EGRET} sample \citep{nandikotkur07}: it is roughly symmetric and centered at $\gamma$= 2.25.  The corresponding distributions for FSRQs and BL Lacs are shown in Fig. \ref{fig:sp_dist} middle and bottom respectively. These distributions appear clearly distinct, with little overlap between them. This is a remarkable feature, given that the statistical uncertainty typically amounts to 0.1 for most sources. The distributions have (mean, rms)=(1.99, 0.22) for BL Lacs and (2.40, 0.17) for FSRQs.  We used a Kolmogorov-Smirnov (KS) test to test the null hypothesis that both index samples are drawn from the same underlying distribution and found a probability of 2 $\times$ 10$^{-12}$ \footnote{We are  aware of the fact that the KS test is not optimal for binned data, but it is accurate enough to reject the null hypothesis}. Although indications for the existence of two spectrally distinct populations (BL Lacs and FSRQs) in the  {\it EGRET} blazar sample were mentioned in the literature \citep{pohl97,vp07}, this is the first time that the distinction appears so clearly. The mean photon index of the 10 HBLs included in the LBAS is 1.76, i.e significantly  lower (sources are harder) than the mean of the whole BL Lac subset as expected for these high-energy peaked sources. 

To  infer physical properties of the blazar populations from the observed photon index distributions, possible instrumental and/or statistical effects have to be assessed. A systematic bias may indeed arise in the likelihood analysis of sources with low photon statistics. To quantify this possible bias we performed a simulation study with the \textit{gtobssim} tool which is part of the {\it ScienceTools}.  This tool allows observations to be simulated using the instrument response functions and the real orbit/attitude parameters. Both instrumental and diffuse backgrounds were modeled on the basis of the real backgrounds observed by the LAT. 
\begin{enumerate}
\item Samples of sources (100 FSRQs and 100  BL Lacs) with random positions in the $|b|>$10$^\circ$ sky were simulated.
\item  The real spacecraft orbit and attitude  profiles spanning  94 days  starting from Aug 4 2008 were used.
\item The sources were assumed to have a power-law energy distributions. The photon index was drawn from a gaussian distribution with (mean, sigma)= (2.0,0.3) for BL Lacs and (2.3,0.3) for FSRQs.  These distributions are referred to as "input" probability distribution functions (pdfs).
\item Fluxes were generated  according to a lognormal distribution
$f(x)=\frac{1}{x \sigma \sqrt{2 \pi}} \exp{\frac{-(\ln{x}-\mu)^2}{2\sigma^2}} $ with $\mu=\ln{10^{-7}}$ and $\sigma=0.4$
\item A likelihood analysis was performed for all sources. The pdfs of the
spectral indices and fluxes were built for sources with TS$>$100 (``like'' pdfs). The TS cut was also applied to the "input" pdfs.
\end{enumerate}

Possible bias arising from the likelihood analysis as well as the robustness of the separation between BL Lac and FSRQ ``like `` pdfs were studied by means of  KS tests. "Input" and "like" pdfs were found to be consistent with a probability of $~99.5\%$, $~88.4\%$ for BLLacs and FSRQs respectively, excluding any sizeable bias coming from the likelihood analysis. The TS cut was observed to only affect the distribution tails. Concerning the separation between BL Lacs and FSRQs, the  KS test returned that the probability for the two distributions to result from the same parent distribution is $~7\times10^{-7}$. 

\section{Sources already detected by {\it EGRET}}

After an elapsed time of about 10 years, it is interesting to look at the fraction of the AGNs that were active in the {\it EGRET} era and are detected again by the LAT with a comparable flux. Out of  116 sources in the {\it Fermi}-LAT sample, 3 sources have positions compatible with sources in the Third {\it EGRET} Catalog. Two additional sources, 0FGL
J1802.6$-$3939 and  0FGL J1833.4$-$2106   located at $|$b$|<$10$^\circ$ fulfills this condition as well. The 35 sources are listed in Table 4, along with the mean fluxes and photon indices measured by the {\it Fermi}-LAT and {\it EGRET} as well as the AGN class. These 35 AGNs are composed of  20 FSRQs, 11 BL Lacs, 3 of uncertain type and  1 AGN (Cen A).  The BL Lacs are again overrepresented (with a fraction of 31\%) as compared to the 1st year sky survey {\it EGRET} sample (14 out of 60, i.e. 23\%). The (non-simultaneous) fluxes and indices measured by both intruments are compared in Fig. \ref{fig:fermi_egret}. The large scatter observed when comparing  the fluxes (Fig. \ref{fig:fermi_egret} left) can be expected from the variable nature of the blazar emission. The  scatter observed when comparing the photon indices is more moderate, as could be expected from the fairly strong correlation between photon index and blazar class mentioned above.  For many sources, and most especially for BL Lacs, the indices are measured by the {\it Fermi}-LAT with a much better accuracy.

\section{Radio gamma-ray connection}

With 116/125 high $|b|$, non-pulsar LAT bright sources associated with radio
sources in the CRATES/CGRaBS and the {\it Roma}-BZCAT lists, we confirm the findings
of the 3EG catalog. In particular, 98/106 ($\sim92\%$) of our high confidence
associations have flux density above 100 mJy at 8.4 GHz. In terms of the radio
luminosity $L_r=\nu L(\nu)$, the sources in the present sample with a measured
redshift span the range $10^{39.09} < L_r < 10^{45.33}$ erg\,s$^{-1}$. As shown
by the histogram in Fig.\ \ref{fig:radiopower}, BL Lacs and FSRQ are not
uniformly distributed in this interval, with the former on average at lower
radio luminosities (Log\,$L_{r,\,\mathrm{BL\,Lacs}}=42.8\pm1.1$
[erg\,s$^{-1}$]) than the latter (Log\,$L_{r,\,\mathrm{FSRQ}}=44.4\pm0.6$
[erg\,s$^{-1}$]). Blazars of uncertain type generally lack a redshift. Of the
two radio galaxies associated with objects in the LBAS, NGC\,1275 is similar to
BL Lacs ($L_r=10^{42.21}$ erg\,s$^{-1}$), while Cen\,A lies at the very lower
end of the radio power distribution, with $L_r=10^{39.09}$ erg\,s$^{-1}$.

Cen\,A, the source associated with 0FGL J1325.4$-$4303, is also the only source
showing a significant amount of extended radio emission at low frequency
($S_{8.4}/S_\mathrm{low}=0.005$). For all other sources with a low frequency
(typically, 365 MHz from the Texas survey, 325 MHz from the WENSS, or 408 MHz
from the B2) and a high frequency, high resolution (typically at 8.4 GHz from
CRATES) flux density measurement, we find little or no evidence of significant
deviation from $L_\mathrm{low} = L_\mathrm{8.4}$. Therefore, we find not only
that all the sources in our sample are radio emitters, but that they also
possess compact cores with flat radio spectral index and much higher luminosity
than those of radio galaxies of similar or larger power \citep{giovannini88}.

Thanks to the comparatively large number of LBAS sources, it is worthwhile to
perform a statistical comparison of their properties in the gamma-ray and radio
bands. Previous studies based on EGRET data for 38 extragalactic point sources
have been reported \citep{muecke97}, which did not support claims of
correlations between radio and gamma-ray luminosities. In particular, the
analysis of possible correlations needs to be treated with care, because of the
many biases that can arise, e.g.\ from the common redshift dependence when one
considers luminosities, or from the reduced dynamical range when one considers
mean flux densities, just to name a few.

We have therefore looked at several possible pairs of observables, and we
summarize our results in Table \ref{tab:radiogamma}. In general, we apply the
K-correction to the luminosities but not to the fluxes, since this would
introduce a bias for the sources without a known redshift. We show in
Fig.\ \ref{fig:radiogamma} (left panel) the peak gamma-ray flux
$S_{E>100\,\mathrm{MeV}}$ vs the radio flux density $S_{8.4\,\mathrm{GHz}}$
from CRATES (or NED, in the few cases in which the source is not in the CRATES
list). In general, BL Lacs tend to populate the low flux region, and FSRQs the
high flux region.  Such a constellation is prone to create correlations
artificially from purely combining both populations. Given their different
redshift distributions, this would be even more apparent in the luminosity
plane. For this reason, it is necessary to consider the two populations
separately (see Table \ref{tab:radiogamma}). Indeed, the results of our
analysis show the significance of a radio-to-gamma-ray connection to be
marginal at most on the basis of the present data, in particular for the
FSRQs. Clearly, there is need for a deeper analysis on an enlarged sample
regarding this issue, including Monte-Carlo simulations, which we defer to a
forthcoming paper.

Finally, we show in the right panel of Fig.\ \ref{fig:radiogamma} the radio
luminosity vs.\ gamma-ray spectral index plane. Thanks to the large LAT energy
range, the separation between BL Lacs and FSRQs is readily seen, showing a trend
of softer spectral indices for more luminous radio sources. Moreover, this plot
seems quite effective at finding sources of a different nature, such as the radio
galaxy Cen\,A, whose gamma-ray index is much softer than that of other low
power radio sources. For instance,
0FGL\,J00174$-$0503 is a FSRQ at $z=0.227$ \citep{cgrabs} with index =
2.71 and radio luminosity $L_r=10^{42.36}$ erg\,s$^{-1}$, which could then be a
rare case of low-energy peak and low radio luminosity blazar. The other source
with large photon index (2.60) and comparatively low radio luminosity
($L_r=10^{43.22}$ erg\,s$^{-1}$) is associated with the peculiar source PMN
J0948+0022.

%%%%%%%%%%%%%%%%%%%%%%%%%%%%%%%%%%%%%%%%%%%%%%%%%%%%%%%%%%%%%%%

\section{Population Studies}

%%%%%%%%%%%%%%%%%%%%%%%%%%%%%%%%%%%%%%%%%%%%%%%%%%%%%%%%%%%%

 As described before, the LBAS includes 57 FSRQs, 42 BL Lac objects,  5 blazars of uncertain type and 2 radio galaxies. Ten other sources have lower confidence associations with known blazars. This sample is already comparable with that provided by EGRET and can be used to derive some {\it early} results about the redshift and source count distributions and the luminosity function of blazars.

\subsection{Redshifts}

Fig. \ref{fig:z_dist_fsrq} and Fig. \ref{fig:z_dist_bllac}  display the redshift distributions for FSRQs and BL Lac objects, respectively, and their comparison with those of the parent distributions in the  BZCat catalog. Please note that 12 of 42 BL Lacs have no measured redshifts.   BL Lac objects are generally found at low, $z\lesssim 0.5$,  redshift, whereas the peak of the FSRQ redshift distribution is  around $z\cong 1$. Similar distributions were observed for the {\it EGRET} blazars \citep{mukherjee97}. In the future, as fainter sources become visible, detection of  additional nearby radio-galaxies will enhance the number of very low redshift objects in the AGN redshift distributions measured with the {\it Fermi} LAT. 

Fig. \ref{fig:LumRed}  shows the luminosities of the detected sources plotted as a function of their redshifts. The isotropic gamma-ray luminosity $L_{\gamma}$ was derived using:
\begin{equation}
L_\gamma = 4\pi S d_L^2 / (1+z)^{1-\alpha}\,.
\end{equation}
Here, $S$ is the $\gamma$-ray energy flux  (E $>$ 100 MeV),  $\alpha$ is the energy index  and $d_L$ is the luminosity distance.  A beaming  factor $\delta=1$ was assumed.
The solid curve corresponds to a flux limit of \hfill {\mbox F$_{100}$ = $ 4\times 10^{-8}$ ph cm$^{-2}$s$^{-1}$}.

\subsection{log N - log S}

\subsubsection{Monte Carlo Simulations}

%\subsection{Monte Carlo Simulations}

Proper population studies must rely on a thorough understanding
of the properties of the survey where these objects have been
detected. In order to properly estimate the source-detection efficiency
and biases, we performed detailed Monte Carlo simulations.
The method we adopted is the 
one developed for ROSAT analysis \citep{hasinger93} and lately used \citep{cappelluti07}
for the analysis of the XMM-COSMOS data.
For each source population (blazars, FSRQs and BL Lacs) we created a set
of $>$20 LAT all-sky images with background patterns resembling as close 
as possible the observed ones. The simulations were performed
using  a similar method as that described in section 4.3. An extragalactic  population of pointlike sources 
was added to each simulated observation.  
The coordinates of each source were randomly drawn in order
to produce an isotropic distribution on the sky. Source fluxes were
randomly drawn from a standard log $N$--log $S$ distribution with parameters
similar to the one observed by LAT (see next section). Even though the method
we adopt to derive the survey sensitivity does not depend on the 
normalization or the slope of the input log $N$--log $S$, using the real
distribution allows us to produce simulated observations which closely resemble the
real LAT sky. The photon index of each source was also drawn from
a Gaussian distribution with observed mean and 1\,$\sigma$ width consistent with
 the 
real population. Thus,  for the three simulation sets we adopted the following
photon indices similar to the measured ones:
\begin{itemize}
\item{2.24$\pm$0.25 for the total blazar population;}
\item{2.41$\pm$0.17 for the FSRQ population;}
\item{1.98$\pm$0.22  for the BL Lac population.}
\end{itemize}

More than 30000 sources were simulated for each population. The mock
observations were processed applying the same filtering criteria used for 
real in-flight data. Source detection was performed on E$>$200\,MeV photons
with a simplified version of the detection algorithm\footnotemark{}.
\footnotetext{The complexity of the official detection algorithm makes it
virtually impossible to apply it to a large number of data sets. 
We tested on real data that, for the scope of this investigation,
our simplified detection algorithm 
produces results consistent with more elaborate ones.}
For every pair of input-output sources, we computed the quantity:
\begin{equation}
R^2 = \left( \frac{x-x_0}{\sigma_x} \right)^2 + 
\left( \frac{y-y_0}{\sigma_y} \right)^2 + 
\left( \frac{S-S_0}{\sigma_S} \right)^2
\end{equation}
where $x$, $y$ and $S$ are the source coordinates and flux 
of the detected sources while $x_0$, $y_0$ and $S_0$ are the corresponding
values of the input sources and  $\sigma_x$, $\sigma_y$, $\sigma_S$ the associated statistical uncertainties. We then flagged those with the minimum value of R$^2$ as the  most likely associations. Only  sources at $|$b$|$$\geq10^{\circ}$ are retained.

The goal of these simulations is to derive the probability of detecting
a source (with given mean properties, e.g. photon index and flux)
in the LAT survey as a function of source flux. This can be computed
from the simulations reported above 
as the ratio between the number of  detected and input sources 
in a given flux bin.
The detection efficiencies for the three source populations are reported in
Fig.~\ref{fig:skycov}. A few things can be noted readily. First, 
the bias of the LAT survey against soft sources (i.e. FSRQs) is apparent. Second, the 
LAT $|$b$|\geq$10$^{\circ}$   survey  becomes {\it complete} for
F($>$100\,MeV)$\geq$2$\times10^{-7}$\,ph cm$^{-2}$ s$^{-1}$   
irrespective of the source photon index or its location in the sky.
Multiplying these functions 
by the  solid angle $\Omega$ of the survey (34089.45\,deg$^{2}$
in case of  a $|$b$|$$\geq10^{\circ}$ cut) yields the so called sky coverage
which is used for the statistical studies reported in the next sections.

%%%%%%%%%%%%%%%%%%%%%%%%%%%%%%%%%%%%%%%%%%%%%%%%%%%%%%%%%%%%
%%%%%%%%%%%%%%%%%%%%%%%%%%%%%%%%%%%%%%%%%%%%%%%%%%%%%%%%%%%%
\subsubsection{Incompleteness of the Extragalactic Sample} 

We report in Table ~\ref{tab:sampleb10} the composition of the 
$|$b$|\geq$10$^{\circ}$ sample.
The number of sources with high-confidence
associations is 106. Of these 57 are FSRQs and 42 are BL Lacs.
As already shown in the previous
sections, FSRQs and BL Lacs are represented in almost equal fractions
in the LAT survey.
The 5 blazars with uncertain classifications are likely
split between these two categories as the redshift-luminosity plane 
(Fig.~\ref{fig:LumRed}) shows. 
The incompleteness factor varies as a function of the sample under
study. When considering the non-pulsar part of the high-confidence
sample, the incompleteness is given by low confidence and unassociated
objects. This turns out to be  $\sim$11\,\%. %(15/132)
However, when considering the FSRQ and BL Lac samples separately  one must
also include the sources with uncertain classifications.
Thus the incompleteness factor of the FSRQ and BL Lac samples 
rises to $\sim$15\,\%.
A reasonable and simple hypothesis is one which assumes 
that these sources reflect the composition of the identified portion of the
sample. This would mean that there are an additional $\sim$9 FSRQs and $\sim$7 BL Lacs are hiding among
the unidentified/unassociated/low-confidence sources. 
These uncertainties will be used  in the next sections.

Since the uncertainty due to the incompleteness is relatively large, we will
use a {\it flux-limited} sample to verify the results derived from the main
sample. Indeed, for F$_{100}\geq1.25\times 10^{-7}$\,ph cm$^{-2}$ s$^{-1}$
the number of uncertain, unassociated, and 
low-confidence sources falls to 2 (and 2 are anti-associated).
Above this flux limit, the sample contains 44 sources of which 29 and 9
are FSRQs and BL Lacs respectively,  while 2 are Radio galaxies.
Moreover, all but one BL Lac have a  measured redshift. Thus,
while low numbers penalize this {\it flux-limited} sample, its incompleteness is $<$5\,\%.

%%%%%%%%%%%%%%%%%%%%%%%%%%%%%%%%%%%%%%%%%%%%%%%%%%%%%%%%%%%%%%%
%%%%%%%%%%%%%%%%%%%%%%%%%%%%%%%%%%%%%%%%%%%%%%%%%%%%%%%%%%%%%%%
\subsubsection{Source Counts Distributions}

The source counts distribution, also known as the log $N$--log $S$, flux, or
size distribution,
is readily computed once the sky coverage is known through the expression:
\begin{equation}
N(>S) = \sum_{i=1}^{N_S} \frac{1}{\Omega_i} {\rm deg^{-2}},
\end{equation}
where $N_S$ is the total number of detected sources with fluxes
greater than $S$, and $\Omega_i$ (i.e. Fig.~\ref{fig:skycov} 
multiplied by the solid angle) is solid angle associated
with the flux of the $ith$ source. The variance of the source number
counts is defined as
\begin{equation}
\sigma_i^2 = \sum_{i=1}^{N_S} \left( \frac{1}{\Omega_i} \right)^2.
\end{equation}
In building the source counts distributions, we used the source
flux averaged over the three month timescale. The log $N$--log $S$
of the entire extragalactic sample (excluding pulsars) is shown
in Fig.~\ref{fig:logn_all}.

 We fitted the source counts distribution
with  a power-law model of the type:
\begin{equation}
\frac{dN}{dS}= n(S)= A\left( \frac{S}{10^{-7}} \right)^{-\alpha}.
\end{equation}
A common practice in this case \citep[e.g., see][]{ajello08} is to fit
the unbinned dataset employing a maximum likelihood (ML) algorithm.
For this purpose the ML estimator can be written as
\begin{equation}
\mathcal{L}= -2\sum_i {\rm ln} \frac{ n(S_i) \Omega(S_i)}{\int n(S) \Omega(S) dS}\;,
\label{eq:ml}
\end{equation}
where $i$ runs over the detected sources. 
The 1\,$\sigma$ error associated to the fitted parameters 
(in this case $\alpha$) is   computed by varying the parameter
of interest, while the others are allowed to float, until an increment
of  $\Delta \mathcal{L}$=1 is achieved. This gives an estimate of the 68\,\%
confidence region for the parameter of interest \citep{avni76}.
In this formulation of the ML 
function, the normalization $A$ is not a parameter of the problem.
Once the slope $\alpha$ is determined, the normalization is derived 
as the value which reproduces the number of observed sources.
An estimate of its statistical error is given by the Poisson error
of sources used to build the log $N$--log $S$.

Since the sky coverage
is somewhat uncertain at very low fluxes, the fit is performed above
F($>$100 MeV)$=7\times10^{-8}$\,ph cm$^{-2}$ s$^{-1}$ even though all the data are displayed. The results of the
best fits to the different source counts distributions are summarized in
Table ~\ref{tab:logn}. It is clear that all distributions are compatible,
within their errors, with a Euclidean distribution ($\alpha = 2.5$).
In order to check the stability of our results we have shifted
the sky coverage of Fig.~\ref{fig:skycov} by 20\,\% on either side.
Taking the whole extragalactic population as an example (see first
line of Table \ref{tab:logn}) we get that the best fit values
of the slope are 2.47 and 2.62 for the $-20$\,\% and +20\,\% case respectively.
These values are consistent within the error (e.g. 2.59$\pm0.12$), showing
that at bright fluxes our analysis does not suffer from major systematic
uncertainties in the sky coverage. The same result holds for the
other log $N$--log $S$ distributions reported in   Table \ref{tab:logn}.

The log $N$-- log$S$ distributions for FSRQs and BL Lacs are shown
in Fig.~\ref{fig:logn_fsrq} and \ref{fig:logn_bllac}. We do not 
find any indication of a break in the source counts distributions of the
two populations. As the fitting results of Table \ref{tab:logn} show,
there might be an intrinsic difference between the log $N$--log $S$
of both populations, with the source counts distribution of 
BL Lacs being flatter than that of
of FSRQs. However, both of them are compatible within
1\,$\sigma$ errors with the Euclidean value of 2.5.
Moreover, the analysis of the {\it flux-limited} sample 
(see bottom part of Table \ref{tab:logn}) confirms the 
results of the main sample, showing that incompleteness
is not a main issue in this study.

For the {\it EGRET} sample, a surface density
for F$_{100}\geq10^{-7}$\,ph cm$^{-2}$ s$^{-1}$
of FSRQs and BL Lacs of 3.31\,sr$^{-1}$ and 0.83\,sr$^{-1}$, respectively,
 is reported \citep{mp00}. From LAT we derive that the surface density 
(above F$_{100}\geq10^{-7}$\,ph cm$^{-2}$ s$^{-1}$) 
of FSRQs and BL Lacs is 
4.41$\pm0.72$\,sr$^{-1}$ and 1.01$\pm0.17$\,sr$^{-1}$ respectively.
Thus the LAT  results are in good agreement with {\it EGRET}.

A measurement of the number fluence using the average three-month fluxes
of bright {\it Fermi} blazars of different classes is
readily obtained from the log $N$--log $S$ distributions through the expression:
\begin{equation}
F_{diffuse} = \int_{f_{min}}^{f_{max}} \frac{dN}{dS}\ S\ dS \;.
\label{eq:diffuse}
\end{equation}
Unless otherwise stated, we adopt a value for $f_{min}$ of 
4$\times10^{-8}$\,ph cm$^{-2}$ s$^{-1}$.
To compare with the {\it EGRET} results, the upper
limit of integration cannot be set to infinity.  Indeed,
all point sources detected above F$_{100}\sim10^{-7}$\,ph cm$^{-2}$ s$^{-1}$
in the Second EGRET Catalog \citep[2EG;][]{thompson95}
were subtracted in the  measurement of the extragalactic diffuse
$\gamma$-ray background  (EDGB)  \citep{sreekumar98}. Thus, we set $f_{max}$ to $10^{-7}$\,ph cm$^{-2}$ s$^{-1}$.
The integral in Eq.~\ref{eq:diffuse} yields a total flux of 
1.06($\pm0.09$)$\times10^{-6}$\,ph cm$^{-2}$ s$^{-1}$ sr$^{-1}$
This can be compared with the intensity of the EDGB, as measured by {\it EGRET},
of 1.45$\times10^{-5}$\,ph cm$^{-2}$ s$^{-1}$ sr$^{-1}$.
Already in this small flux range, LAT
is resolving into pointlike sources $\sim$7\,\% of the {\it EGRET}  EDGB.
Preliminary analysis of the log $N$--log $S$ distributions shows that LAT is 
expected to resolve a much larger fraction of the EDGB within the 
next few months of observation.

%%%%%%%%%%%%%%%%%%%%%%%%%%%%%%%%%%%%%%%%%%%%%%%%%%%%%%%%%%%%%
%%%%%%%%%%%%%%%%%%%%%%%%%%%%%%%%%%%%%%%%%%%%%%%%%%%%%%%%%%%%%
\subsection{Evolution of Blazars}

%%%%%%%%%%%%%%%%%%%%%%%%%%%%%%%%%%%%%%%%%%%%%%%%%%%%%%%%%%%%%
\subsubsection{Evolutionary Test}

A simple and robust test of evolution is the 
 $V/V_{\rm MAX}$ test \cite{schmidt68}.
The quantity $V/V_{\rm MAX}$  is the ratio between
the (comoving) volume within which the source has been detected and 
the maximum comoving volume available for its detection.
For a given source, $V/V_{\rm MAX}$ is expected to be uniformly distributed between 0 and 1.
For a population uniformly distributed in Euclidean
space (and with constant properties with $z$)
and non-evolving,  the average $V/V_{\rm MAX}$
should be consistent with a value of 0.5. The error on the average 
value is $\sigma=1/(12N)^{1/2}$ for $N$ sources.
A value of  $<$$V/V_{\rm MAX}$$>$ $>0.5$ 
indicates positive evolution (more sources or brighter sources at earlier
times), and the opposite indicates negative evolution.

The comoving volume for a {\it ith} source is given by
\begin{equation}
V = \int_{z=0}^{z=z_{i}} \frac{dV}{dz}\ \Omega(L_i,z) dz,
\label{eq:vvmax}
\end{equation}
where $dV/dz$ is the comoving volume element per unit redshift and 
unit solid angle \citep[see e.g.][]{hogg99} and $\Omega(L_i,z)$
is the aforementioned sky coverage for the source with 
rest-frame luminosity $L_i$ at redshift $z$. We note
that the definition of the $V/V_{\rm MAX}$ reported in Eq.~\ref{eq:vvmax}
encompasses also the definition of the $V_e/V_a$ test
\citep{ab80}, which for the purposes here are formally equivalent.

We computed the average $<$$V/V_{\rm MAX}$$>$ for FSRQs, BL Lacs and 
all sources in the high-confidence sample with measured redshift 
(these includes the sources with uncertain classification).
The results are summarized in Table \ref{tab:vvmax}.
All 57 FSRQs present in the extragalactic sample 
(see Table \ref{tab:sampleb10}) have a measured redshift. The 
$V/V_{\rm MAX}$ shows that the population of FSRQs detected by LAT
evolves positively (i.e. there were more FSRQs in the past 
or they were more luminous) at the 3\,$\sigma$ level.
This result is also confirmed by the analysis of the 29 FSRQs which
constitute a {\it flux-limited} sample 
(see lower part of  Table \ref{tab:sampleb10}).

Only 31 out of the 42 BL Lac objects have a measured redshift.
The $V/V_{\rm MAX}$ test
is compatible within $\sim1\sigma$  with no evolution.
Assigning the mean  redshift value of the BL Lac sample (i.e. $\langle z\rangle$=0.38) to those objects  without a
without a redshift  
produces a value of $\langle V/V_{\rm MAX}\rangle$=0.472$\pm0.046$.
The result does not change if the redshift is drawn from 
a Gaussian distribution with mean and dispersion consistent
with the observed redshift distribution  of BL Lacs.
However, it is difficult to assess  the validity of 
both these hypotheses.
Indeed, the fact that these objects
show a featureless continuum might suggest that their redshifts could
be the largest in the sample \citep{padovani07}. 
In this case, their true redshift would produce
a larger value of the $V/V_{\rm MAX}$ statistic.
The  $V/V_{\rm MAX}$ of all the objects
with a measured redshift in the high-confidence sample is compatible
with no evolution.

%%%%%%%%%%%%%%%%%%%%%%%%%%%%%%%%%%%%%%%%%%%%%%%%%%%%%%
%%%%%%%%%%%%%%%%%%%%%%%%%%%%%%%%%%%%%%%%%%%%%%%%%%%%%%
\subsubsection{Luminosity Function of FSRQs}

We estimate the gamma-ray luminosity function (GLF) in fixed
redshift bins using the $1/V_{MAX}$ method (equivalent in our
formalism to the $1/V_{a}$ method). For each bin of redshift the
GLF can be expressed as
\begin{equation}
\Phi(L_{\gamma},z) = \frac{dN}{dL_{\gamma}} = \frac{1}{\Delta\ L_{\gamma}}\
\sum_{i=1}^{N} \frac{1}{V_{MAX,i}}
\end{equation}
where $V_{MAX,i}$ is the maximum comoving volume associated with the 
$i_{th}$ source (see Eq.~\ref{eq:vvmax}).
The cumulative and differential
 luminosity functions  of FSRQs, in three redshift bins, are
reported in Fig.~\ref{fig:glf_fsrq}. One thing is readily apparent 
from this figure. FSRQs are strongly evolving. A non-evolving population
would have GLFs which are continuous across different redshift bins. 
In the case of FSRQs we note a change in space
density (or luminosity) with redshift. Also, one can see that
the space density of intermediate-luminosity FSRQs 
(e.g. L$_{\gamma}\sim$10$^{47}$\,erg s$^{-1}$) is increasing with redshift. On the
other hand, the most luminous FSRQs have an almost 
constant  space density with redshift. This might be a sign of a cut-off
in the evolution of FSRQs. 
A decline in the space density of luminous FSRQs has also been determined
at radio and X-ray energies \citep[e.g.,][]{wall05,padovani07,ajello09}.
We derive from the GLF that the space density of FSRQs with
L$_{\gamma}>7\times10^{45}$\,erg s$^{-1}$ is 1.05$\pm0.13$\,Gpc$^{-3}$.

We made a Maximum Likelihood fit to the three unbinned datasets
using a simple GLF model defined as 
\begin{equation}
\Phi(L_{\gamma},z)\propto L_{\gamma}^{-\beta}\;.
\end{equation}
The ML estimator can be expressed similarly to
Eq.~\ref{eq:ml} by the expression
\begin{equation}
\mathcal{L}= -2\sum_i {\rm ln} \frac{ \Phi(L_{\gamma,i},z_i) V(L_{\gamma,i},z_i)}
{\int  \Phi(L_{\gamma},z)  V(L_{\gamma},z) dL_{\gamma}}\ .
\end{equation}
The results of the ML fits to the GLF of FSRQs are summarized in
Table \ref{tab:fsrq_fits}. For z $\geq1$,
the GLF can be successfully parametrized by a single power-law model.
The slope is compatible with the canonical value of 2.5--2.8 determined for
X--ray selected samples of  radio-quiet AGNs \citep{ueda03,hasinger05,silverman08}. 
This indicates that at high redshifts the {\it Fermi}-LAT is sampling the bright end
 of the luminosity distribution of FSRQs. For $z\leq 1$,
the best-fit value of the slope $\beta$ is 1.56$\pm0.10$, compatible
with luminosity function slopes found in radio/X-ray 
selected samples \citep{padovani07}. This
is much flatter than the canonical value of $\beta = 2.5$. 
As the cumulative GLF shows
(left panel of Fig.~\ref{fig:glf_fsrq}), there might be a hint of 
a break with respect to a simple power-law model  in the GLF. 
%However,  GLFs built using  1/V$_{MAX}$ method
%are not sensitive to evolution within a bin of redshift. 
%Thus what we see in the first redshift bin might be the
%joint effect of cosmic evolution and of LAT sampling the low-luminosity
%part of the GLF. 
A more detailed analysis, comparing different methods
to derive the GLF and its evolution, will be considered in future publications.

%%%%%%%%%%%%%%%%%%%%%%%%%%%%%%%%%%%%%%%%%%%%%%%%%%%%%%
%%%%%%%%%%%%%%%%%%%%%%%%%%%%%%%%%%%%%%%%%%%%%%%%%%%%%%
\subsubsection{Luminosity Function of BL Lacs}

The luminosity function of BL Lacs, reported in Fig. \ref{fig:glf_bllac},
is in agreement with the results of the $V/V_{\rm MAX}$ test.
Indeed, sub-dividing the entire BL Lac sample in two bins of redshift
produces two GLFs which connect smoothly to each other.
A simple power-law GLF describes the entire dataset well. The GLF
slope is $\beta = 2.17 \pm0.05$ and is well in agreement with 
the value of 2.12$\pm0.16$ reported  for 
a radio/X--ray selected sample of BL Lacs \citep{padovani07}. 
The GLF of 12 EGRET BL Lac objects in a recent study
\citep{bsm08} was found to show no significant evidence for
evolution, with a GLF slope $\beta  = 2.37 \pm0.3$.
Past claims \citep[e.g.,][]{rector00,beckmann03} 
of negative evolution of BL Lac objects, selected
mainly in the X--ray band, are not confirmed by our data.
The dynamical range of the LAT GLF samples 4 decades
in luminosity and nearly 8 in space density. From our  GLF we derive  that the density of BL Lac objects with
L$_{\gamma}>$3$\times10^{44}$\,erg s$^{-1}$  
is 1.9($\pm0.4$)$\times 10^{-7}$\,Mpc$^{-3}$.

Above a  luminosity of L$_{\gamma}\sim 10^{47}$erg s$^{-1}$,
the cumulative density of BL Lacs and FSRQs is comparable, with
BL Lacs being $\sim3$ times less numerous than FSRQs. However,
given the fact that they reach lower luminosities, BL Lacs
are $\sim 200$ times more abundant than FSRQs above their respective
limiting luminosities.

%%%%%%%%%%%%%%%%%%%%%%%%%%%%%%%%%%%%%%%%%%%%%%%%%%%%%%%%%%%%%%

\section{Discussion}

The value TS $ >100$ defining the detection significance for bright sources corresponds to $\gtrsim 10\,\sigma$ significance, or a limiting flux over the entire high-latitude sky of $\approx$ (3 -- $10)\times 10^{-8}$ ph($>100$ MeV) cm$^{-2}$ s$^{-1}$ during the three-month sky survey. In comparison, {\it EGRET} reached a $5\sigma$ high-confidence on-axis flux limit of $\approx 15\times 10^{-8}$ ph($>100$ MeV) cm$^{-2}$ s$^{-1}$ for a two-week pointing over $\approx 0.5$ sr of the sky, only becoming complete at  
$S$
$F_{100}$$\approx 25\times 10^{-8}$ ph
(%$>100$ MeV) 
cm$^{-2}$ s$^{-1}$ \citep{dermer07}.  Of the 66 high-confidence and 27 lower-confidence AGN associations in the 3EG catalog \citep{3EGcatalog}, 32 sources in the {\it Fermi}-LAT sample  were also detected with {\it EGRET}. An additional source is detected at $|$b$|$$<$ 10$^\circ$. Many of the other high-confidence {\it EGRET} sources are detected with {\it Fermi}-LAT at $TS < 100$, reflecting the rapid variability and periods of activity of $\gamma$-ray blazars on timescales of years or longer.

During the 18-month {\it EGRET} all-sky survey when exposure to all parts of the sky was relatively uniform compared to the remainder of the mission, 60 high-confidence blazars consisting of 14 BL Lacs and 46 FSRQs were found \citep{fichtel94}. Compared with  $\approx 23$\% of {\it EGRET} blazars being BL Lac objects, nearly $40$\% of the {\it Fermi}-LAT blazars are BL Lac objects. The larger fraction of BL Lac objects in the {\it Fermi} bright AGN sample is partly a consequence of the good sensitivity to high-energy emission by {\it Fermi}-LAT, whereas self-vetoing in {\it EGRET} reduced its effective area to photons  with energies $\gtrsim 5$ GeV\citep{thompson93}. Consequently, dim hard-spectrum sources are favored to be detected with the {\it Fermi}-LAT compared to {\it EGRET}. 

A clear separation between the spectral indices of FSRQs and BL Lacs is found in the {\it Fermi}-LAT data (Fig. \ref{fig:sp_dist}), with mean photon indices of $\Gamma = 2.40\pm 0.17$ (rms)  for FSRQs and $\Gamma = 1.99\pm 0.22$ (rms)  for BL Lac objects. A KS test gives a probability of 2 $\times$ 10$^{-12}$ for the two index samples to be drawn from the same parent distribution.  Moreover, the SEDs of bright flaring blazars in the cases of  3C 454.3 and AO 0235+164 show a spectral softening at $E \gtrsim 2$ GeV. If this behavior persists in weaker FSRQs, then an even greater fraction of BL Lac objects will be found in {\it Fermi}-LAT analyses over longer times, because signal-to-noise detection significance for weak hard-spectrum sources becomes better than for weak soft-spectrum sources due to the reduced background at higher photon energies. 

Another reason for the  larger fraction of BL Lac objects in the {\it Fermi}-LAT blazars could be related to the redshift distribution of the bright AGNs. The BL Lac objects are  dominated by low-redshift, $z\lesssim 0.5$ blazars, with a tail extending to $z \approx 1$, 
whereas the FSRQs have a broad distribution peaking at $z \approx 1$ and extending to $z \approx 3$ (see Fig.\ \ref{fig:z_dist_fsrq} and Fig. \ref{fig:z_dist_bllac}). These distributions are similar to the distribution of {\it EGRET} blazars \citep{mukherjee97}.  Because the peak of the {\it EGRET} FSRQ redshift distribution is already at $z \approx 1$, detection of higher redshift FSRQs with the more sensitive {\it Fermi}-LAT would be impeded by cosmological factors that strongly reduce the received fluxes. Moreover, the period of dominant AGN activity was probably at $z \approx 1$ or 2. The increased  sensitivity for the BL Lac objects with {\it Fermi}-LAT, on the other hand, allows it to probe beyond the low-redshift population of BL Lac objects detected with {\it EGRET} where the detectable volume is still rapidly increasing with $z$. The likelihood of detecting $z \approx 1$ BL Lac objects does depend, however, on their evolution.

The simplest index of population evolution is the $V/V_{MAX}$ test. We found $\langle V/V_{MAX}\rangle = 0.43\pm 0.055$ for the BL Lac objects with redshift in the LBAS (Table 8), so that the BL Lac objects are within $\approx 1 \sigma$ of showing no evidence for evolution. For the FSRQs in the LBAS, by contrast, we found $\langle V/V_{MAX}\rangle = 0.64\pm 0.04$, so that the FSRQs exhibit strong positive evolution. The strong positive evolution of FSRQs and weakly negative or no evolution of BL Lac objects in the LBAS is contrary, however, to our reasoning that population evolution of the lower redshift BL Lac objects explains the larger fraction of BL Lacs in the LBAS compared with the BL Lac fraction observed with EGRET.  As indicated by the indices of the $\log N$ -- $\log S$ (eq.\ 6 and Table 7), which show much weaker evidence for evolution than given by the $V/V_{MAX}$ test, the actual situation may be more complicated and depend on both density and luminosity evolution. 

The BL Lac objects are found to display systematically harder spectra, with $\nu F_\nu$ spectra rising at GeV energies, compared to the powerful FSRQs where the peak of the $\nu F_\nu$ spectrum is at photon energies $\lesssim 100$ MeV -- GeV. This is generally attributed to a different dominant radiation process; self-Compton scattering of the jet electron's synchrotron emission in the case of BL Lac objects, and Compton scattering of external radiation fields in the case of FSRQs if leptonic processes dominate the radiation output \citep[recently reviewed in][]{bottcher07}. The excellent sensitivity and full-sky coverage of the {\it Fermi} LAT is, for the first time, giving us  broadband evolving SEDs from the radio to the $\gamma$-ray regime in sources like 3C 454.3, PKS 2155-304,  and others that will require detailed spectral modeling to assess the relative importance of self-Compton and external Compton scattering processes in the different blazar classes.

Such results will be important to determine whether FSRQs and BL Lac objects may have a direct evolutionary relationship, or instead represent separate unrelated tracks of supermassive black hole fueling and growth. A scenario whereby BL Lac objects are the late stages of FSRQs, as the gas and dust produced in a galaxy merger or tidal interaction fuels the supermassive black hole \citep{bd02,ce02}, provides a framework to understand the blazar phenomenology and makes definite predictions about the relative black hole masses in the two classes. The more abundant scattered radiation and fueling in the evolution from FSRQ to BL Lac object would then lead to a blazar sequence like behavior \citep{fossati98,ghisellini98} if the amount of accreting matter controls black hole power and the surrounding radiation field.

It is still premature to compare the number of blazars in this bright source list with prelaunch predictions \citep{mp00,ss01,nt06,dermer07,it08} made on the basis of differing assumptions, to sensitivities $\approx 5\sigma$ rather than $10\sigma$, and over different spans of time. Nevertheless, nearly complete surveys with far more sources than detected with {\it EGRET} are now available for calculating luminosity and number evolution, with implications that can be compared with results from the {\it EGRET} era.

This study can be used to examine the observational basis for assuming an underlying radio/$\gamma$-ray connection used to calculate the blazar contribution to the $\gamma$-ray background \citep{ss96,giommi06,nt06}. Figure \ref{fig:radiogamma} shows that except for a (at most) weak correlation of the brightest $\gamma$-ray blazars with the most radio-bright blazars, the $\gamma$-ray and radio fluxes display a large amount of scatter. Whether a stronger correlation can be found by comparing mean $\gamma$-ray fluxes with radio fluxes will require further study. But even at this early stage of the {\it Fermi} mission, we find that the bright sources can already comprise about $7$\% of the diffuse extragalactic $\gamma$-ray background flux measured with {\it EGRET} \citep{sreekumar98}.

We conclude this study by noting that the {\it Fermi}-LAT results imply the
non-thermal luminosity density of AGNs on various size scales. 
A $\gamma$-ray blazar makes a contribution to the non-thermal emissivity
$\propto L/V$ in terms of $\gamma$-ray luminosity $L_\gamma$ and 
injection volume $V$ derived from redshift.  The {\it Fermi}-LAT results from Table 2 show that
BL Lac objects provide local emissivities $\ell_{BL} \gtrsim 10^{31}$ W Mpc$^{-3}$, 
whereas FSRQs have $\ell_{FSRQ} \approx 10^{30}$ W Mpc$^{-3}$. Cen A, because of 
its proximity at $d\cong 3.5$ Mpc, dominates the non-thermal luminosity, with
$\ell_{Cen A } \approx 3\times 10^{31}$ W Mpc$^{-3}$ \citep{drfa09}. Sources of 
UHECRs must have a luminosity density within the GZK radius, $\approx 100$ Mpc, of 
$\ell_{UHECR} \approx 3\times 10^{29}$ W Mpc$^{-3}$ or 
$\ell_{UHECR}\approx 10^{44}$ ergs Mpc$^{-3}$ yr$^{-1}$ \citep{wb99}.
To have sufficient emissivity within the GZK radius,
if AGNs are the sources of the UHECRs \citep{pao07},  the {\it Fermi}-LAT results would therefore seem to 
favor BL Lac objects over FSRQs as the source of the UHECRs.

\section{Summary}

We have presented a list of 116 bright, $\gtrsim 10\sigma$ sources at
$|b|\geq 10\degr$ taken from the list of bright sources
\citep{SourceList} observed with the {\it Fermi}-LAT in its initial
three-month observing period extending from August 4 to October 30 of
2008. Of these sources, 106 are associated with blazars with high
confidence and compose the LBAS. The number of low-confidence AGN
associations is 11 (one source having two possible associations - one high and one low confidence). At
$|b|\geq 10\degr$, 5 sources out of a total of 125 non-pulsar sources
remain unidentified. Two of the AGNs are associated with radio galaxies.
The purpose of this work is to present the key properties of the AGN
population of this bright GeV source list. The main results are
summarized as follows:

\begin{enumerate}

\item With a $\sim 90\%$ success rate from correlating the bright
gamma-ray source list with AGN radio catalogs (CRATES/CGRABS, BZCAT) the
bright extragalactic gamma-ray sky continues to be dominated by
radio-bright AGNs.

\item The number of HBLs in the LBAS detected at GeV energies (even when
not flaring) has risen to at least 10 (out of 42 BL Lacs) as compared
to one (out of 14 BL Lacs) detected by {\it EGRET}. Seven LBAS
HBLs are known TeV-blazars.

\item Only $\sim 30\%$ of the bright Fermi AGN list were also detected
by {\it EGRET}. This may be a consequence of the duty cycle and
variability behavior of GeV blazars.

\item BL Lac objects make up almost half of the bright Fermi AGN sample
(consisting of 57 FSRQs, 42 BL Lac objects, 2 radio galaxies, and only 5
AGN remain unclassified), while the BL Lac fraction in the 3EG catalog
was only $\sim 23\%$. This feature most probably arises from the different instrument responses of the LAT and {\it EGRET}.

\item The mean flux distribution of the {\it Fermi} AGN remains similar to the
corresponding one based on the {\it EGRET} sample, while the peak flux
distributions differ appreciably.

\item We find a spectral separation between BL Lacs and FSRQs in the GeV
gamma-ray band with FSRQs having significantly softer
spectra
%\footnote{Spectral indices are derived from a single power-law fit in the range 200~MeV-100~GeV, while the quoted fluxes represent the sum from fitting a power-law spectrum independently in 100-1000 MeV and 1-100 GeV bands.} 
than BL
Lac objects. This confirms earlier indications for the existence of
spectrally distinct populations in the {\it EGRET} blazar sample. The
average photon index is $1.99\pm 0.22$ (rms) for BL Lacs, with a
tendency of HBLs displaying even harder spectra, and $2.40\pm 0.17$ (rms)
for FSRQs.  A KS test gives a probability of 2 $\times$ 10$^{-12}$ for the two index samples to be drawn from the same parent distribution.

\item {\it Fermi} FSRQs in the bright source list are on average more luminous
and more distant than the {\it Fermi}-detected BL Lac objects in that list.
I.e., FSRQs exhibit a broad redshift distribution, starting with
$z=0.158$ (3C~273), peaking at $z\approx 1$ and extending up to
$z\approx 3$ while BL Lacs are mostly found in the $\sim 0.1$ redshift
bin with a tail extending up to $z\approx 1$. No significant relation
between the gamma-ray photon index and redshift is found within each
source class, in agreement with corresponding studies based on the {\it
EGRET} AGN samples.

\item The peak gamma-ray flux is at best only weakly related to the 8.4 GHz
radio flux, with the brightest gamma-ray AGNs having the largest radio
flux densities.

\item Using mean fluxes the Log N-Log S distribution of all the bright
sources (except the
pulsars) appears compatible with
an Euclidean distribution without any breaks. This is also true within
$1\sigma$ for the source counts distributions of the FSRQ and BL Lac
sample separately.
Surface densities of $4.28\pm 0.72$ sr$^{-1}$ and $1.01\pm 0.17$
sr$^{-1}$ ($F_{100}\geq 10^{-7}$ph cm$^{-2}$
s$^{-1}$) for FSRQs and BL Lacs, respectively, are reached.

\item
The combined emission in the flux range $F_{100,\rm mean}\approx
(7-10)\times 10^{-8}$ph cm$^{-2}$ s$^{-1}$ observed from these
individually resolved AGN during this three-month period already
corresponds to $\sim 7\%$ of the {\it EGRET} detected extragalactic
diffuse gamma-ray background.

\item A $V/V_{max}$ analysis shows positive evolution at the 3 $\sigma$
level for the bright {\it Fermi}-detected FSRQs with the most luminous FSRQs
having an almost constant space density with redshift, while for the
{\it Fermi}-detected BL Lacs no evolution within one $\sigma$ is apparent.

\item The gamma-ray luminosity function of bright FSRQs can be described
by a single power-law with index $\sim 2.5$ and $\sim 1.5$ for the high
($\geq 0.9$) and low ($\leq 0.9$) redshift range, respectively, while
the BL Lac gamma-ray luminosity function follows a power law with index
$\sim 2.1$. The space density of gamma-ray emitting BL Lacs of $\sim
190$ Gpc$^{-3}$ above their limiting luminosity, $\sim 3\times
10^{44}$erg s$^{-1}$, is a factor $\sim 200$ larger than for the {\it Fermi}-detected FSRQ population above their limiting luminosity, $\sim 7\times
10^{45}$erg s$^{-1}$. Thus, within the {\it Fermi} bright AGN list BL Lacs are
intrinsically more numerous than FSRQs. Bright Fermi detected BL Lacs and FSRQs display comparable cumulative
number counts above $\sim 10^{47}$erg s$^{-1}$, with BL Lacs being $\sim
3$ times more numerous than FSRQs.
\end{enumerate}

These early results from the first three months of the science mission of the {\it Fermi Gamma ray Space Telescope} demonstrate its exceptional capabilities to provide important new knowledge about $\gamma$-ray emission from active galactic nuclei and blazars. As the {\it Fermi}-LAT data accumulate, many more AGNs at lower flux levels will likely be detected- as well as flaring AGNs at brighter fluxes than yet observed -  helping to refine these results and improve our understanding of supermassive black holes.

\section{Acknowledgments}
\acknowledgments
The \emph{Fermi} LAT Collaboration acknowledges the generous support of a number of agencies and institutes that have supported the ${\it Fermi}$ LAT Collaboration. These include the National Aeronautics and Space Administration and the Department of Energy in the United States, the Commissariat \`a l'Energie Atomique and the Centre National de la Recherche Scientifique / Institut National de Physique Nucl\'eaire et de Physique des Particules in France, the Agenzia Spaziale Italiana and the Istituto Nazionale di Fisica Nucleare in Italy, the Ministry of Education, Culture, Sports, Science and Technology (MEXT), High Energy Accelerator Research Organization (KEK) and Japan Aerospace Exploration Agency (JAXA) in Japan, and the K.\ A.\ Wallenberg Foundation, the Swedish Research Council and the Swedish National Space Board in Sweden. 

Additional support for science analysis during the operations phase from the following agencies is also gratefully acknowledged: the Istituto Nazionale di Astrofisica in Italy and the K.~A. Wallenberg Foundation in Sweden for providing a grant in support of a Royal Swedish Academy of Sciences Research fellowship for JC. 

MA acknowledges N. Cappelluti for extensive discussion about the sky coverage.

{\it Facilities:} \facility{{\it Fermi} LAT}.

\bibliography{Fermi_detected_blazars.bib}

\clearpage
\begin{figure}
\centering
\resizebox{10cm}{!}{\rotatebox[]{0}{\includegraphics{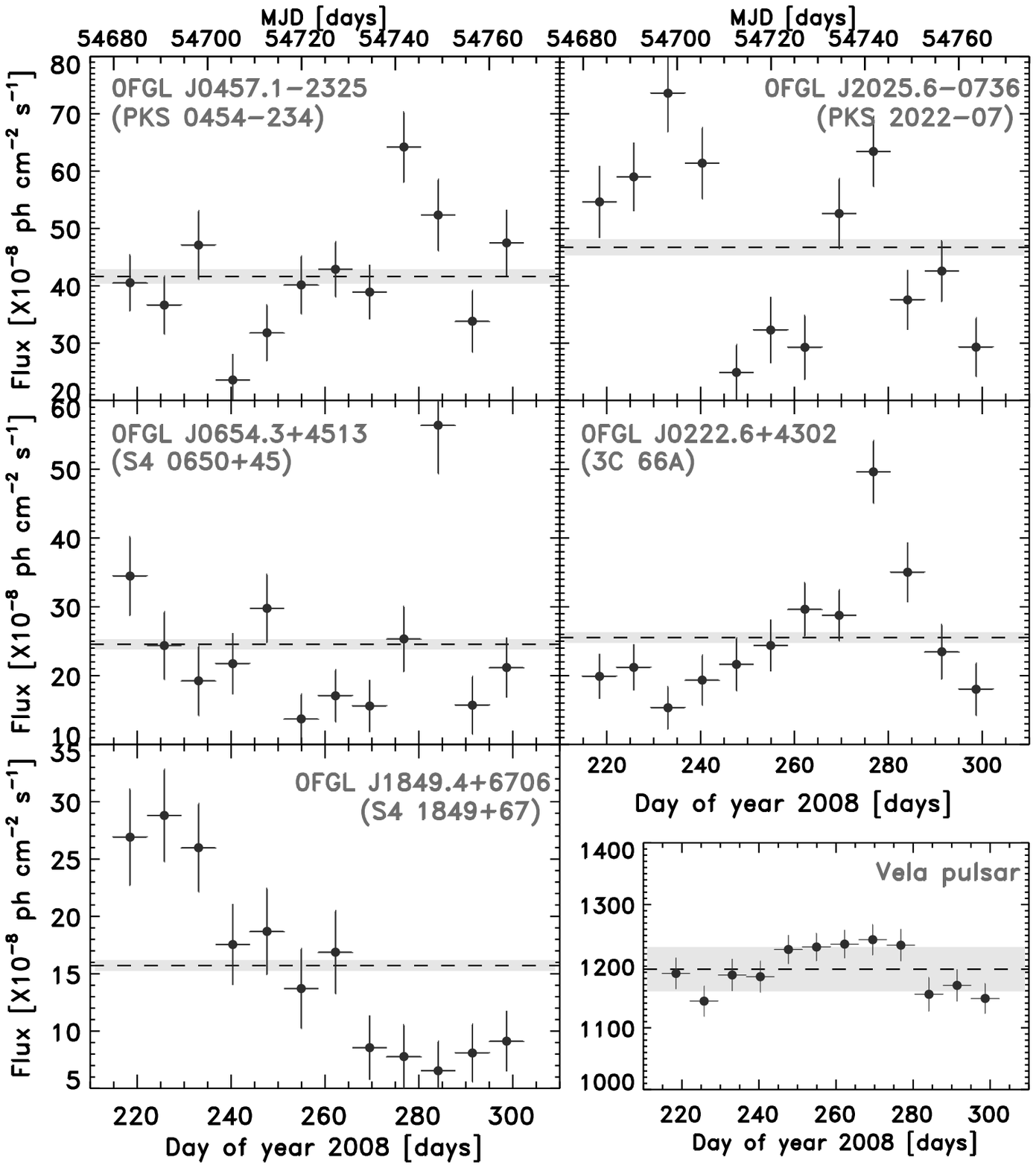}}}
\caption{Examples of weekly light curves for five bright blazars detected by {\it Fermi}-LAT and
the Vela light curve for comparison (flux unit: $\times 10^{-8}$ photons cm$^{-2}$
s${-1}$, please note the different scales). The dashed line is the average value and the grey area
shows the 3\% systematic error we have adopted. Different flux variability amplitudes and timescales are clearly visible in the blazar light curves. }
\label{fig:lcurves}
\end{figure} 
% Figures

\begin{figure}
\epsscale{.80}
\plotone{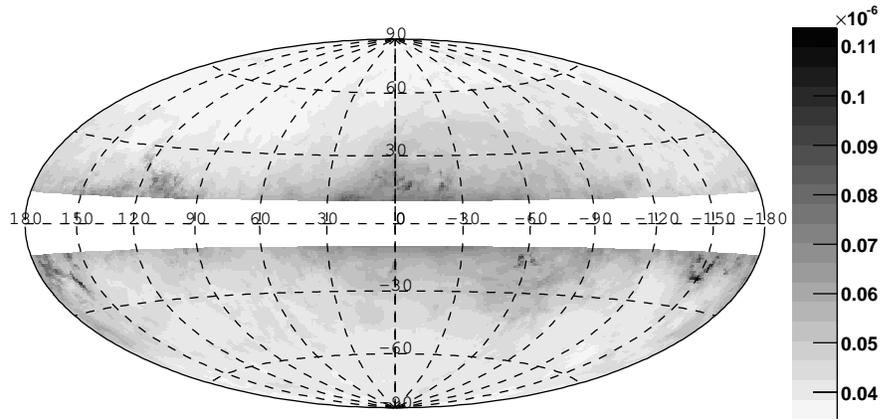}
\caption{ Flux limit [E$>$100 MeV] (ph cm$^{-2}$s$^{-1}$) as a function of sky location (in galactic coordinates), for a photon index=2.2} 
\label{fig:coverage}
\end{figure}

\begin{figure}
\epsscale{1.}
\plotone{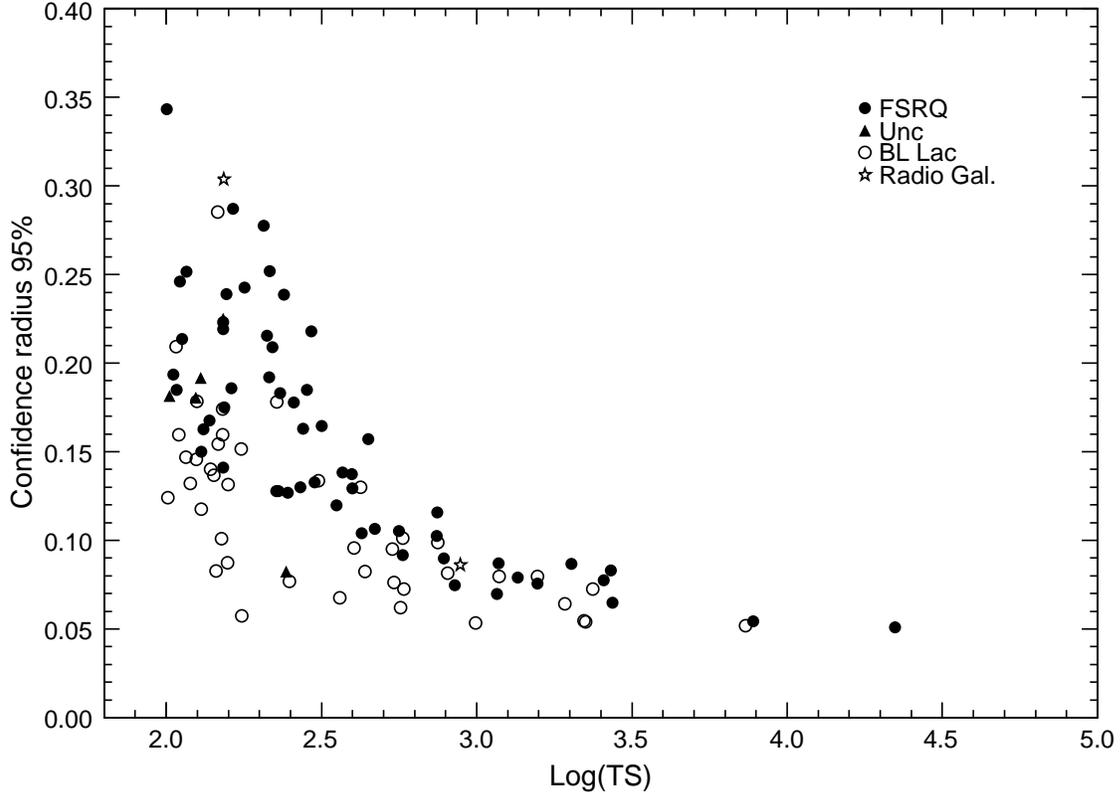}
\caption{95\% error radius as a function of TS for the sources presented in this paper. FSRQs: closed  circles, BL Lacs: open circles, Uncertain type: closed triangles, Radio galaxies: open stars. } 
\label{fig:r95_ts}
\end{figure}

\begin{figure}
\epsscale{.70}
\plotone{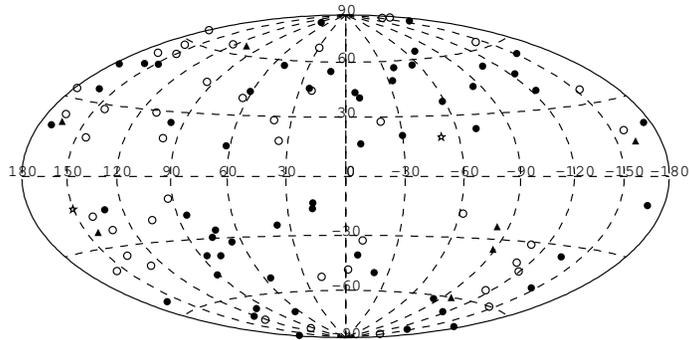}
\caption{Location of the LBAS sources. FSRQs: closed circles, BL Lacs: open circles, Uncertain type: closed triangles, RG: open stars.} 
\label{fig:sky_map}
\end{figure}

\begin{figure}
\epsscale{.70}
\plotone{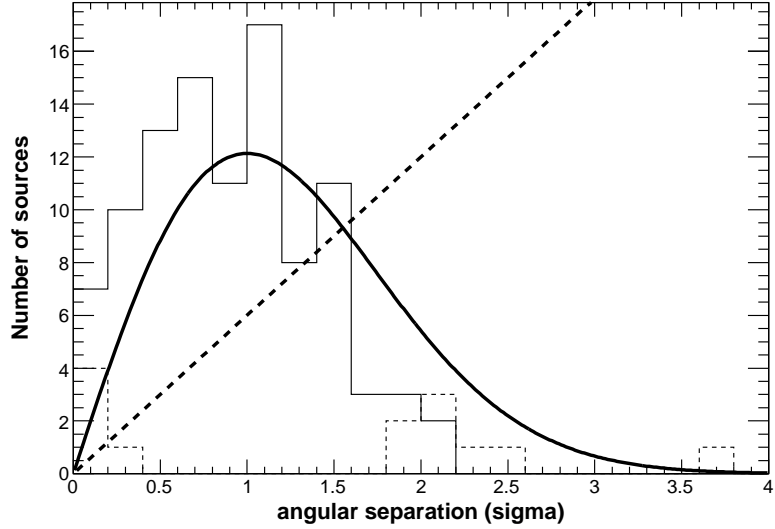}
\caption{ Normalized angular separation between the {\it Fermi}-LAT location and that of the counterpart. The solid (dashed) histogram corresponds to the sources with high-(low-) confidence associations. The solid curve corresponds to the expected distribution ($\chi^2$ distribution with 2 d.o.f.) for real associations, the dashed one for accidental associations. } 
\label{fig:ang_sep}
\end{figure}

\begin{figure}
\centering
\plotone{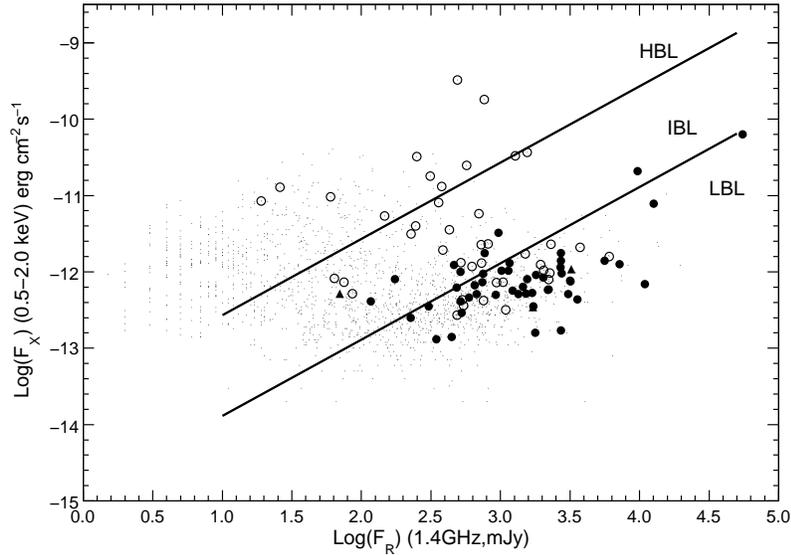}
%\resizebox{7cm}{!}{\rotatebox[]{-90}{\includegraphics{fermi_fxfr.ps}}}
\caption{The $X$-ray (0.5 - 2.0 keV) vs radio flux density (1.4 GHz) plot of all $X$-ray detected blazars in the BZCAT catalog (small dots) and  the {\it Fermi}-LAT detected blazars (BL Lacs: open circles,  FSRQs: filled circles, blazars of uncertain type: triangles, radio galaxies: stars). 
\label{fig:ps_rx} }
\end{figure}

\begin{figure}
\epsscale{.70}
\plotone{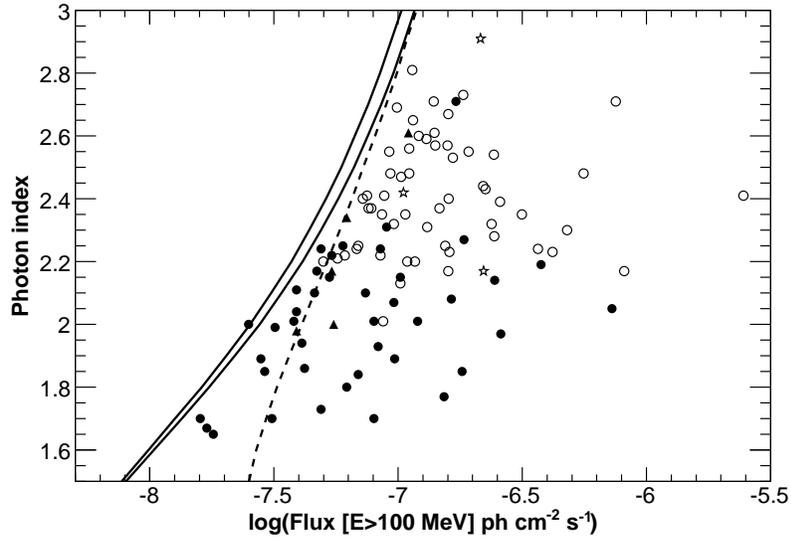}
\caption{ Flux [E$>$100 MeV] vs photon index for the 116 sources. FSRQs: closed  circles, BL Lacs: open circles, Uncertain type: closed triangles, Radio galaxies: open stars. The solid curves represent the TS=100 limit estimated for two galactic latitudes b=20$^\circ$ and b=80$^\circ$ . The dashed curve represents the TS=100 limit for b=80$^\circ$ and 0.2 $<$E$<$ 3 GeV. } 
\label{fig:flux_index_lim}
\end{figure}
\clearpage

\begin{figure}
\epsscale{1.}
\plotone{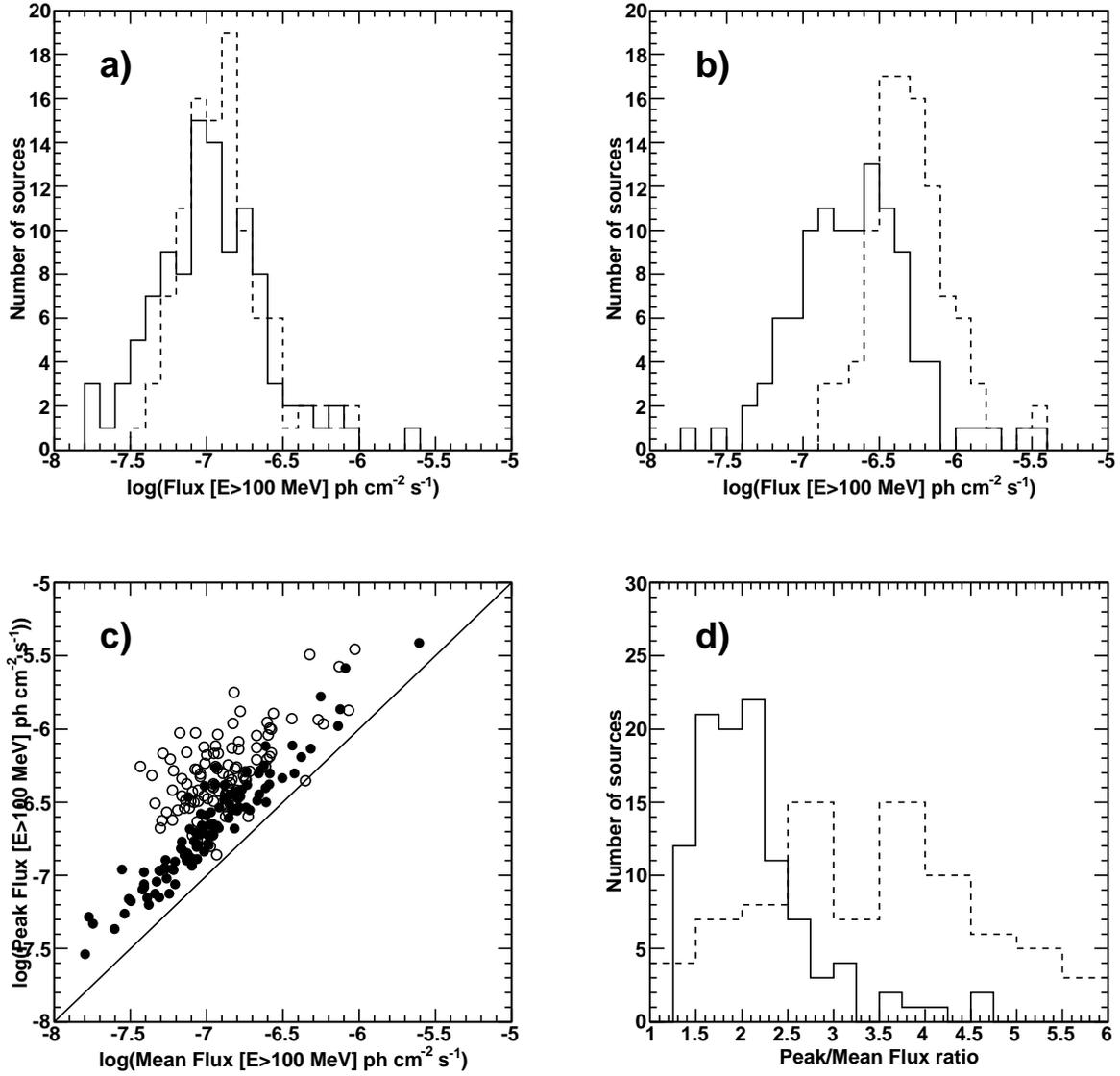}
\caption{a) Comparison of mean flux distribution for blazars detected by {\it Fermi}-LAT (solid) and {\it EGRET} (dashed). b) same of as a), for the peak flux distribution. c) Peak flux as a function of  mean flux, for the {\it Fermi}-LAT (closed circles) and {\it EGRET} (open circles) AGNs. d) same as a), for the peak/mean flux ratio.} 
\label{fig:flux_dis}
\end{figure}

\begin{figure}
\epsscale{1.}
\plotone{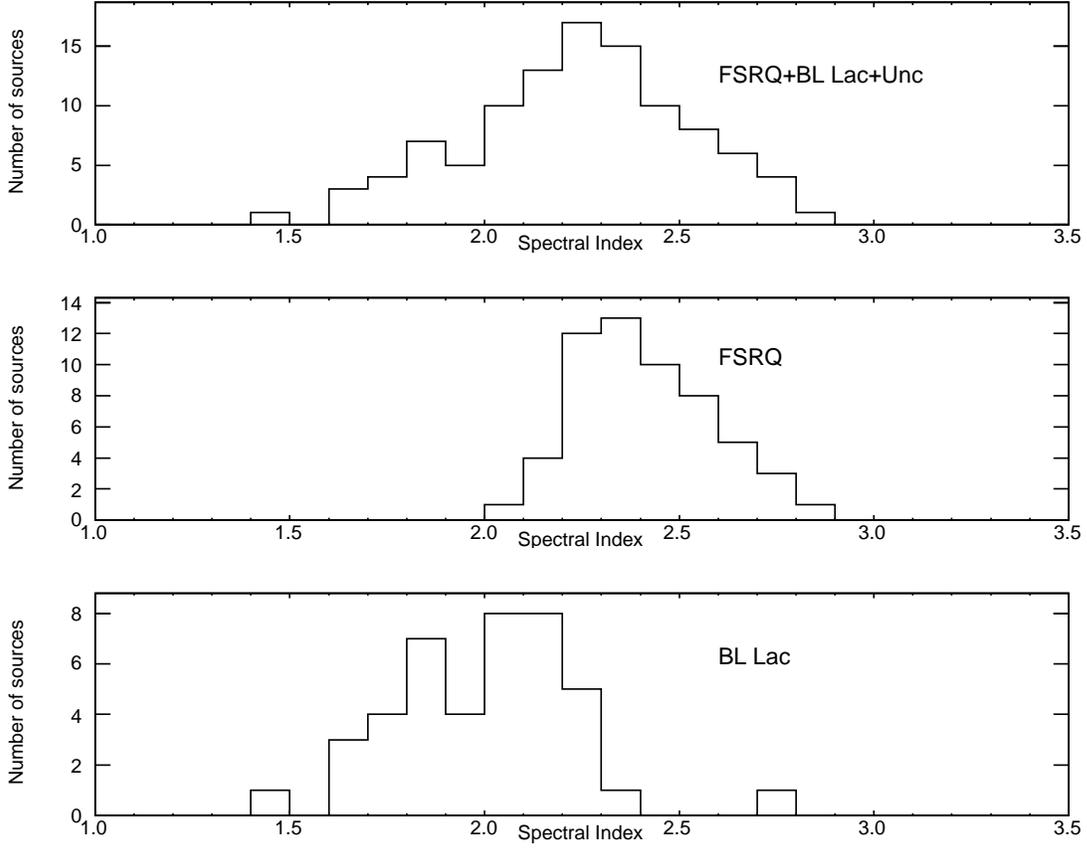}
\caption{Photon index distributions for the LBAS  blazars. Top: All sources. Middle: FSRQs. Bottom: BL Lacs.} 
\label{fig:sp_dist}
\end{figure}

\begin{figure}
\plotone{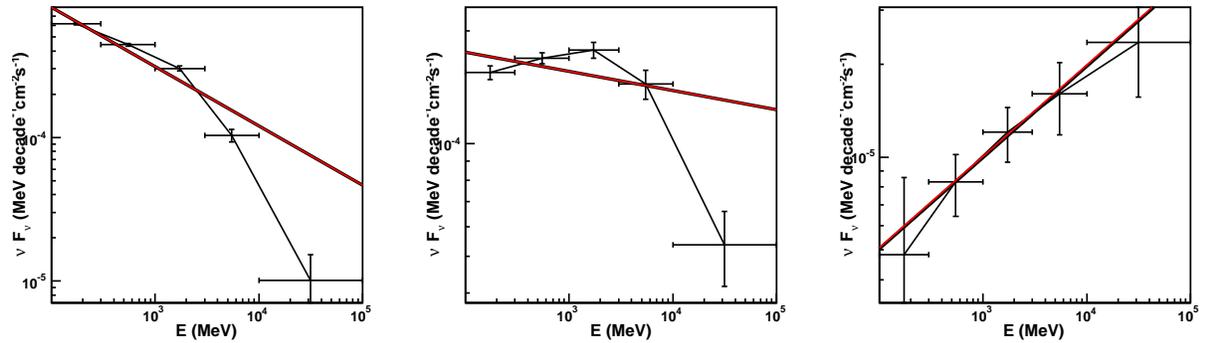}
\caption{ Gamma-ray SED of 3 bright blazars calculated in five energy bands, compared with the power law fitted over the whole energy range. Left: 3C454.3 (FSRQ), middle: AO 0235+164 (IBL), right: Mkn 501 (HBL) } 
\label{fig:bright_sed}
\end{figure}

\begin{figure}
\epsscale{1.}
\plotone{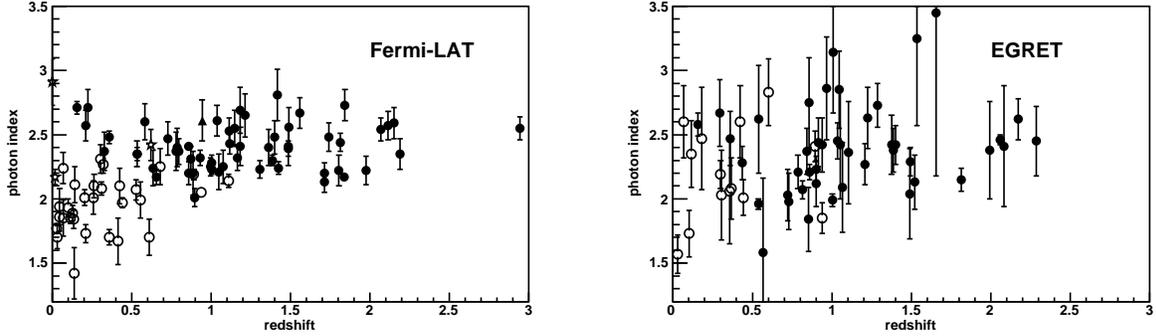}
\caption{Left: LBAS photon index as a function of redshift. Same symbols as before. Right: same as left, for the {\it EGRET} sample.} 
\label{fig:index_redshift}
\end{figure}

\begin{figure}
\plotone{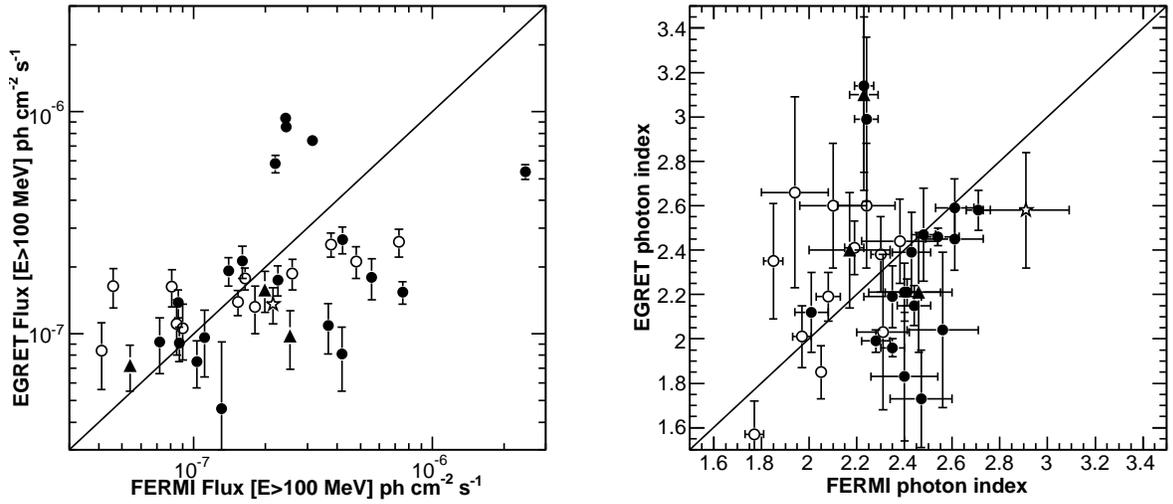}
\caption{ Left: {\it Fermi}-LAT vs {\it EGRET} mean flux for the 33 AGNs present in both samples FSRQs: closed  circles, BL Lacs: open circles, Uncertain type: closed triangles. Right:  same as left, for photon  index.} 
\label{fig:fermi_egret}
\end{figure}

\clearpage

\begin{figure}
\plotone{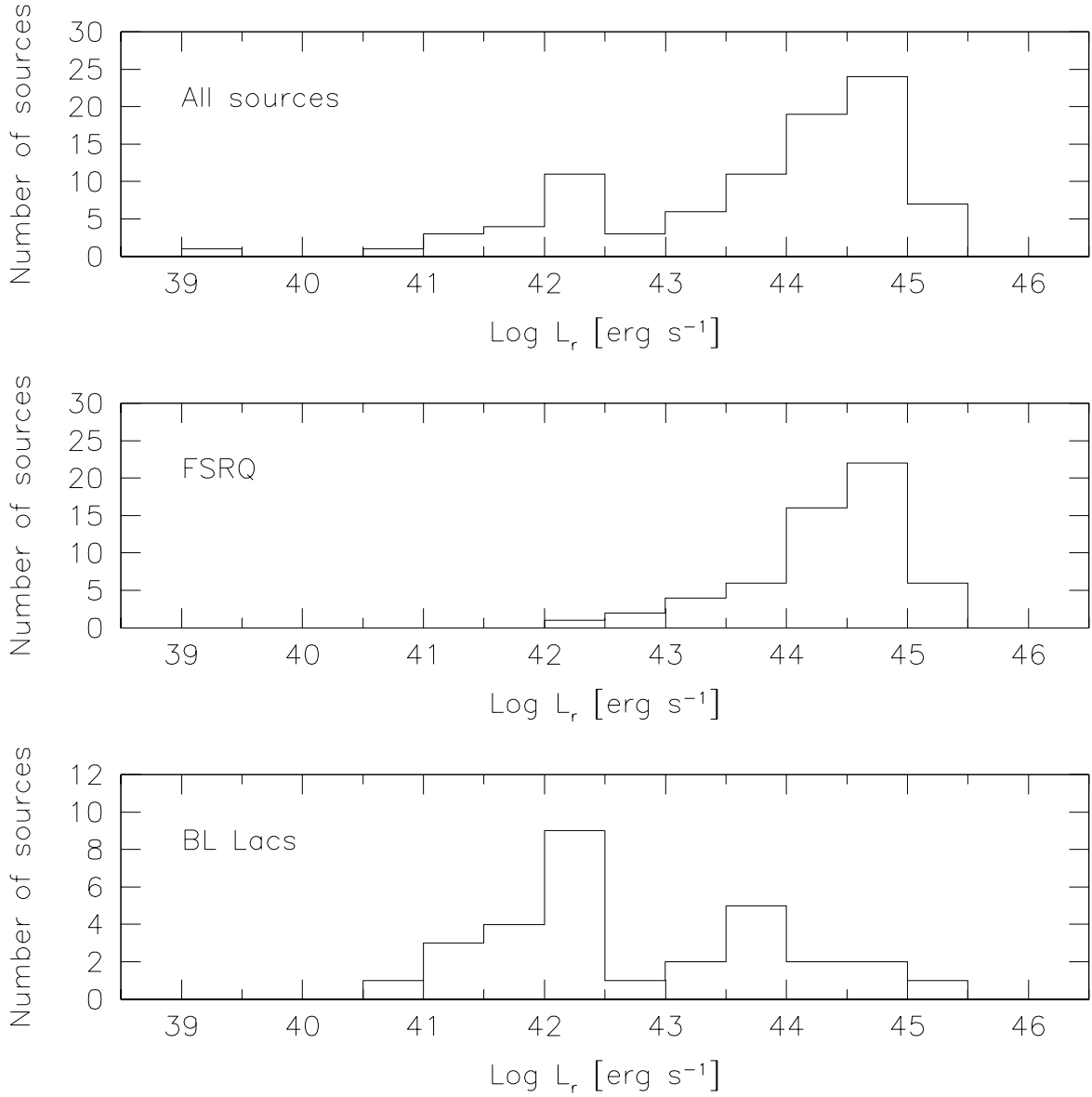}
\caption{Histogram of the radio power distribution for LBAS sources, for all sources (upper panel), FSRQs (middle), and BL Lacs (bottom) only. \label{fig:radiopower} }
\end{figure}

\begin{figure}
\plottwo{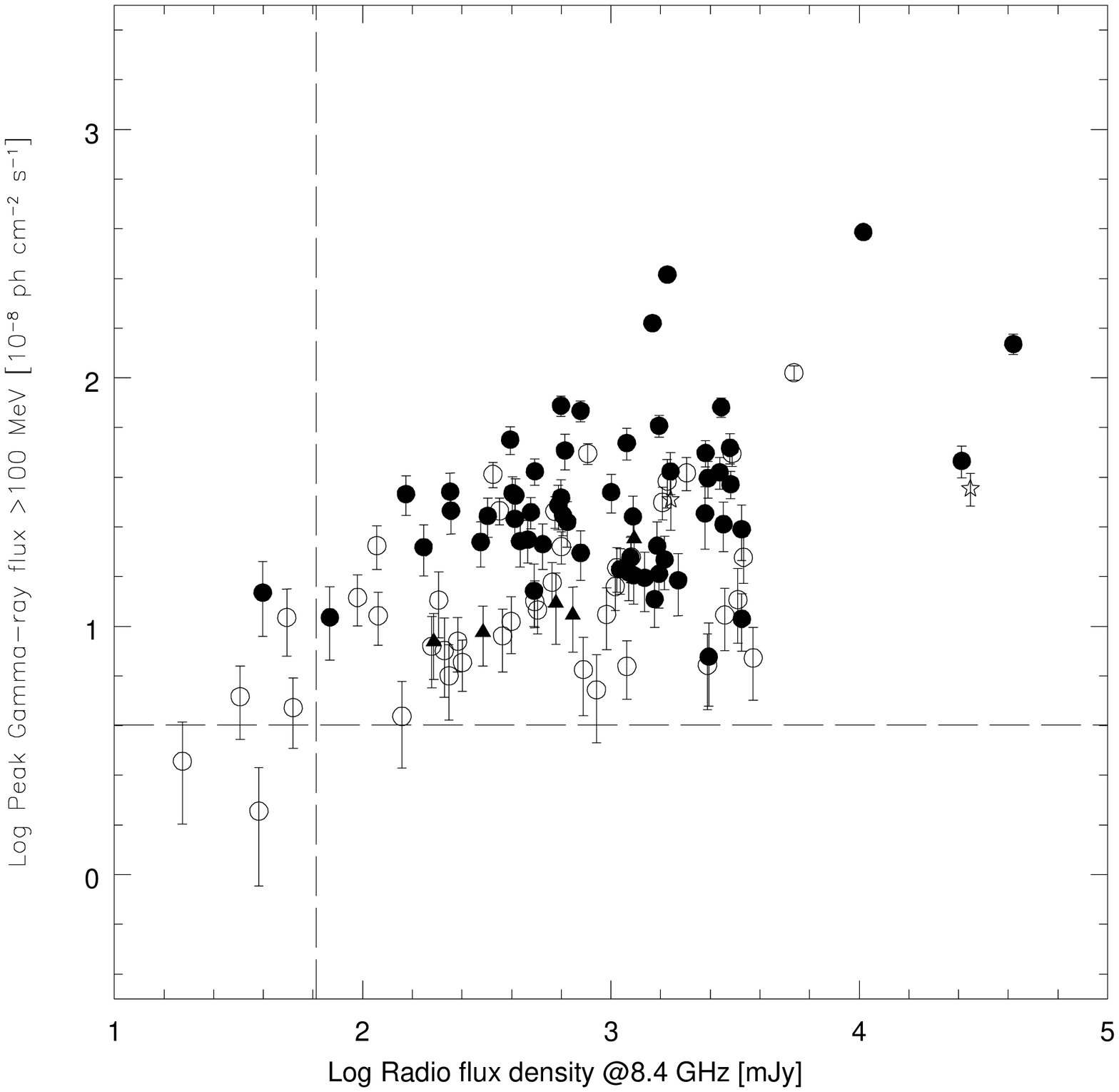}{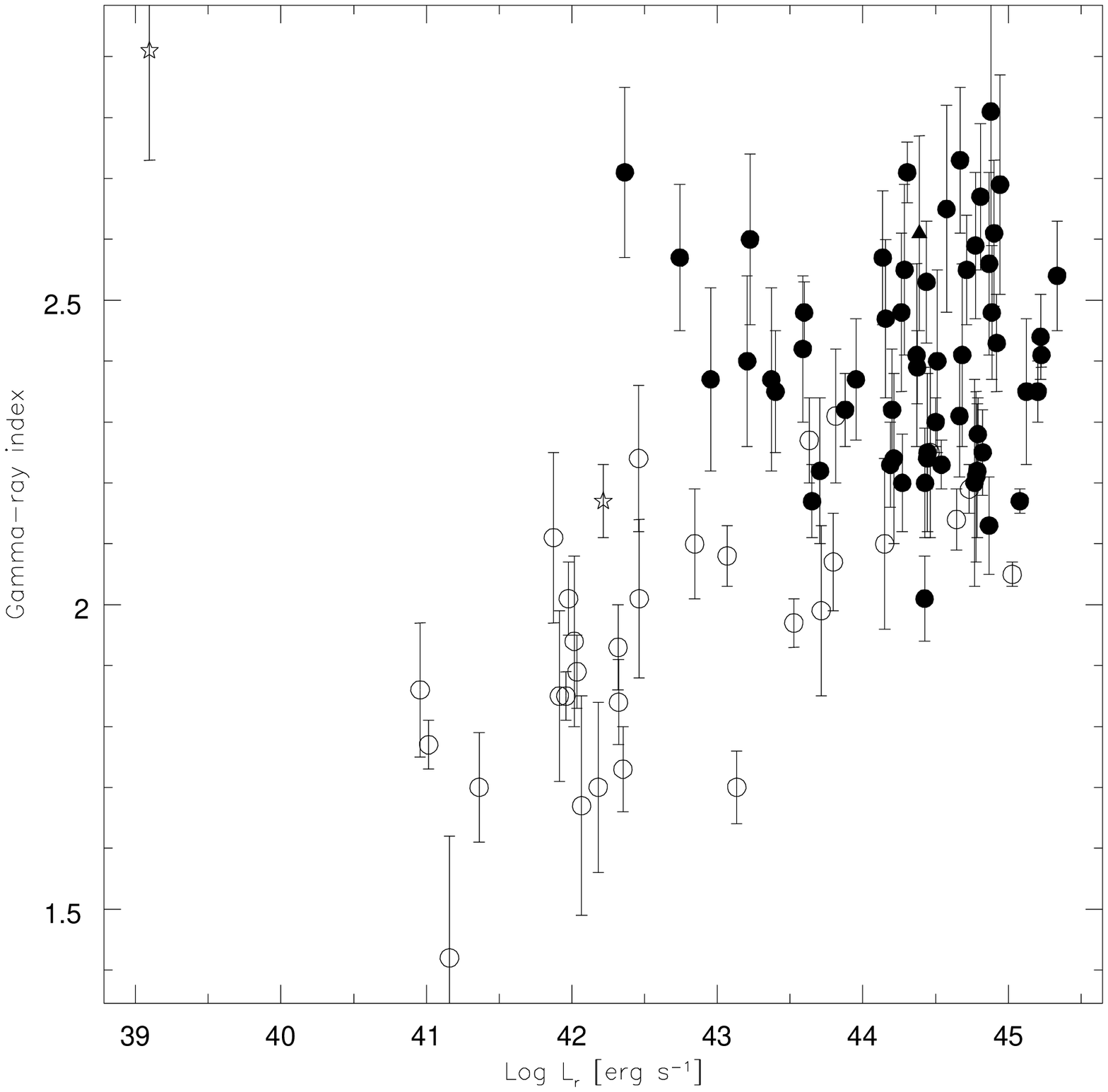}
\caption{Radio vs.\ gamma-ray properties. Left: peak gamma-ray flux vs.\ radio flux density at 8.4 GHz; the dashed lines show the CRATES flux density limit and the typical LAT detection threshold. Right: gamma-ray photon index vs.\ radio luminosity. \label{fig:radiogamma} }
\end{figure}
\clearpage

\begin{figure}
\epsscale{1.}
\plotone{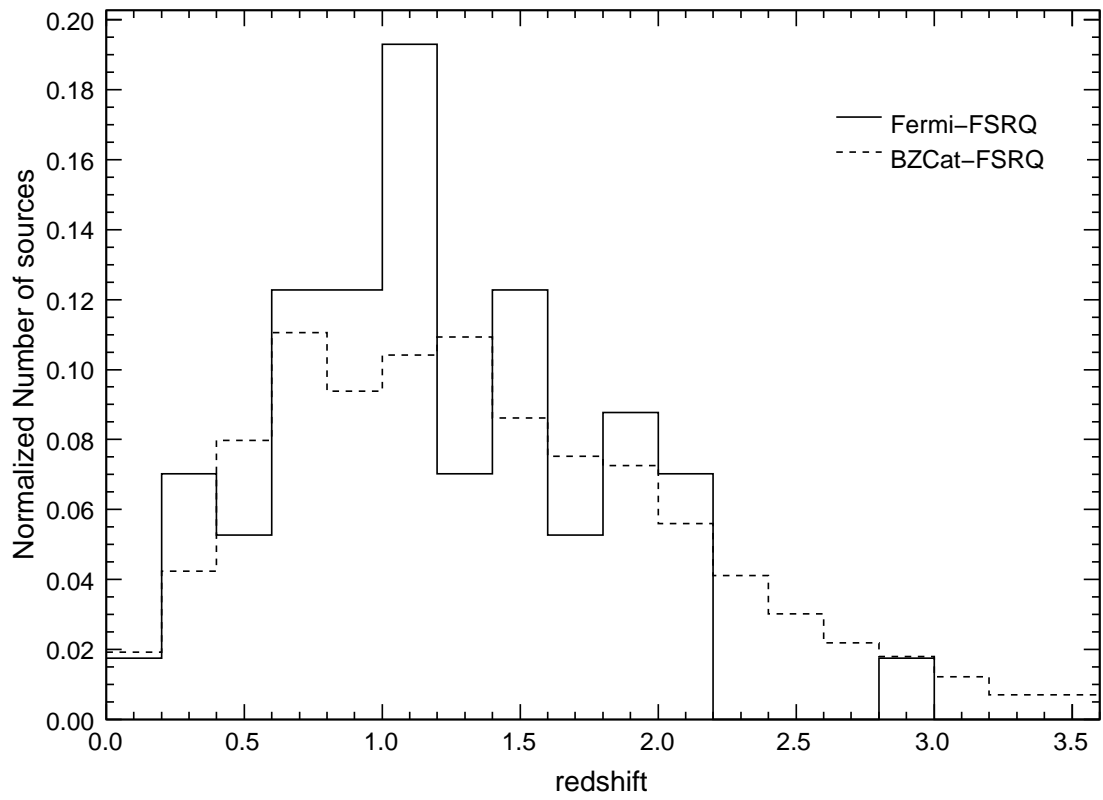}
\caption{Redshift distribution for the FSRQs in the LBAS (solid) and in the BZCat catalog (dashed).} 
\label{fig:z_dist_fsrq}
\end{figure}

\begin{figure}
\epsscale{1.}
\plotone{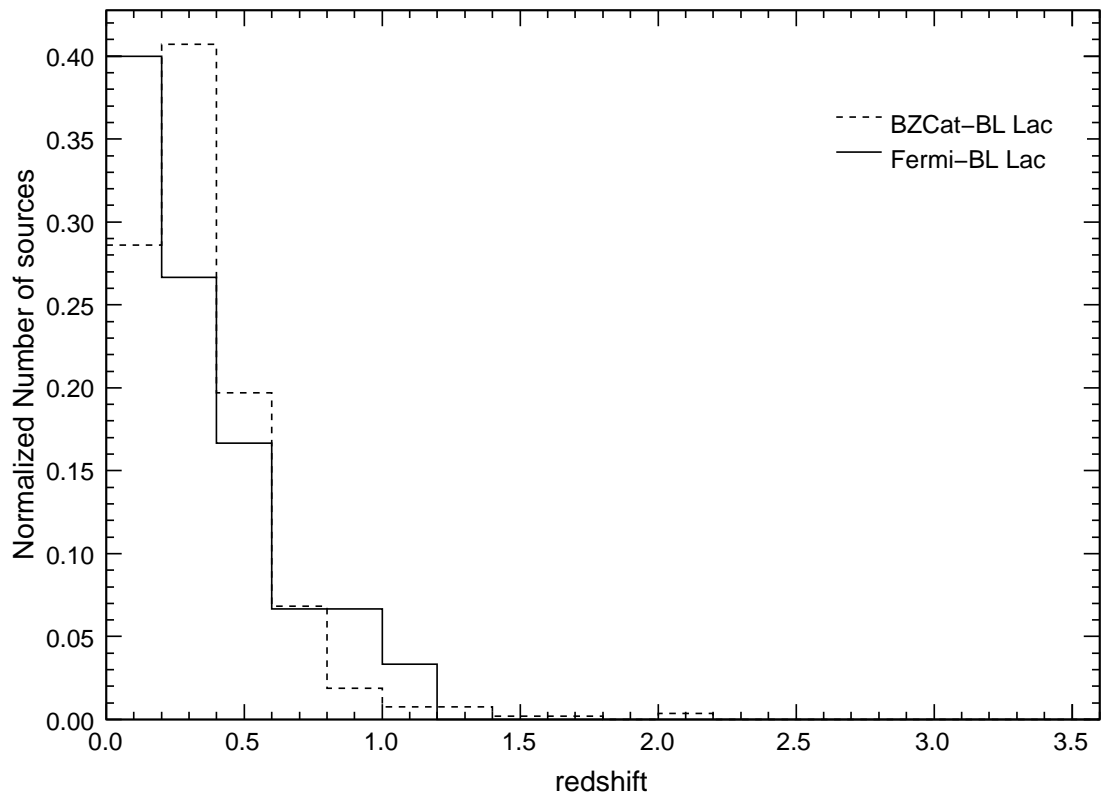}
\caption{Redshift distribution for the BL Lacs in the LBAS (solid) and in the BZCat catalog (dashed).}  
\label{fig:z_dist_bllac}
\end{figure}
\clearpage
\begin{figure}
\epsscale{1.}
\plotone{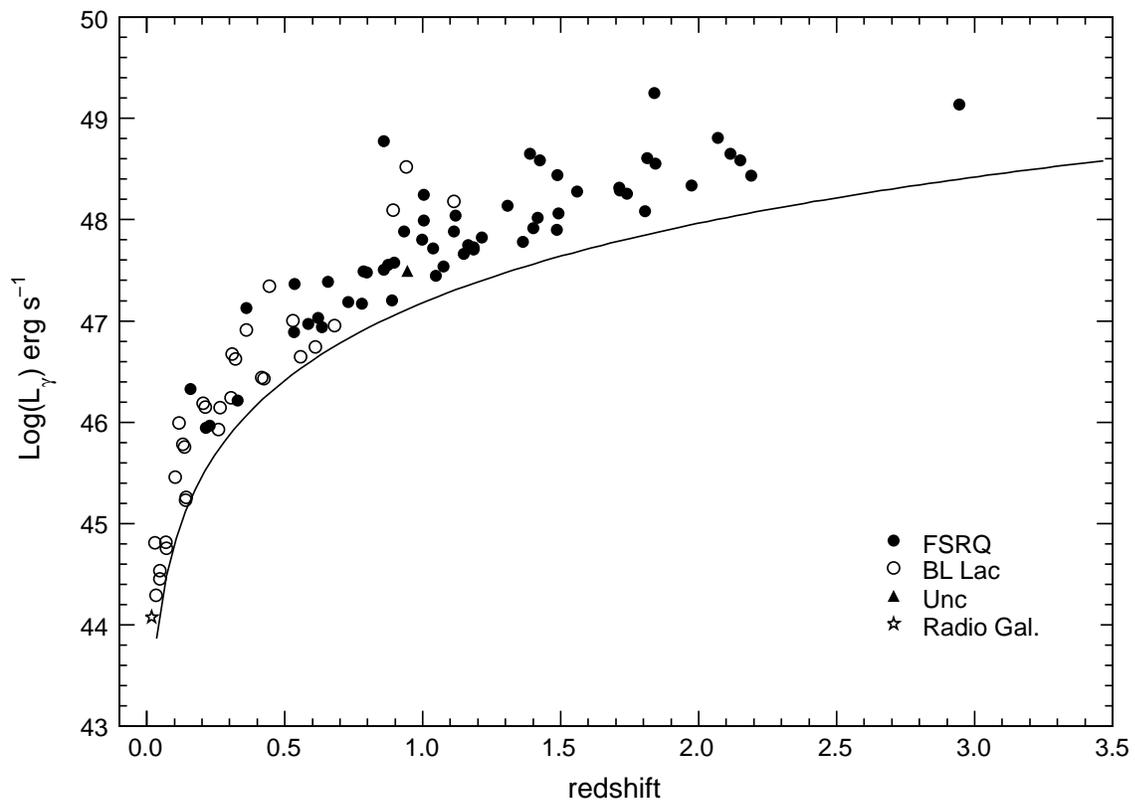}
 \caption{Gamma-ray Luminosity vs redshift for the LBAS. The solid line was drawn using a $F_{100}=4\times10^{-8}$\,\pflux and a photon index of 2.2.}
\label{fig:LumRed}
\end{figure}

\begin{figure}[ht!]
  \begin{center}
  	 \includegraphics[scale=0.7]{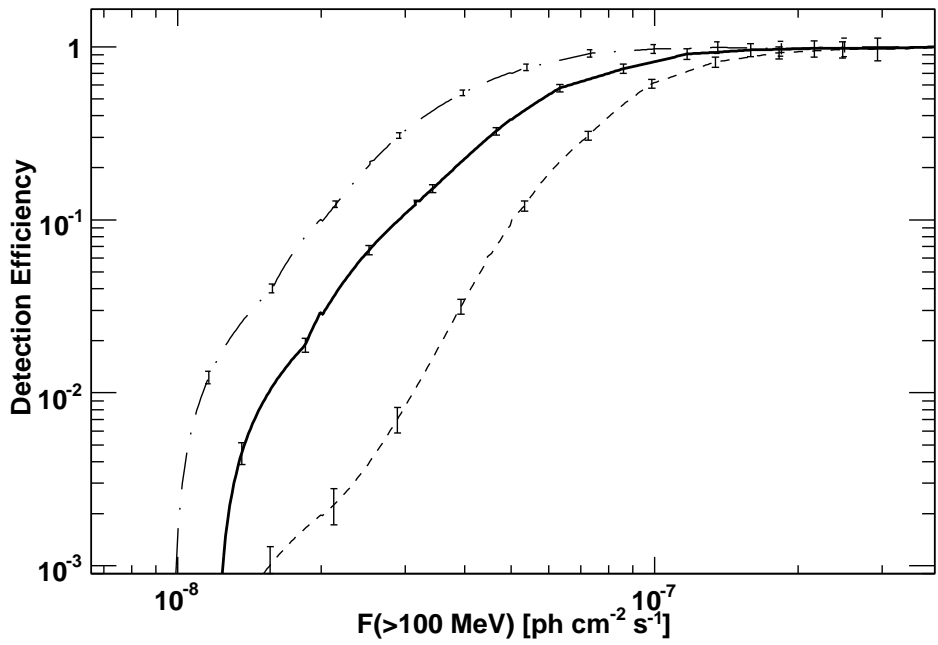}
  \end{center}
  \caption{Detection efficiencies in the LAT $|$b$|\geq$10$^{\circ}$ survey
as a function of flux. The solid line is for the entire blazar population
while the dashed and long-dashed are for the FSRQs and BL Lacs respectively.
The errors on the detection efficiency are due to the counting statistics
in our Monte Carlo simulations.
\label{fig:skycov}}
\end{figure}

\begin{figure}[ht!]
  \begin{center}
  	 \includegraphics[scale=0.8]{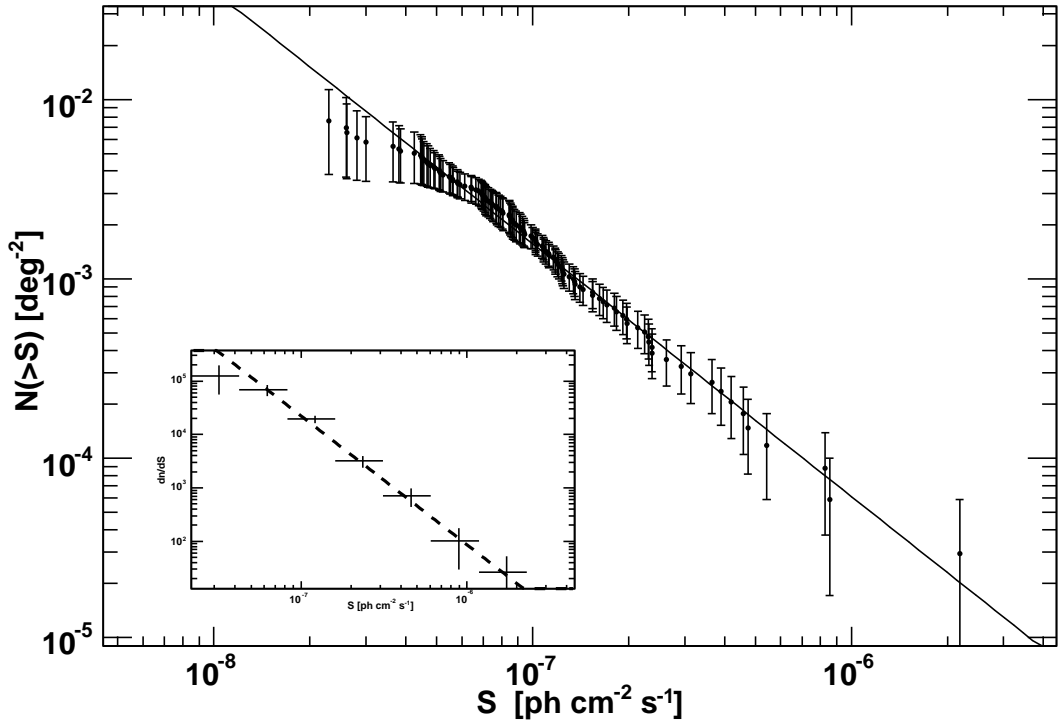}
  \end{center}
  \caption{Source count distribution for the whole extragalactic population
(excluding the pulsars). The dashed line is the best  power-law  fit to the 
F($>$100 MeV)$\geq7\times10^{-8}$\,ph cm$^{-2}$ s$^{-1}$ data. The inset
shows the differential distribution.
\label{fig:logn_all}}
\end{figure}

\begin{figure}[ht!]
  \begin{center}
  	 \includegraphics[scale=0.8]{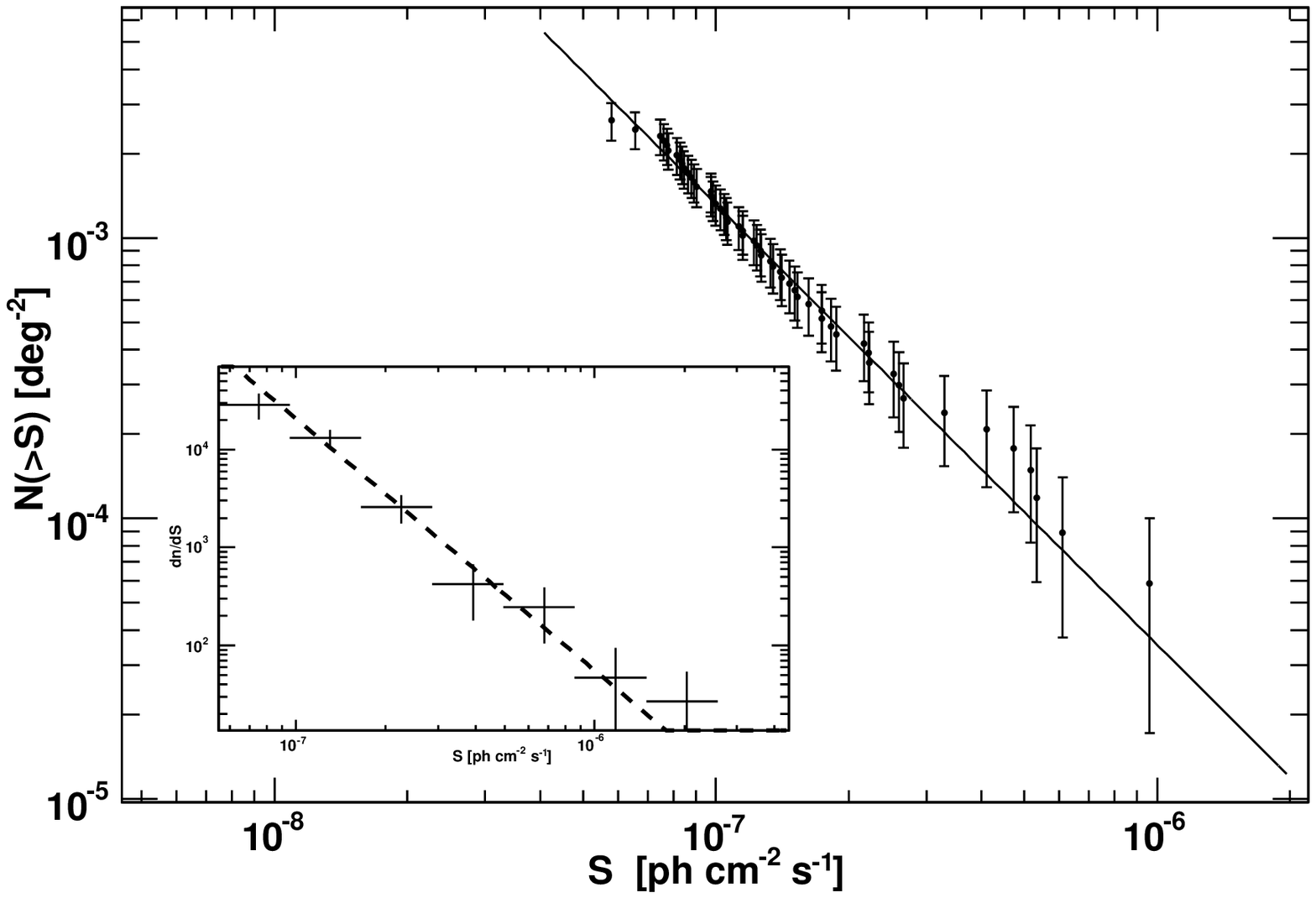}
  \end{center}
  \caption{Source count distribution for FSRQs. 
The dashed line is the best  power-law  fit to the 
F($>$100 MeV)$\geq7\times10^{-8}$\,ph cm$^{-2}$ s$^{-1}$ data. The inset
shows the differential distribution.
\label{fig:logn_fsrq}}
\end{figure}

\begin{figure}[ht!]
  \begin{center}
  	 \includegraphics[scale=0.8]{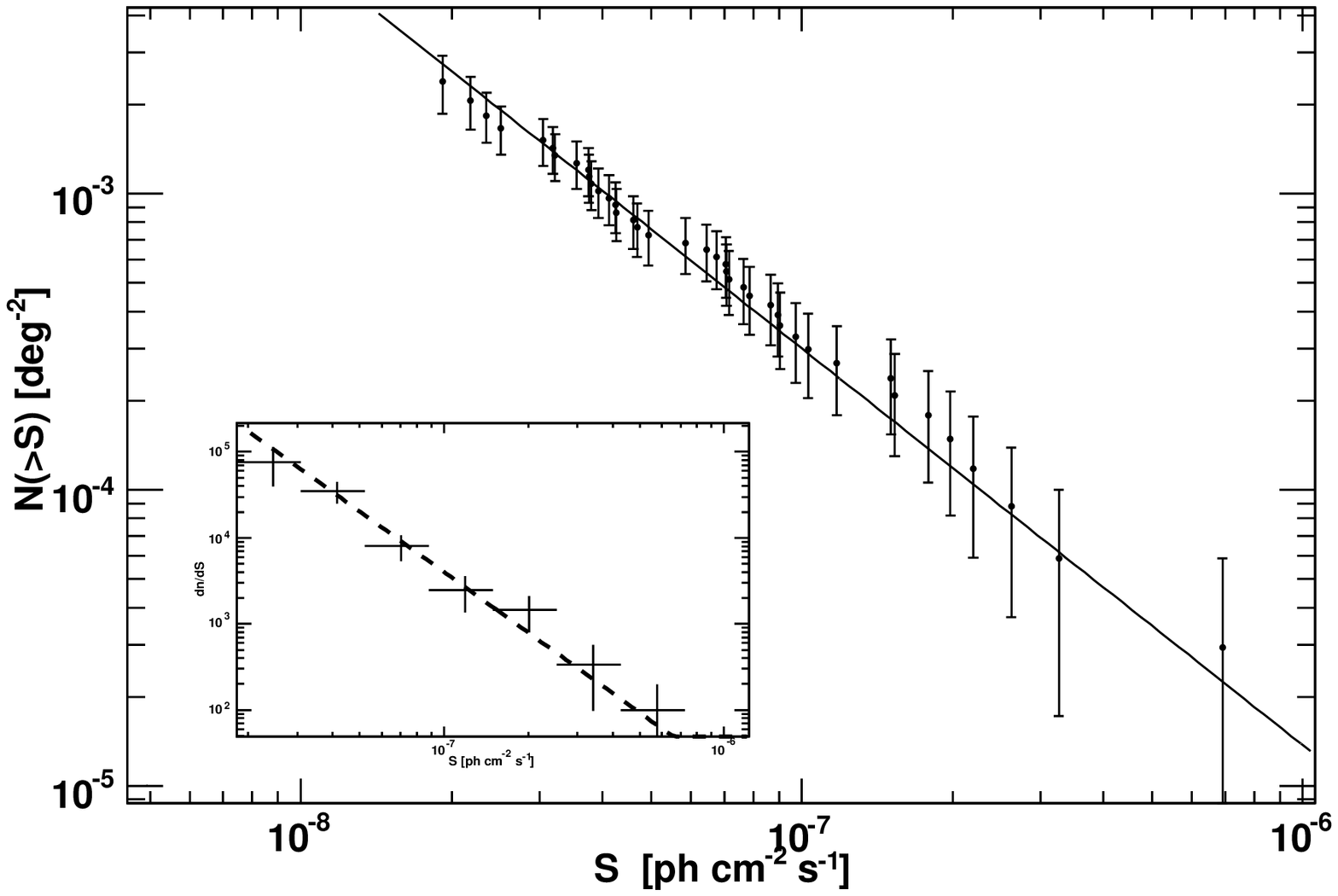}
  \end{center}
  \caption{Source count distribution for BL Lacs. 
The dashed line is the best  power-law  fit to the 
F($>$100 MeV)$\geq3\times10^{-8}$\,ph cm$^{-2}$ s$^{-1}$ data. The inset
shows the differential distribution.
\label{fig:logn_bllac}}
\end{figure}

\begin{figure*}[ht!]
  \begin{center}
  \begin{tabular}{cc}
    \includegraphics[scale=0.43]{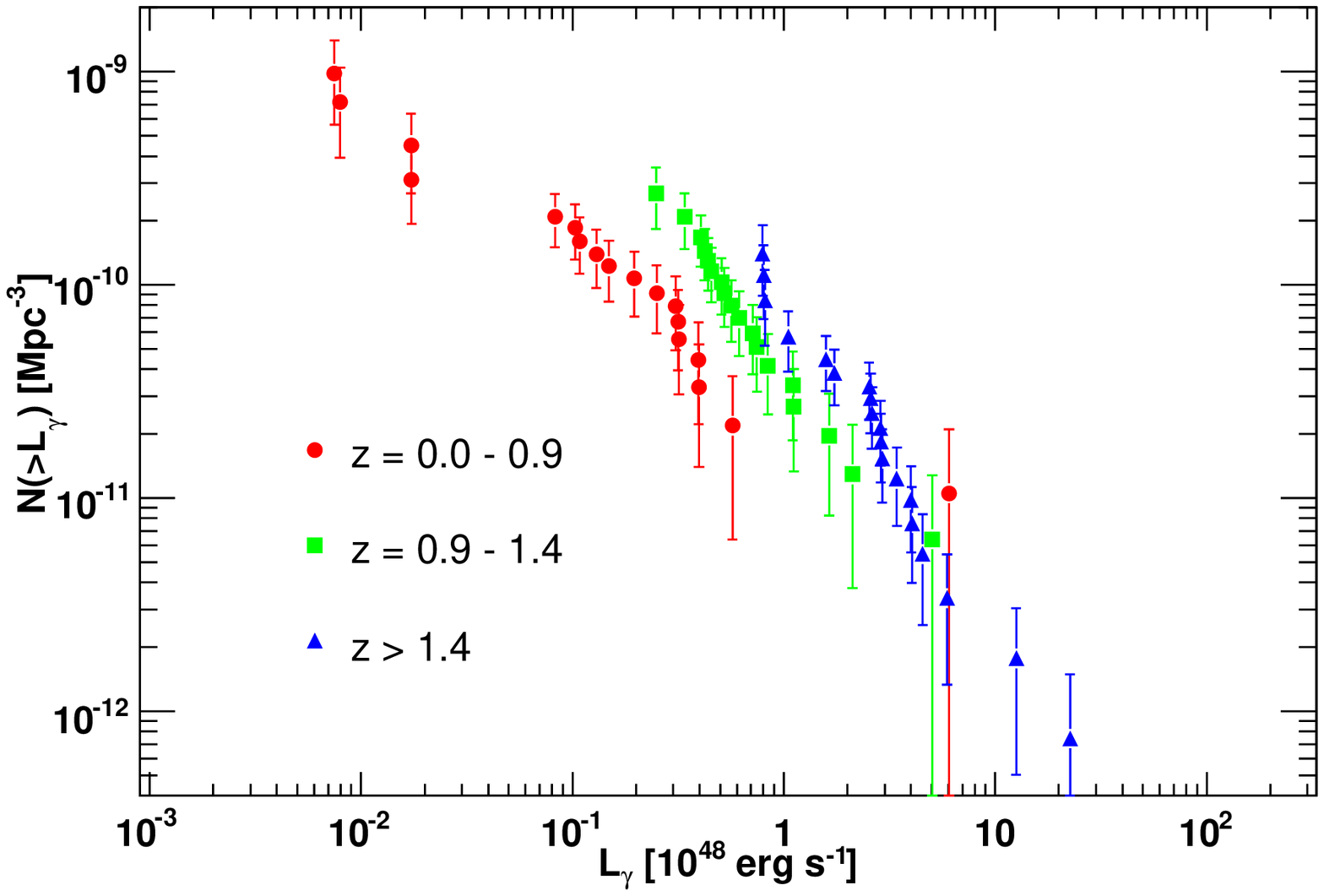} 
  	 \includegraphics[scale=0.43]{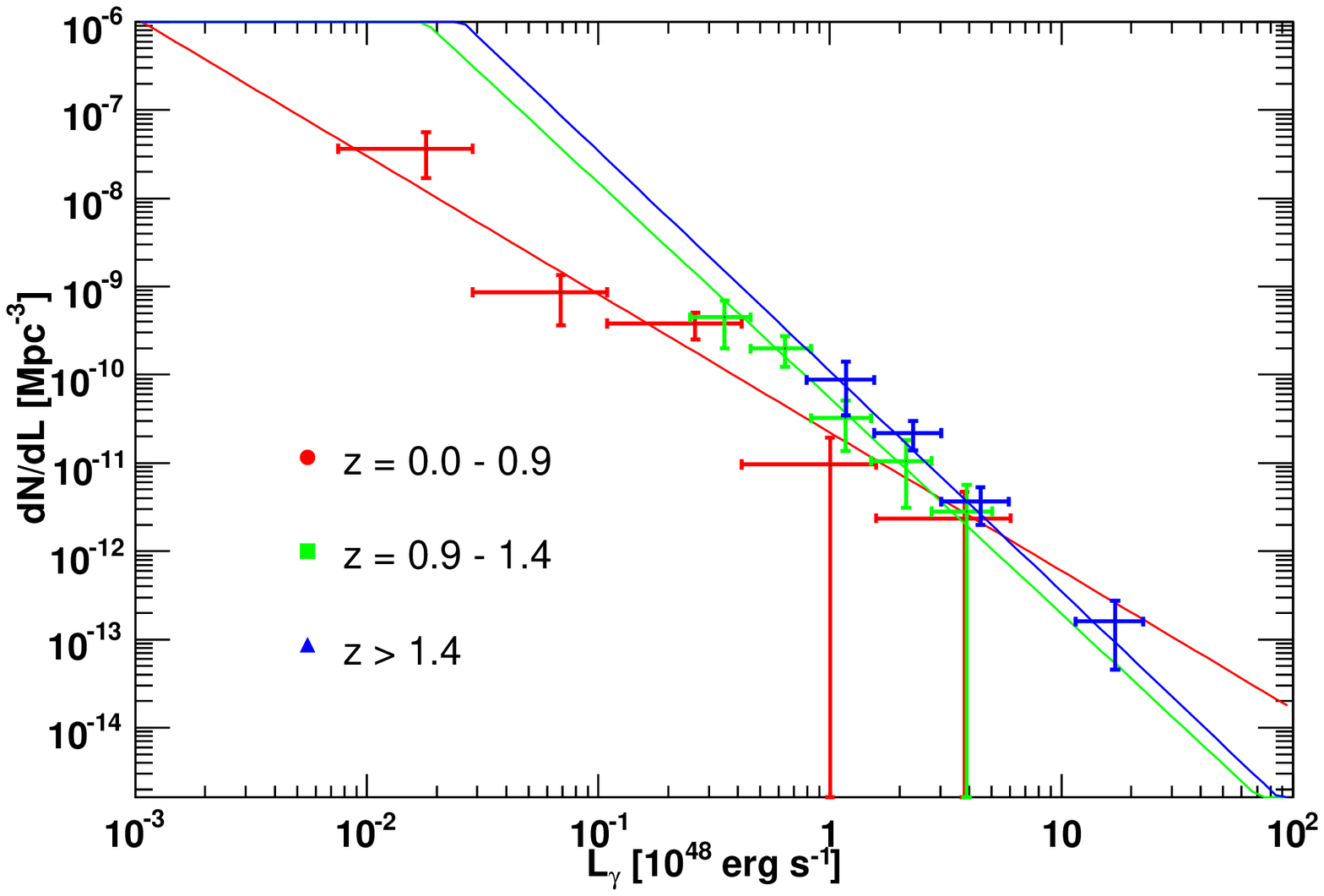}\\
\end{tabular}
  \end{center}
  \caption{Luminosity functions of FSRQs in bins of redshift. The cumulative
and differential distributions are shown, respectively, 
on the left and on the right panel. The (color-coded) 
solid lines are the ML fits to the  3 different datasets using a simple
power law to model the GLF.
\label{fig:glf_fsrq}}
\end{figure*}

\begin{figure*}[ht!]
  \begin{center}
  \begin{tabular}{cc}
    \includegraphics[scale=0.43]{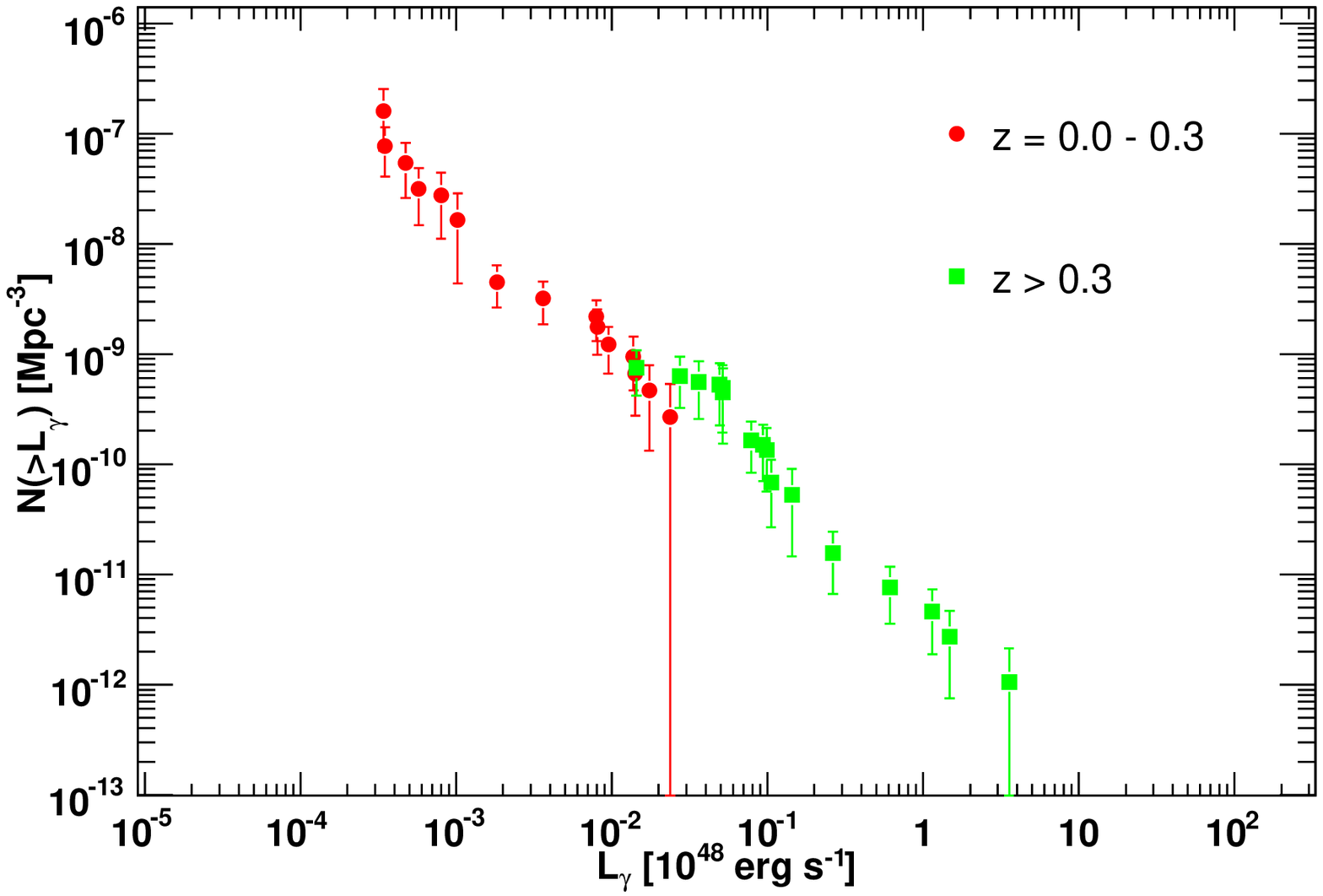} 
  	 \includegraphics[scale=0.43]{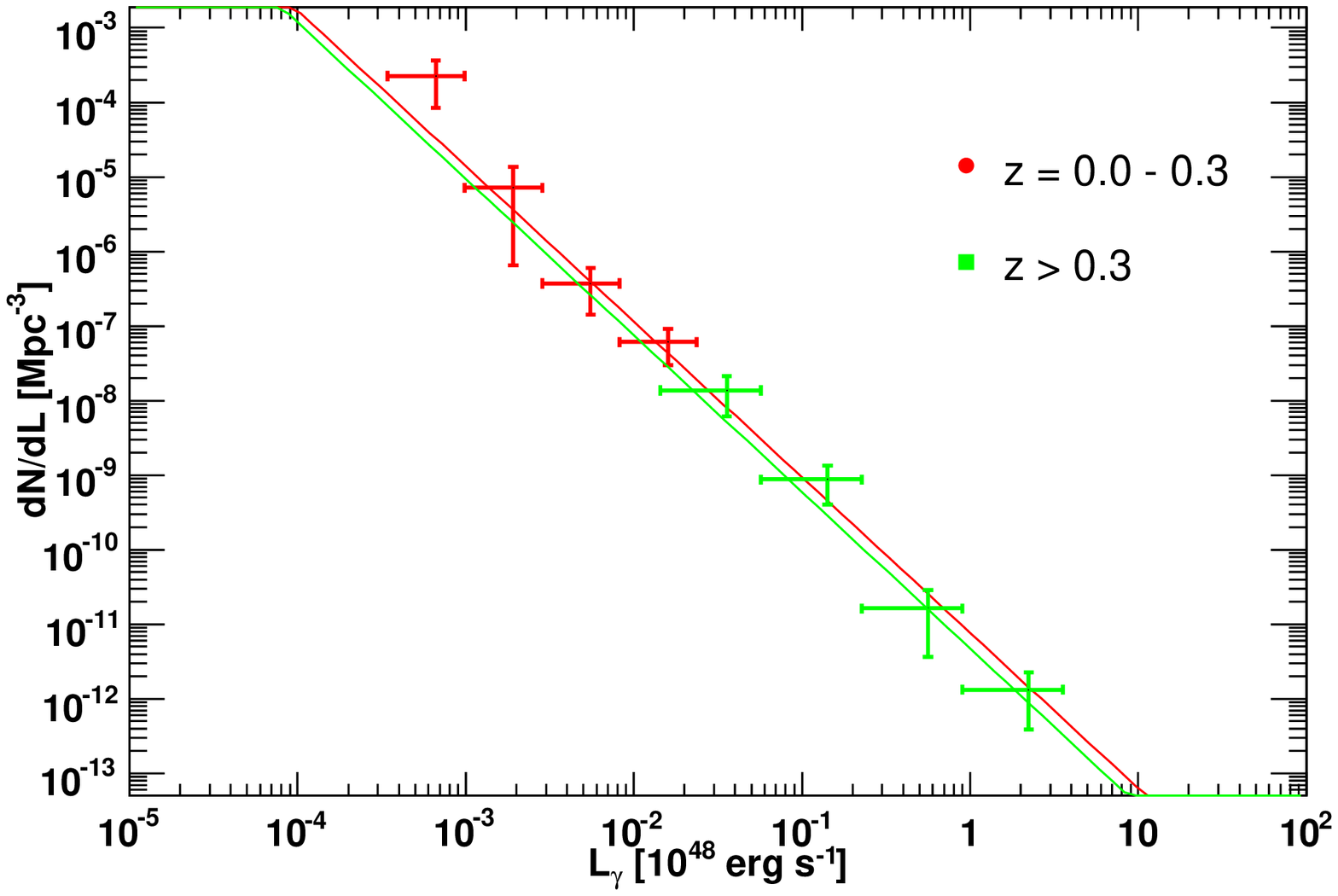}\\
\end{tabular}
  \end{center}
  \caption{Luminosity functions of BL Lacs in bins of redshift. The cumulative
and differential distributions are shown, respectively, 
on the left and on the right panel. The (color-coded) 
solid lines are the ML fits to the  2 different datasets using a simple
power law to model the GLF.
\label{fig:glf_bllac}}
\end{figure*}

\begin{deluxetable}{llllllllll}
\tablecolumns{9} 
\setlength{\tabcolsep}{0.04in}
\tablewidth{0pc} 
\tabletypesize{\tiny}
\tablecaption{The high-confidence association Bright AGN List\label{table1)}}
\tablenum{1}
%\rotate
\tablehead{
\colhead{} & \multicolumn{3}{c}{FoM} & \colhead{} & \multicolumn{2}{c}{{\it gtsrcid}} & \colhead{} & \colhead{}  & \colhead{} \\ \cline{2-4} \cline{6-7} \\
\colhead{LAT Name} & 
\colhead{Source Name} &  
\colhead{FoM} & 
\colhead{Prob.} & 
\colhead{} & 
\colhead{Source Name} & 
\colhead{Prob.} & 
\colhead{Other Names} & 
\colhead{$z$} & 
\colhead{Class} \\ 
} 
\startdata
0FGL J0017.4$-$0503 & CGRaBS J0017$-$0512 & 16.20 & 0.93 &   & CGRaBS J0017$-$0512 &0.92 &\nodata &0.227 & FSRQ
 \\
0FGL J0033.6$-$1921 & \nodata & \nodata &  \nodata  &   & BZB J0033$-$1921 &0.99 &KUV 00311$-$1938 &0.610 & BLLac
 \\
0FGL J0050.5$-$0928 & CGRaBS J0050$-$0929 & 61.01 & 1.00 &   & CRATES J0050$-$0929 &0.99 &PKS 0048$-$097 &\nodata & BLLac \\
0FGL J0051.1$-$0647 & CGRaBS J0051$-$0650 & 42.95 & 0.98 &   & CRATES J0051$-$0650 &0.99 &PKS 0048$-$071 &1.975 & FSRQ
 \\
0FGL J0112.1+2247 & CGRaBS J0112+2244 & 48.96 & 0.98 &   & S2 0109+22 &1.00 &S2 0109+22 &0.265 & BLLac
 \\
0FGL J0118.7$-$2139 & CGRaBS J0118$-$2141 & 35.02 & 0.97 &   & CGRaBS J0118$-$2141 &0.99 &PKS 0116$-$219 &1.165 & FSRQ
 \\
0FGL J0120.5$-$2703 & CGRaBS J0120$-$2701 & 60.25 & 1.00 &   & PKS 0118$-$272 &1.00 &PKS 0118$-$272 &0.557 & BLLac
 \\
0FGL J0136.6+3903 & BZB J0136+3905 & 12.45 & 0.91 &   & B3 0133+388 &1.00 &B3 0133+388 &\nodata & BLLac
 \\
0FGL J0137.1+4751 & CGRaBS J0136+4751 & 52.21 & 0.99 &   & CGRaBS J0136+4751 &0.99 &DA 55 &0.859 & FSRQ
 \\
0FGL J0144.5+2709 & CRATES J0144+2705 & 30.93 & 0.96 &   & CRATES J0144+2705 &0.58 &TXS 0141+268 &\nodata & BLLac
 \\
0FGL J0145.1$-$2728 & CGRaBS J0145$-$2733 & 37.41 & 0.97 &   & CGRaBS J0145$-$2733 &0.96 &PKS 0142$-$278 &1.148 & FSRQ
 \\
0FGL J0204.8$-$1704 & CGRaBS J0204$-$1701 & 55.22 & 0.99 &   & CGRaBS J0204$-$1701 &0.96 &PKS 0202$-$17 &1.740 & FSRQ
 \\
0FGL J0210.8$-$5100 & CGRaBS J0210$-$5101 & 69.87 & 1.00 &   & PKS 0208$-$512 &1.00 &PKS 0208$-$512 &1.003 & FSRQ
 \\
0FGL J0217.8+0146 & CGRaBS J0217+0144 & 52.67 & 0.99 &   & CGRaBS J0217+0144 &1.00 &PKS 0215+015 &1.715 & FSRQ
 \\
0FGL J0220.9+3607 & CGRaBS J0221+3556 & 7.66 & 0.89 &   & CGRaBS J0221+3556 &0.95 &B2 0218+35 &0.944 & Unc\tablenotemark{***}  \\
0FGL J0222.6+4302 & BZB J0222+4302 & 23.18 & 0.95 &   & 3C 66A &1.00 &3C 66A &0.444 & BLLac
 \\
0FGL J0229.5$-$3640 & BZQ J0229$-$3643 & 29.20 & 0.96 &   & BZQJ0229$-$3643 &0.94 &PKS 0227$-$369 &2.115 & FSRQ
 \\
0FGL J0238.6+1636 & CGRaBS J0238+1636 & 60.54 & 1.00 &   & CGRaBS J0238+1636 &1.00 &AO 0235+164 &0.940 & BLLac
 \\
0FGL J0245.6$-$4656 & CRATES J0246$-$4651 & 23.23 & 0.95 &   & CRATES J0246$-$4651 &0.54 &PKS 0244$-$470 &\nodata & Unc\tablenotemark{***}  \\
0FGL J0303.7$-$2410 & CRATES J0303$-$2407 & 9.32 & 0.90 &   & PKS 0301$-$243 &1.00 &PKS 0301$-$243 &0.260 & BLLac
 \\
0FGL J0320.0+4131 & CGRaBS J0319+4130 & 33.67 & 0.97 &   & 0316+413 &1.00 &NGC 1275 &0.018 & RG
 \\
0FGL J0334.1$-$4006 & CGRaBS J0334$-$4008 & 63.24 & 1.00 &   & PKS 0332$-$403 &1.00 &PKS 0332$-$403 &\nodata & BLLac
 \\
0FGL J0349.8$-$2102 & CGRaBS J0349$-$2102 & 47.40 & 0.98 &   & CGRaBS J0349$-$2102 &0.99 &PKS 0347$-$211 &2.944 & FSRQ
 \\
0FGL J0428.7$-$3755 & CGRaBS J0428$-$3756 & 54.09 & 0.99 &   & CGRaBS J0428$-$3756 &1.00 &PKS 0426$-$380 &1.112 & BLLac
 \\
0FGL J0449.7$-$4348 & CRATES J0449$-$4350 & 5.52 & 0.81 &   & PKS 0447$-$439 &1.00 &PKS 0447$-$439 &0.205 & BLLac
 \\
0FGL J0457.1$-$2325 & CGRaBS J0457$-$2324 & 35.74 & 0.97 &   & CGRaBS J0457$-$2324 &1.00 &PKS 0454$-$234 &1.003 & FSRQ
 \\
0FGL J0507.9+6739 & BZB J0507+6737 & 4.74 & 0.76 &   & 1ES 0502+675 &1.00 &1ES 0502+675 &0.416 & BLLac
 \\
0FGL J0516.2$-$6200 & CGRaBS J0516$-$6207 & 12.04 & 0.91 &   & CGRaBS J0516$-$6207 &0.94 &PKS 0516$-$621 &\nodata & Unc\tablenotemark{***}  
\\
0FGL J0531.0+1331 & CGRaBS J0530+1331 & 65.48 & 1.00 &   &  CRATES J0530+1331 &1.00 &PKS 0528+134 &2.070 & FSRQ
 \\
0FGL J0538.8$-$4403 & CRATES J0538$-$4405 & 53.80 & 0.99 &   & BZBJ0538$-$4405 &0.99 &PKS 0537$-$441 &0.892 & BLLac
 \\
0FGL J0654.3+4513 & CGRaBS J0654+4514 & 42.13 & 0.98 &   & CGRaBS J0654+4514 &1.00 &B3 0650+453 &0.933 & FSRQ
 \\
0FGL J0654.3+5042 & CGRaBS J0654+5042 & 49.98 & 0.99 &   & CGRaBS J0654+5042 &1.00 &\nodata &\nodata & Unc\tablenotemark{***}  \\
0FGL J0700.0$-$6611 & CRATES J0700$-$6610 & 33.82 & 0.97 &   & CRATES J0700$-$6610 &0.64 &PKS 0700$-$661 &\nodata & Unc\tablenotemark{***}  \\
0FGL J0712.9+5034 & CGRaBS J0712+5033 & 44.20 & 0.98 &   & CGRaBS J0712+5033 &0.99 &\nodata &\nodata & BLLac
 \\
0FGL J0714.2+1934 & CLASS J0713+1935 & 20.54 & 0.94 &   & \nodata &\nodata &\nodata &0.534 & FSRQ \\
0FGL J0719.4+3302 & CRATES J0719+3307 & 14.33 & 0.92 &   & BZUJ0719+3307 &0.89 &TXS 0716+332 &0.779 & FSRQ
 \\
0FGL J0722.0+7120 & CGRaBS J0721+7120 & 66.40 & 1.00 &   & CRATES J0721+7120 &1.00 &S5 0716+71 &0.310 & BLLac
 \\
0FGL J0738.2+1738 & CGRaBS J0738+1742 & 25.45 & 0.95 &   & PKS 0735+17 &1.00 &PKS 0735+178 &0.424 & BLLac
 \\
0FGL J0818.3+4222 & CGRaBS J0818+4222 & 61.26 & 1.00 &   & OJ 425 &1.00 &OJ 425 &0.530 & BLLac
 \\
0FGL J0824.9+5551 & CGRaBS J0824+5552 & 57.80 & 0.99 &   & CGRaBS J0824+5552 &0.98 &TXS 0820+560 &1.417 & FSRQ
 \\
0FGL J0855.4+2009 & CGRaBS J0854+2006 & 8.67 & 0.90 &   & OJ 287 &0.99 &OJ 287 &0.306 & BLLac
 \\
0FGL J0921.2+4437 & CGRaBS J0920+4441 & 13.49 & 0.92 &   & CGRaBS J0920+4441 &0.95 &RGB J0920+446 &2.190 & FSRQ
 \\
0FGL J0948.3+0019 & CGRaBS J0948+0022 & 18.64 & 0.93 &   & CGRaBS J0948+0022 &0.94 &PMN J0948+0022 &0.585 & FSRQ
 \\
0FGL J0957.6+5522 & CRATES J0957+5522 & 50.91 & 0.99 &   & BZQJ0957+5522 &0.96 &4C +55.17 &0.896 & FSRQ
 \\
0FGL J1012.9+2435 & CRATES J1012+2439 & 13.63 & 0.92 &   & \nodata &\nodata &\nodata &1.805 & FSRQ
 \\
0FGL J1015.2+4927 & CGRaBS J1015+4926 & 18.06 & 0.93 &   & 1ES 1011+496 &1.00 &1ES 1011+496 &0.212 & BLLac
 \\
0FGL J1015.9+0515 & CRATES J1016+0513 & 28.86 & 0.96 &   & CRATES J1016+0513 &0.78 &PMN J1016+0512 &1.713 & FSRQ
 \\
0FGL J1034.0+6051 & CGRaBS J1033+6051 & 52.57 & 0.99 &   & CGRaBS J1033+6051 &0.98 &S4 1030+61 &1.401 & FSRQ
 \\
0FGL J1053.7+4926 & BZB J1053+4929 & 11.55 & 0.91 &   & MS 1050.7+4946 &1.00 &MS 1050.7+4946 &0.140 & BLLac
 \\
0FGL J1054.5+2212 & CLASS J1054+2210 & 16.20 & 0.93 &   & \nodata &\nodata &\nodata &\nodata & BLLac
 \\
0FGL J1057.8+0138 & CGRaBS J1058+0133 & 3.71 & 0.68 &   & CGRaBS J1058+0133 &0.93 &PKS 1055+018 &0.888 & FSRQ
 \\
0FGL J1058.9+5629 & CGRaBS J1058+5628 & 24.66 & 0.95 &   & RXS J10586+5628 &1.00 &RXS J10586+5628 &0.143 & BLLac
 \\
0FGL J1100.2$-$8000 & CGRaBS J1058$-$8003 & 53.65 & 0.99 &   & CGRaBS J1058$-$8003 &0.99 &PKS 1057$-$79 &\nodata & BLLac
 \\
0FGL J1104.5+3811 & CGRaBS J1104+3812 & 35.10 & 0.97 &   & Mrk 421 &1.00 &Mrk 421 &0.030 & BLLac
 \\
0FGL J1129.8$-$1443 & CRATES J1130$-$1449 & 27.54 & 0.96 &   & BZQ J1130$-$1449 &0.84 &PKS 1127$-$14 &1.184 & FSRQ
 \\
0FGL J1146.7$-$3808 & CGRaBS J1147$-$3812 & 45.04 & 0.98 &   & CGRaBS J1147$-$3812 &0.99 &PKS 1144$-$379 &1.048 & FSRQ
 \\
0FGL J1159.2+2912 & CGRaBS J1159+2914 & 39.38 & 0.97 &   & CGRaBS J1159+2914 &0.98 &4C 29.45 &0.729 & FSRQ
 \\
0FGL J1218.0+3006 & CGRaBS J1217+3007 & 31.80 & 0.96 &   & B2 1215+30 &1.00 &B2 1215+30 &0.130 & BLLac
 \\
0FGL J1221.7+2814 & CGRaBS J1221+2813 & 36.82 & 0.97 &   & W Com &1.00 &W Com &0.102 & BLLac
 \\
0FGL J1229.1+0202 & CGRaBS J1229+0203 & 73.53 & 1.00 &   & 3C 273 &1.00 &3C 273 &0.158 & FSRQ
 \\
0FGL J1246.6$-$2544 & CGRaBS J1246$-$2547 & 43.45 & 0.98 &   & CGRaBS J1246$-$2547 &0.99 &PKS 1244$-$255 &0.635 & FSRQ
 \\
0FGL J1253.4+5300 & CRATES J1253+5301 & 43.34 & 0.98 &   & S4 1250+53 &1.00 &S4 1250+53 &\nodata & BLLac
 \\
0FGL J1256.1$-$0547 & CGRaBS J1256$-$0547 & 71.21 & 1.00 &   & 3C279 &1.00 &3C 279 &0.536 & FSRQ
 \\
0FGL J1310.6+3220 & CGRaBS J1310+3220 & 55.91 & 0.99 &   & CGRaBS J1310+3220 &0.99 &B2 1308+32 &0.997 & FSRQ
 \\
0FGL J1325.4$-$4303 & BZU J1325$-$4301 & 75.23 & 1.00 &   & NGC 5128 &1.00 &NGC 5128, Cen A &0.002 & RG
 \\
0FGL J1331.7$-$0506 & CGRaBS J1332$-$0509 & 44.64 & 0.98 &   & CGRaBS J1332$-$0509 &0.93 &PKS 1329$-$049 &2.150 & FSRQ
 \\
0FGL J1333.3+5058 & CLASS J1333+5057 & 21.52 & 0.94 &   & \nodata &\nodata &\nodata &1.362 & FSRQ  \\
0FGL J1355.0$-$1044 & CRATES J1354$-$1041 & 22.52 & 0.94 &   & BZUJ1354$-$1041 &0.84 &PKS 1352$-$104 &0.330 & FSRQ
 \\
0FGL J1427.1+2347 & CRATES J1427+2347 & 19.69 & 0.94 &   & PKS 1424+240 &1.00 &PKS 1424+240 &\nodata & BLLac
 \\
0FGL J1457.6$-$3538 & CGRaBS J1457$-$3539 & 26.03 & 0.95 &   & CGRaBS J1457$-$3539 &0.99 &PKS 1454$-$354 &1.424 & FSRQ
 \\
0FGL J1504.4+1030 & CGRaBS J1504+1029 & 48.85 & 0.98 &   & CGRaBS J1504+1029 &1.00 &PKS 1502+106 &1.839 & FSRQ
 \\
0FGL J1511.2$-$0536 & PKS 1508$-$05 & 10.27 & 0.90 &   & BZQJ1510$-$0543 &0.73 &PKS 1508$-$05 &1.185 & FSRQ
 \\
0FGL J1512.7$-$0905 & PKS 1510$-$08 & 74.49 & 1.00 &   & BZQJ1512$-$0905 &0.98 &PKS 1510$-$08 &0.360 & FSRQ
 \\
0FGL J1517.9$-$2423 & CGRaBS J1517$-$2422 & 19.18 & 0.94 &   & AP Lib &1.00 &AP Lib &0.048 & BLLac
 \\
0FGL J1522.2+3143 & CGRaBS J1522+3144 & 51.06 & 0.99 &   & CGRaBS J1522+3144 &1.00 &TXS 1520+319 &1.487 & FSRQ
 \\
0FGL J1543.1+6130 & CRATES J1542+6129 & 45.22 & 0.98 &   & RXS J15429+6129 &1.00 &RXS J15429+6129 &\nodata & BLLac
 \\
0FGL J1553.4+1255 & CRATES J1553+1256 & 26.38 & 0.95 &   & PKS 1551+130 &0.85 &PKS 1551+130 &1.308 & FSRQ
 \\
0FGL J1555.8+1110 & CGRaBS J1555+1111 & 44.23 & 0.98 &   & PG 1553+11 &1.00 &PG 1553+11 &0.360 & BLLac
 \\
0FGL J1625.8$-$2527 & CGRaBS J1625$-$2527 & 56.82 & 0.99 &   & PKS 1622$-$253 &0.99 &PKS 1622$-$253 &0.786 & FSRQ
 \\
0FGL J1635.2+3809 & CGRaBS J1635+3808 & 54.10 & 0.99 &   & CRATESJ1635+3808	 &0.99 &4C +38.41 &1.814 & FSRQ
 \\
0FGL J1653.9+3946 & CGRaBS J1653+3945 & 59.08 & 0.99 &   & Mrk 501 &1.00 &Mrk 501 &0.033 & BLLac
 \\
0FGL J1719.3+1746 & CGRaBS J1719+1745 & 40.87 & 0.98 &   & PKS 1717+177 &1.00 &PKS 1717+177 &0.137 & BLLac
 \\
0FGL J1751.5+0935 & CGRaBS J1751+0939 & 19.73 & 0.94 &   & CGRaBS J1751+0939 &0.99 &OT 081 &0.322 & BLLac
 \\
0FGL J1802.2+7827 & CGRaBS J1800+7828 & 28.07 & 0.96 &   & CGRaBS J1800+7828 &0.99 &S5 1803+78 &0.680 & BLLac
 \\
0FGL J1847.8+3223 & CGRaBS J1848+3219 & 12.76 & 0.92 &   & CGRaBS J1848+3219 &0.94 &TXS 1846+322 &0.798 & FSRQ
 \\
0FGL J1849.4+6706 & CGRaBS J1849+6705 & 53.89 & 0.99 &   & CGRaBS J1849+6705 &1.00 &S4 1849+67 &0.657 & FSRQ
 \\
0FGL J1911.2$-$2011 & CGRaBS J1911$-$2006 & 23.51 & 0.95 &   & CGRaBS J1911$-$2006 &0.97 &PKS 1908$-$201 &1.119 & FSRQ
 \\
0FGL J1923.3$-$2101 & CGRaBS J1923$-$2104 & 37.72 & 0.97 &   & CGRaBS J1923$-$2104 &0.97 &TXS 1920$-$211 &0.874 & FSRQ
 \\
0FGL J2000.2+6506 & CGRaBS J1959+6508 & 19.12 & 0.94 &   & 1ES 1959+650 &1.00 &1ES 1959+650 &0.047 & BLLac
 \\
0FGL J2009.4$-$4850 & CGRaBS J2009$-$4849 & 72.13 & 1.00 &   & PKS 2005$-$489 &1.00 &PKS 2005$-$489 &0.071 & BLLac
 \\
0FGL J2025.6$-$0736 & CRATES J2025$-$0735 & 42.71 & 0.98 &   & BZQJ2025$-$0735 &0.98 &PKS 2022$-$07 &1.388 & FSRQ
 \\
0FGL J2056.1$-$4715 & CGRaBS J2056$-$4714 & 67.00 & 1.00 &   & CRATES J2055$-$4716 &1.00 &PKS 2052$-$47 &1.491 & FSRQ
 \\
0FGL J2139.4$-$4238 & CRATES J2139$-$4235 & 13.48 & 0.92 &   & MH 2136$-$428 &1.00 &MH 2136$-$428 &\nodata & BLLac
 \\
0FGL J2143.2+1741 & CGRaBS J2143+1743 & 36.88 & 0.97 &   & CGRaBS J2143+1743 &0.96 &OX 169 &0.213 & FSRQ 
 \\
0FGL J2147.1+0931 & CGRaBS J2147+0929 & 53.97 & 0.99 &   & CGRaBS J2147+0929 &0.99 &PKS 2144+092 &1.113 & FSRQ
 \\
0FGL J2157.5+3125 & CGRaBS J2157+3127 & 54.48 & 0.99 &   & CGRaBS J2157+3127 &0.97 &B2 2155+31 &1.486 & FSRQ
 \\
0FGL J2158.8$-$3014 & CGRaBS J2158$-$3013 & 54.87 & 0.99 &   & CGRaBS J2158$-$3013 &1.00 &PKS 2155$-$304 &0.116 & BLLac
 \\
0FGL J2202.4+4217 & BZB J2202+4216 & 45.62 & 0.98 &   &  BZB J2139$-$4239  &1.00 &BL Lacertae &0.069 & BLLac
 \\
0FGL J2203.2+1731 & CGRaBS J2203+1725 & 23.91 & 0.95 &   & CGRaBS J2203+1725 &0.93 &PKS 2201+171 &1.076 & FSRQ
 \\
0FGL J2207.0$-$5347 & CGRaBS J2207$-$5346 & 39.56 & 0.97 &   & CGRaBS J2207$-$5346 &0.99 &PKS 2204$-$54 &1.215 & FSRQ
 \\
0FGL J2229.8$-$0829 & CGRaBS J2229$-$0832 & 42.99 & 0.98 &   & CGRaBS J2229$-$0832 &0.99 &PHL 5225 &1.560 & FSRQ
 \\
0FGL J2232.4+1141 & BZQ J2232+1143 & 45.97 & 0.98 &   & BZQ J2232+1143 &1.00 &CTA 102 &1.037 & FSRQ
 \\
0FGL J2254.0+1609 & CGRaBS J2253+1608 & 70.34 & 1.00 &   & CGRaBS J2253+1608 &1.00 &3C 454.3 &0.859 & FSRQ
 \\
0FGL J2325.3+3959 & CRATES J2325+3957 & 29.25 & 0.96 &   & B3 2322+396 &1.00 &B3 2322+396 &\nodata & BLLac
 \\
0FGL J2327.3+0947 & CGRaBS J2327+0940 & 21.12 & 0.94 &   & CGRaBS J2327+0940 &0.93 &PKS 2325+093 &1.843 & FSRQ
 \\
0FGL J2345.5$-$1559 & CGRaBS J2345$-$1555 & 30.19 & 0.96 &   & CGRaBS J2345$-$1555 &0.93 &PMN J2345$-$1555 &0.621 & FSRQ
 \\
\enddata
\tablenotetext{***}{All these source have a flat radio spectrum but there are no other data reported in literature which allow to classify them either as FSRQs or BL Lacs.}
\end{deluxetable}

\clearpage
\begin{deluxetable}{llllllllll}
\tablecolumns{10} 
\setlength{\tabcolsep}{0.04in}
\tablewidth{0pc} 
\tabletypesize{\tiny}
\tablecaption{The low-confidence association Bright AGN  List\label{table2)}}
\tablenum{2}
%\rotate
\tablehead{
\colhead{} & \multicolumn{3}{c}{FoM} & \colhead{} & \multicolumn{2}{c}{{\it gtsrcid}} & \colhead{} & \colhead{}  & \colhead{} \\ \cline{2-4} \cline{6-7} \\
\colhead{LAT Name} & 
\colhead{Source Name} &  
\colhead{FoM} & 
\colhead{Prob.} & 
\colhead{} & 
\colhead{Source Name} & 
\colhead{Prob.} & 
\colhead{Other Names} & 
\colhead{$z$} & 
\colhead{Class} \\ 
}
\startdata
0FGL J0100.2+0750 & CRATES J0100+0745 & 5.12 & 0.78 &  &\nodata & \nodata & \nodata & 0.000 &  Unc\tablenotemark{***}  \\
0FGL J0238.4+2855 & CGRaBS J0237+2848 & 7.67 & 0.89 &   & CGRaBS J0237+2848 &0.88 &B2 0234+28 &1.213 & FSRQ
 \\
0FGL J0407.6$-$3829 & CRATES J0406$-$3826 & 3.00 & 0.61 &   & \nodata &\nodata & PKS 0405$-$385 &1.285 & Unc\tablenotemark{***}  \\
0FGL J0412.9$-$5341 & CRATES J0413$-$5332 & 1.92 & 0.46 &   & \nodata &0.00 &\nodata &\nodata & Unc\tablenotemark{***}  \\
0FGL J0423.1$-$0112 & CGRaBS J0423$-$0120 & 4.26 & 0.72 &   & CRATESJ0423$-$0120 &0.84 &PKS 0420$-$014 &0.915 & FSRQ
 \\
0FGL J0909.7+0145 & CGRaBS J0909+0200 & 4.16 & 0.71 &   & PKS 0907+022 &0.87 &PKS 0907+022 &\nodata & BLLac
 \\
0FGL J1034.0+6051 & CRATES J1032+6051 & 5.22 & 0.79 &   & \nodata &\nodata &\nodata &1.064 & FSRQ
 \\
0FGL J1248.7+5811 & \nodata & \nodata & \nodata &   & PG 1246+586 &0.86 &\nodata &\nodata & BLLac\\
0FGL J1625.9$-$2423 & CRATES J1627$-$2426 & 2.33 & 0.53 &   & \nodata &\nodata &\nodata &\nodata & Unc\tablenotemark{***}  \\
0FGL J1641.4+3939 & CLASS J1641+3935 & 6.22 & 0.85 &   & \nodata &\nodata &\nodata &0.539 & FSRQ\\
0FGL J2017.2+0602 & CLASS J2017+0603 & 7.03 & 0.88 &   & \nodata &\nodata &\nodata &\nodata & Unc\tablenotemark{***}  \\
\enddata
%%\tablenotetext{*}{These sources can be considered intermediate between BL Lac and FSRQ objects because of their peculiar optical spectra. }
%%\tablenotetext{**}{The optical spectra of these sources showing only one broad emission line without a measured EW. } 
\tablenotetext{***}{All these sources have a flat radio spectrum but there are no other data reported in literature which allow to classify them either as FSRQs or BL Lacs.}
\end{deluxetable}

\begin{deluxetable}{lrrrrrrrrrc}
\tablecolumns{11}
\tablewidth{0pc}
\tabletypesize{\tiny}
\tablecaption{The {\it Fermi}\,$-$\,LAT $|{b}|>10^{\circ}$ Bright AGN List\label{table3}}
\tablenum{3}
%\rotate
\tablehead{
%\colhead{ASO Name} &
\colhead{LAT Name} &
\colhead{R.A.} &
\colhead{Dec} &
\colhead{$l$} &
\colhead{$b$} &
\colhead{$\sqrt{TS}$} &
\colhead{$\Gamma$\tablenotemark{a}} &
\colhead{$F_{100}$\tablenotemark{b}} &
\colhead{$F_{peak}$\tablenotemark{c}} &
\colhead{$F_{25}$\tablenotemark{d}} &
\colhead{Var.}  \\ }
\startdata
0FGL  J0017.4$$-$$0503  &  4.358 & $-$5.054 & 101.273 & $-$66.485 &  14.7 &  2.71\,$\pm$\,0.14 & 13.9\,$\pm$\,2.4 & 34.8\,$\pm$\,6.5 & 12.1\,$\pm$\,1.4 & T \\
0FGL  J0033.6$$-$$1921  &  8.401 & $-$19.360 & 94.215 & $-$81.220 &  10.7 &  1.70\,$\pm$\,0.14 & 1.6\,$\pm$\,0.4 & 2.9\,$\pm$\,1.3 & 0.4\,$\pm$\,0.1$\tablenotemark{\dagger}$ & \nodata \\
0FGL  J0050.5$$-$$0928  &  12.637 & $-$9.470 & 122.209 & $-$72.341 &  20.5 &  2.15\,$\pm$\,0.08 & 10.2\,$\pm$\,1.4 & 19.0\,$\pm$\,4.0 & 8.8\,$\pm$\,1.3 & T \\
0FGL  J0051.1$$-$$0647  &  12.796 & $-$6.794 & 122.751 & $-$69.666 &  15.7 &  2.22\,$\pm$\,0.11 & 8.5\,$\pm$\,1.5 & 19.7\,$\pm$\,4.4 & 7.2\,$\pm$\,1.4 & T \\
{\it 0FGL  J0100.2+0750}  &  15.051 & 7.844 & 126.716 & $-$54.963 &  11.1 &  1.80\,$\pm$\,0.16 & 1.9\,$\pm$\,0.7 & 3.9\,$\pm$\,1.7 & 0.3\,$\pm$\,0.1$\tablenotemark{\dagger}$ & \nodata \\
0FGL  J0112.1+2247  &  18.034 & 22.789 & 129.148 & $-$39.832 &  17.6 &  2.10\,$\pm$\,0.09 & 7.4\,$\pm$\,1.2 & 12.6\,$\pm$\,2.7 & 6.0\,$\pm$\,0.7 & \nodata \\
0FGL  J0118.7$$-$$2139  &  19.676 & $-$21.656 & 172.990 & $-$81.728 &  17.8 &  2.32\,$\pm$\,0.10 & 9.6\,$\pm$\,1.4 & 21.4\,$\pm$\,4.5 & 7.6\,$\pm$\,1.1 & T \\
0FGL  J0120.5$$-$$2703  &  20.128 & $-$27.056 & 213.951 & $-$83.529 &  11.8 &  1.99\,$\pm$\,0.14 & 3.2\,$\pm$\,0.8 & 6.7\,$\pm$\,2.3 & 2.6\,$\pm$\,0.8 & \nodata \\
0FGL  J0136.6+3903  &  24.163 & 39.066 & 132.446 & $-$22.969 &  12.5 &  1.65\,$\pm$\,0.13 & 1.8\,$\pm$\,0.5 & 4.7\,$\pm$\,1.5 & 0.5\,$\pm$\,0.1$\tablenotemark{\dagger}$ & \nodata \\
0FGL  J0137.1+4751  &  24.285 & 47.854 & 130.818 & $-$14.317 &  18.8 &  2.20\,$\pm$\,0.09 & 10.9\,$\pm$\,1.7 & 18.6\,$\pm$\,4.5 & 10.8\,$\pm$\,1.6 & T \\
0FGL  J0144.5+2709  &  26.142 & 27.159 & 137.248 & $-$34.231 &  10.4 &  2.22\,$\pm$\,0.14 & 5.4\,$\pm$\,1.3 & 12.7\,$\pm$\,3.8 & 2.0\,$\pm$\,0.5 & \nodata \\
0FGL  J0145.1$$-$$2728  &  26.289 & $-$27.478 & 217.694 & $-$78.067 &  13.4 &  2.55\,$\pm$\,0.14 & 9.2\,$\pm$\,1.7 & 26.3\,$\pm$\,5.4 & 9.4\,$\pm$\,1.3 & T \\
0FGL  J0204.8$$-$$1704  &  31.219 & $-$17.068 & 186.072 & $-$70.274 &  16.6 &  2.48\,$\pm$\,0.11 & 11.1\,$\pm$\,1.7 & 18.9\,$\pm$\,3.9 & 10.7\,$\pm$\,1.3 & \nodata \\
0FGL  J0210.8$$-$$5100  &  32.706 & $-$51.013 & 276.083 & $-$61.776 &  34.1 &  2.28\,$\pm$\,0.06 & 24.4\,$\pm$\,2.0 & 76.2\,$\pm$\,6.9 & 22.8\,$\pm$\,1.2 & T \\
0FGL  J0217.8+0146  &  34.467 & 1.768 & 162.139 & $-$54.389 &  21.7 &  2.13\,$\pm$\,0.08 & 10.2\,$\pm$\,1.3 & 16.5\,$\pm$\,3.8 & 9.8\,$\pm$\,1.2 & T \\
0FGL  J0220.9+3607  &  35.243 & 36.121 & 142.504 & $-$23.325 &  12.3 &  2.61\,$\pm$\,0.16 & 11.0\,$\pm$\,2.4 & 22.5\,$\pm$\,6.1 & 10.9\,$\pm$\,1.3 & \nodata \\
0FGL  J0222.6+4302  &  35.653 & 43.043 & 140.132 & $-$16.763 &  47.4 &  1.97\,$\pm$\,0.04 & 25.9\,$\pm$\,1.6 & 49.6\,$\pm$\,4.8 & 26.6\,$\pm$\,1.4 & T \\
0FGL  J0229.5$$-$$3640  &  37.375 & $-$36.681 & 243.801 & $-$67.189 &  19.2 &  2.57\,$\pm$\,0.11 & 15.8\,$\pm$\,2.1 & 34.1\,$\pm$\,6.2 & 14.1\,$\pm$\,1.5 & T \\
{\it 0FGL  J0238.4+2855}  &  39.600 & 28.923 & 149.521 & $-$28.368 &  10.9 &  2.49\,$\pm$\,0.15 & 9.0\,$\pm$\,2.0 & 24.7\,$\pm$\,5.9 & 8.6\,$\pm$\,1.6 & \nodata \\
0FGL  J0238.6+1636  &  39.663 & 16.613 & 156.775 & $-$39.112 &  85.7 &  2.05\,$\pm$\,0.02 & 72.6\,$\pm$\,2.5 & 104.8\,$\pm$\,7.1 & 67.6\,$\pm$\,2.2 & T \\
0FGL  J0245.6$$-$$4656  &  41.423 & $-$46.934 & 262.019 & $-$60.098 &  11.4 &  2.34\,$\pm$\,0.15 & 6.2\,$\pm$\,1.5 & 12.4\,$\pm$\,4.0 & 5.6\,$\pm$\,0.8 & \nodata \\
0FGL  J0303.7$$-$$2410  &  45.940 & $-$24.176 & 214.764 & $-$60.119 &  12.3 &  2.01\,$\pm$\,0.13 & 3.8\,$\pm$\,0.9 & 8.0\,$\pm$\,2.8 & 2.9\,$\pm$\,0.9 & \nodata \\
0FGL  J0320.0+4131  &  50.000 & 41.524 & 150.601 & $-$13.230 &  29.7 &  2.17\,$\pm$\,0.06 & 22.1\,$\pm$\,1.9 & 35.9\,$\pm$\,5.3 & 18.2\,$\pm$\,1.4 & T \\
0FGL  J0334.1$$-$$4006  &  53.546 & $-$40.107 & 244.710 & $-$54.088 &  13.2 &  2.15\,$\pm$\,0.12 & 5.3\,$\pm$\,1.1 & 11.2\,$\pm$\,3.1 & 4.9\,$\pm$\,1.4 & \nodata \\
0FGL  J0349.8$$-$$2102  &  57.465 & $-$21.046 & 214.385 & $-$49.035 &  21.2 &  2.55\,$\pm$\,0.09 & 19.2\,$\pm$\,2.3 & 27.8\,$\pm$\,5.0 & 17.3\,$\pm$\,1.6 & \nodata \\
{\it 0FGL  J0407.6$$-$$3829}  &  61.923 & $-$38.491 & 241.360 & $-$47.751 &  13.5 &  2.31\,$\pm$\,0.13 & 7.5\,$\pm$\,1.5 & 22.2\,$\pm$\,4.1 & 6.9\,$\pm$\,1.3 & T \\
{\it 0FGL  J0412.9$$-$$5341}  &  63.230 & $-$53.686 & 263.001 & $-$44.716 &  10.7 &  2.30\,$\pm$\,0.15 & 5.4\,$\pm$\,1.3 & 12.3\,$\pm$\,3.8 & 6.0\,$\pm$\,1.3 & \nodata \\
{\it 0FGL  J0423.1$$-$$0112}  &  65.785 & $-$1.204 & 195.131 & $-$33.092 &  11.5 &  2.38\,$\pm$\,0.16 & 8.1\,$\pm$\,2.2 & 13.4\,$\pm$\,4.0 & 10.5\,$\pm$\,3.1 & \nodata \\
0FGL  J0428.7$$-$$3755  &  67.193 & $-$37.923 & 240.689 & $-$43.597 &  39.6 &  2.14\,$\pm$\,0.05 & 24.5\,$\pm$\,1.8 & 31.5\,$\pm$\,4.7 & 23.1\,$\pm$\,1.6 & \nodata \\
0FGL  J0449.7$$-$$4348  &  72.435 & $-$43.815 & 248.780 & $-$39.859 &  28.4 &  2.01\,$\pm$\,0.06 & 12.0\,$\pm$\,1.3 & 21.1\,$\pm$\,4.2 & 12.2\,$\pm$\,1.4 & \nodata \\
0FGL  J0457.1$$-$$2325  &  74.288 & $-$23.432 & 223.739 & $-$34.880 &  52.3 &  2.23\,$\pm$\,0.04 & 41.8\,$\pm$\,2.3 & 64.2\,$\pm$\,6.4 & 36.6\,$\pm$\,1.8 & T \\
0FGL  J0507.9+6739  &  76.985 & 67.650 & 143.772 & 15.905 &  13.2 &  1.67\,$\pm$\,0.18 & 1.7\,$\pm$\,0.8 & 5.2\,$\pm$\,1.7 & 0.3\,$\pm$\,0.1$\tablenotemark{\dagger}$ & \nodata \\
0FGL  J0516.2$$-$$6200  &  79.063 & $-$62.000 & 271.376 & $-$34.834 &  11.2 &  2.17\,$\pm$\,0.17 & 5.4\,$\pm$\,1.7 & 11.1\,$\pm$\,3.3 & 0.4\,$\pm$\,0.1$\tablenotemark{\dagger}$ & \nodata \\
0FGL  J0531.0+1331  &  82.761 & 13.528 & 191.385 & $-$10.992 &  17.3 &  2.54\,$\pm$\,0.09 & 24.3\,$\pm$\,2.9 & 39.5\,$\pm$\,6.7 & 23.6\,$\pm$\,2.1 & T \\
0FGL  J0538.8$$-$$4403  &  84.725 & $-$44.062 & 250.057 & $-$31.075 &  48.6 &  2.19\,$\pm$\,0.04 & 37.6\,$\pm$\,2.2 & 49.7\,$\pm$\,5.6 & 34.3\,$\pm$\,1.8 & T \\
0FGL  J0654.3+4513  &  103.590 & 45.220 & 171.228 & 19.369 &  29.2 &  2.32\,$\pm$\,0.06 & 23.8\,$\pm$\,2.1 & 56.4\,$\pm$\,7.2 & 20.3\,$\pm$\,1.6 & T \\
0FGL  J0654.3+5042  &  103.592 & 50.711 & 165.676 & 21.107 &  15.6 &  2.00\,$\pm$\,0.10 & 5.5\,$\pm$\,1.1 & 9.5\,$\pm$\,2.6 & 4.9\,$\pm$\,1.3 & T \\
0FGL  J0700.0$$-$$6611  &  105.016 & $-$66.199 & 276.778 & $-$23.809 &  10.1 &  1.98\,$\pm$\,0.14 & 3.9\,$\pm$\,1.0 & 8.7\,$\pm$\,2.6 & 0.4\,$\pm$\,0.1 $\tablenotemark{\dagger}$& \nodata \\
0FGL  J0712.9+5034  &  108.231 & 50.575 & 166.688 & 23.900 &  11.2 &  2.04\,$\pm$\,0.14 & 3.9\,$\pm$\,1.1 & 10.5\,$\pm$\,2.7 & 3.3\,$\pm$\,0.7 & \nodata \\
0FGL  J0714.2+1934  &  108.552 & 19.574 & 197.685 & 13.648 &  15.0 &  2.35\,$\pm$\,0.10 & 10.7\,$\pm$\,1.6 & 27.0\,$\pm$\,5.0 & 10.0\,$\pm$\,1.6 & T \\
0FGL  J0719.4+3302  &  109.869 & 33.037 & 185.139 & 19.855 &  12.3 &  2.37\,$\pm$\,0.15 & 7.8\,$\pm$\,1.7 & 20.8\,$\pm$\,4.9 & 7.5\,$\pm$\,1.5 & T \\
0FGL  J0722.0+7120  &  110.508 & 71.348 & 143.976 & 28.029 &  34.4 &  2.08\,$\pm$\,0.05 & 16.4\,$\pm$\,1.4 & 29.0\,$\pm$\,4.2 & 17.0\,$\pm$\,1.6 & T \\
0FGL  J0738.2+1738  &  114.575 & 17.634 & 201.933 & 18.081 &  11.9 &  2.10\,$\pm$\,0.14 & 4.6\,$\pm$\,1.1 & 7.5\,$\pm$\,2.4 & 3.6\,$\pm$\,1.4 & \nodata \\
0FGL  J0818.3+4222  &  124.579 & 42.367 & 178.244 & 33.409 &  20.9 &  2.07\,$\pm$\,0.08 & 9.6\,$\pm$\,1.3 & 14.5\,$\pm$\,2.9 & 7.0\,$\pm$\,1.1 & \nodata \\
0FGL  J0824.9+5551  &  126.239 & 55.859 & 161.981 & 35.142 &  10.6 &  2.81\,$\pm$\,0.20 & 11.4\,$\pm$\,2.9 & 42.0\,$\pm$\,8.1 & 10.8\,$\pm$\,1.3 & T \\
0FGL  J0855.4+2009  &  133.857 & 20.162 & 206.810 & 35.974 &  15.1 &  2.31\,$\pm$\,0.11 & 9.0\,$\pm$\,1.5 & 19.0\,$\pm$\,4.1 & 7.8\,$\pm$\,1.3 & \nodata \\
{\it 0FGL  J0909.7+0145}  &  137.446 & 1.757 & 228.640 & 31.262 &  11.6 &  2.67\,$\pm$\,0.16 & 10.4\,$\pm$\,2.1 & 22.9\,$\pm$\,6.1 & 9.5\,$\pm$\,0.3 & \nodata \\
0FGL  J0921.2+4437  &  140.320 & 44.617 & 175.809 & 44.876 &  15.2 &  2.35\,$\pm$\,0.12 & 8.6\,$\pm$\,1.5 & 15.7\,$\pm$\,4.2 & 9.2\,$\pm$\,1.4 & \nodata \\
0FGL  J0948.3+0019  &  147.077 & 0.317 & 236.530 & 38.549 &  12.8 &  2.60\,$\pm$\,0.14 & 12.1\,$\pm$\,2.2 & 29.2\,$\pm$\,5.7 & 9.1\,$\pm$\,1.4 & T \\
0FGL  J0957.6+5522  &  149.424 & 55.375 & 158.605 & 47.939 &  24.0 &  2.01\,$\pm$\,0.07 & 8.7\,$\pm$\,1.1 & 12.9\,$\pm$\,3.0 & 9.2\,$\pm$\,1.3 & \nodata \\
0FGL  J1012.9+2435  &  153.241 & 24.598 & 207.897 & 54.406 &  12.4 &  2.22\,$\pm$\,0.12 & 6.1\,$\pm$\,1.2 & 10.9\,$\pm$\,3.6 & 4.5\,$\pm$\,0.9 & T \\
0FGL  J1015.2+4927  &  153.809 & 49.463 & 165.473 & 52.727 &  23.8 &  1.73\,$\pm$\,0.07 & 4.9\,$\pm$\,0.7 & 7.1\,$\pm$\,1.7 & 8.9\,$\pm$\,1.5 & \nodata \\
0FGL  J1015.9+0515  &  153.991 & 5.254 & 236.457 & 47.036 &  20.6 &  2.20\,$\pm$\,0.08 & 11.7\,$\pm$\,1.5 & 21.8\,$\pm$\,4.5 & 13.1\,$\pm$\,1.5 & T \\
0FGL  J1034.0+6051  &  158.504 & 60.853 & 147.765 & 49.122 &  14.8 &  2.48\,$\pm$\,0.13 & 9.3\,$\pm$\,1.7 & 22.0\,$\pm$\,4.7 & 7.5\,$\pm$\,1.3 & \nodata \\
0FGL  J1053.7+4926  &  163.442 & 49.449 & 160.309 & 58.263 &  10.1 &  1.42\,$\pm$\,0.20 & 0.5\,$\pm$\,0.3 & 1.8\,$\pm$\,0.9 & 0.2\,$\pm$\,0.1$\tablenotemark{\dagger}$ & \nodata \\
0FGL  J1054.5+2212  &  163.626 & 22.215 & 216.968 & 63.049 &  11.2 &  2.24\,$\pm$\,0.15 & 4.9\,$\pm$\,1.3 & 10.8\,$\pm$\,3.3 & 4.3\,$\pm$\,1.0 & \nodata \\
0FGL  J1057.8+0138  &  164.451 & 1.643 & 251.219 & 52.709 &  10.3 &  2.20\,$\pm$\,0.17 & 5.0\,$\pm$\,1.4 & 10.7\,$\pm$\,2.8 & 9.2\,$\pm$\,1.7 & \nodata \\
0FGL  J1058.9+5629  &  164.731 & 56.488 & 149.521 & 54.442 &  12.0 &  2.11\,$\pm$\,0.14 & 3.9\,$\pm$\,1.0 & 8.3\,$\pm$\,2.7 & 5.0\,$\pm$\,1.8 & \nodata \\
0FGL  J1100.2$$-$$8000  &  165.057 & $-$80.012 & 298.047 & $-$18.212 &  12.1 &  2.71\,$\pm$\,0.16 & 17.1\,$\pm$\,3.8 & 38.4\,$\pm$\,8.5 & 11.1\,$\pm$\,2.2 & T \\
0FGL  J1104.5+3811  &  166.137 & 38.187 & 179.868 & 65.056 &  47.1 &  1.77\,$\pm$\,0.04 & 15.3\,$\pm$\,1.1 & 20.9\,$\pm$\,3.1 & 15.9\,$\pm$\,1.3 & \nodata \\
0FGL  J1129.8$$-$$1443  &  172.454 & $-$14.727 & 275.133 & 43.694 &  10.5 &  2.69\,$\pm$\,0.18 & 9.9\,$\pm$\,2.4 & 25.8\,$\pm$\,5.8 & 10.8\,$\pm$\,1.6 & \nodata \\
0FGL  J1146.7$$-$$3808  &  176.689 & $-$38.149 & 289.170 & 22.988 &  10.4 &  2.21\,$\pm$\,0.14 & 5.7\,$\pm$\,1.4 & 7.5\,$\pm$\,2.8 & 3.5\,$\pm$\,1.2 & \nodata \\
0FGL  J1159.2+2912  &  179.800 & 29.216 & 199.605 & 78.307 &  14.6 &  2.47\,$\pm$\,0.13 & 10.3\,$\pm$\,1.8 & 16.0\,$\pm$\,3.8 & 9.4\,$\pm$\,1.0 & \nodata \\
0FGL  J1218.0+3006  &  184.517 & 30.108 & 188.826 & 82.097 &  27.4 &  1.89\,$\pm$\,0.06 & 9.7\,$\pm$\,1.1 & 40.9\,$\pm$\,4.7 & 10.4\,$\pm$\,1.0 & T \\
0FGL  J1221.7+2814  &  185.439 & 28.243 & 201.593 & 83.336 &  24.0 &  1.93\,$\pm$\,0.07 & 8.3\,$\pm$\,1.1 & 17.2\,$\pm$\,3.5 & 7.5\,$\pm$\,0.9 & T \\
0FGL  J1229.1+0202  &  187.287 & 2.045 & 289.975 & 64.355 &  52.0 &  2.71\,$\pm$\,0.05 & 75.2\,$\pm$\,4.3 & 137.0\,$\pm$\,13.0 & 65.5\,$\pm$\,2.6 & T \\
0FGL  J1246.6$$-$$2544  &  191.655 & $-$25.734 & 301.571 & 37.125 &  11.7 &  2.24\,$\pm$\,0.14 & 6.8\,$\pm$\,1.6 & 15.3\,$\pm$\,4.3 & 7.6\,$\pm$\,1.4 & \nodata \\
{\it 0FGL  J1248.7+5811}  &  192.189 & 58.191 & 123.617 & 58.934 &  14.3 &  1.95\,$\pm$\,0.11 & 3.8\,$\pm$\,0.8 & 8.0\,$\pm$\,2.4 & 7.4\,$\pm$\,1.6 & \nodata \\
0FGL  J1253.4+5300  &  193.369 & 53.001 & 122.229 & 64.125 &  12.1 &  2.17\,$\pm$\,0.14 & 4.7\,$\pm$\,1.1 & 9.1\,$\pm$\,2.6 & 5.6\,$\pm$\,1.5 & \nodata \\
0FGL  J1256.1$$-$$0547  &  194.034 & $-$5.800 & 305.081 & 57.052 &  36.8 &  2.35\,$\pm$\,0.05 & 31.5\,$\pm$\,2.3 & 46.3\,$\pm$\,6.8 & 29.7\,$\pm$\,1.8 & T \\
0FGL  J1310.6+3220  &  197.656 & 32.339 & 85.458 & 83.331 &  27.3 &  2.25\,$\pm$\,0.07 & 15.5\,$\pm$\,1.6 & 37.3\,$\pm$\,4.6 & 16.4\,$\pm$\,1.1 & T \\
0FGL  J1325.4$$-$$4303  &  201.353 & $-$43.062 & 309.501 & 19.376 &  12.4 &  2.91\,$\pm$\,0.18 & 21.5\,$\pm$\,4.5 & 32.3\,$\pm$\,8.0 & 22.2\,$\pm$\,2.4 & \nodata \\
0FGL  J1331.7$$-$$0506  &  202.935 & $-$5.112 & 321.247 & 56.320 &  14.3 &  2.59\,$\pm$\,0.12 & 13.0\,$\pm$\,2.1 & 33.0\,$\pm$\,5.9 & 10.7\,$\pm$\,1.2 & T \\
0FGL  J1333.3+5058  &  203.331 & 50.973 & 107.300 & 64.865 &  12.4 &  2.40\,$\pm$\,0.14 & 7.2\,$\pm$\,1.5 & 13.7\,$\pm$\,4.6 & 9.1\,$\pm$\,1.3 & T \\
0FGL  J1355.0$$-$$1044  &  208.764 & $-$10.735 & 327.221 & 49.113 &  11.5 &  2.37\,$\pm$\,0.15 & 7.6\,$\pm$\,1.8 & 34.4\,$\pm$\,5.5 & 8.7\,$\pm$\,1.3 & T \\
0FGL  J1427.1+2347  &  216.794 & 23.785 & 29.472 & 68.166 &  24.1 &  1.80\,$\pm$\,0.07 & 6.2\,$\pm$\,0.8 & 8.7\,$\pm$\,2.2 & 5.1\,$\pm$\,1.0 & \nodata \\
0FGL  J1457.6$$-$$3538  &  224.407 & $-$35.639 & 329.936 & 20.530 &  39.6 &  2.24\,$\pm$\,0.05 & 36.6\,$\pm$\,2.4 & 77.2\,$\pm$\,7.1 & 32.1\,$\pm$\,0.5 & T \\
0FGL  J1504.4+1030  &  226.115 & 10.505 & 11.409 & 54.577 &  88.2 &  2.17\,$\pm$\,0.02 & 81.4\,$\pm$\,2.7 & 260.0\,$\pm$\,15.0 & 69.3\,$\pm$\,2.1 & T \\
0FGL  J1511.2$$-$$0536  &  227.814 & $-$5.613 & 354.099 & 42.948 &  10.8 &  2.41\,$\pm$\,0.15 & 8.8\,$\pm$\,2.1 & 16.2\,$\pm$\,4.4 & 8.6\,$\pm$\,1.7 & \nodata \\
0FGL  J1512.7$$-$$0905  &  228.196 & $-$9.093 & 351.282 & 40.153 &  45.0 &  2.48\,$\pm$\,0.05 & 55.8\,$\pm$\,3.3 & 165.9\,$\pm$\,11.7 & 50.6\,$\pm$\,2.3 & T \\
0FGL  J1517.9$$-$$2423  &  229.496 & $-$24.395 & 340.724 & 27.521 &  12.3 &  1.94\,$\pm$\,0.14 & 4.1\,$\pm$\,1.2 & 7.0\,$\pm$\,2.4 & 5.2\,$\pm$\,0.6 & \nodata \\
0FGL  J1522.2+3143  &  230.552 & 31.726 & 50.143 & 57.014 &  34.3 &  2.39\,$\pm$\,0.06 & 25.7\,$\pm$\,2.1 & 42.0\,$\pm$\,5.1 & 22.2\,$\pm$\,1.5 & T \\
0FGL  J1543.1+6130  &  235.784 & 61.504 & 95.383 & 45.370 &  10.5 &  2.00\,$\pm$\,0.15 & 2.5\,$\pm$\,0.7 & 4.3\,$\pm$\,1.7 & 3.3\,$\pm$\,1.4 & \nodata \\
0FGL  J1553.4+1255  &  238.368 & 12.922 & 23.746 & 45.225 &  23.7 &  2.23\,$\pm$\,0.07 & 16.1\,$\pm$\,1.8 & 33.6\,$\pm$\,5.6 & 15.6\,$\pm$\,2.2 & T \\
0FGL  J1555.8+1110  &  238.951 & 11.181 & 21.911 & 43.941 &  31.5 &  1.70\,$\pm$\,0.06 & 8.0\,$\pm$\,1.0 & 11.6\,$\pm$\,2.3 & 10.2\,$\pm$\,2.0 & \nodata \\
0FGL  J1625.8$$-$$2527  &  246.470 & $-$25.451 & 352.164 & 16.308 &  11.4 &  2.40\,$\pm$\,0.15 & 16.0\,$\pm$\,4.6 & 28.4\,$\pm$\,8.0 & 19.8\,$\pm$\,1.3 & \nodata \\
{\it 0FGL  J1625.9$$-$$2423}  &  246.494 & $-$24.393 & 353.005 & 16.995 &  10.1 &  2.46\,$\pm$\,0.14 & 19.9\,$\pm$\,5.1 & 32.1\,$\pm$\,8.1 & 10.7\,$\pm$\,0.9 & \nodata \\
0FGL  J1635.2+3809  &  248.821 & 38.158 & 61.118 & 42.333 &  27.3 &  2.44\,$\pm$\,0.07 & 22.0\,$\pm$\,2.2 & 49.8\,$\pm$\,6.0 & 17.6\,$\pm$\,1.3 & T \\
{\it 0FGL  J1641.4+3939}  &  250.355 & 39.666 & 63.239 & 41.239 &  17.7 &  2.43\,$\pm$\,0.10 & 13.4\,$\pm$\,2.1 & 33.7\,$\pm$\,6.3 & 12.7\,$\pm$\,1.4 & T \\
0FGL  J1653.9+3946  &  253.492 & 39.767 & 63.612 & 38.841 &  19.0 &  1.70\,$\pm$\,0.09 & 3.1\,$\pm$\,0.6 & 6.9\,$\pm$\,1.8 & 3.3\,$\pm$\,0.8 & \nodata \\
0FGL  J1719.3+1746  &  259.830 & 17.768 & 39.553 & 28.080 &  23.3 &  1.84\,$\pm$\,0.07 & 6.9\,$\pm$\,0.9 & 15.0\,$\pm$\,3.0 & 6.3\,$\pm$\,1.2 & T \\
0FGL  J1751.5+0935  &  267.893 & 9.591 & 34.867 & 17.614 &  23.1 &  2.27\,$\pm$\,0.07 & 18.4\,$\pm$\,2.1 & 41.4\,$\pm$\,6.3 & 17.8\,$\pm$\,1.9 & T \\
0FGL  J1802.2+7827  &  270.567 & 78.466 & 110.026 & 28.990 &  12.6 &  2.25\,$\pm$\,0.14 & 6.0\,$\pm$\,1.4 & 11.1\,$\pm$\,3.1 & 5.9\,$\pm$\,1.3 & \nodata \\
0FGL  J1847.8+3223  &  281.954 & 32.385 & 62.065 & 14.838 &  16.0 &  2.37\,$\pm$\,0.10 & 14.7\,$\pm$\,2.4 & 28.0\,$\pm$\,4.9 & 9.4\,$\pm$\,0.6 & T \\
0FGL  J1849.4+6706  &  282.365 & 67.102 & 97.503 & 25.027 &  28.0 &  2.17\,$\pm$\,0.06 & 15.9\,$\pm$\,1.5 & 28.8\,$\pm$\,4.1 & 14.9\,$\pm$\,1.5 & T \\
0FGL  J1911.2$$-$$2011  &  287.813 & $-$20.186 & 16.818 & $-$13.266 &  20.0 &  2.43\,$\pm$\,0.08 & 22.5\,$\pm$\,2.7 & 52.3\,$\pm$\,7.2 & 18.7\,$\pm$\,0.8 & T \\
0FGL  J1923.3$$-$$2101  &  290.840 & $-$21.031 & 17.205 & $-$16.199 &  16.4 &  2.31\,$\pm$\,0.10 & 13.1\,$\pm$\,2.0 & 41.6\,$\pm$\,6.1 & 11.3\,$\pm$\,0.6 & T \\
0FGL  J2000.2+6506  &  300.053 & 65.105 & 97.974 & 17.630 &  15.8 &  1.86\,$\pm$\,0.11 & 4.2\,$\pm$\,1.0 & 6.3\,$\pm$\,2.1 & 3.4\,$\pm$\,1.3 & \nodata \\
0FGL  J2009.4$$-$$4850  &  302.363 & $-$48.843 & 350.361 & $-$32.607 &  10.9 &  1.85\,$\pm$\,0.14 & 2.9\,$\pm$\,0.9 & 5.5\,$\pm$\,2.1 & 3.0\,$\pm$\,0.6 & \nodata \\
{\it 0FGL  J2017.2+0602}  &  304.302 & 6.048 & 48.596 & $-$15.991 &  12.7 &  1.87\,$\pm$\,0.12 & 3.7\,$\pm$\,0.9 & 6.6\,$\pm$\,2.3 & 0.6\,$\pm$\,0.1$\tablenotemark{\dagger}$ & \nodata \\
0FGL  J2025.6$$-$$0736  &  306.415 & $-$7.611 & 36.883 & $-$24.389 &  50.6 &  2.30\,$\pm$\,0.04 & 48.0\,$\pm$\,2.6 & 73.6\,$\pm$\,7.1 & 43.0\,$\pm$\,2.0 & T \\
0FGL  J2056.1$$-$$4715  &  314.034 & $-$47.251 & 352.586 & $-$40.358 &  12.5 &  2.56\,$\pm$\,0.15 & 11.1\,$\pm$\,2.3 & 21.1\,$\pm$\,5.2 & 10.7\,$\pm$\,1.7 & \nodata \\
0FGL  J2139.4$$-$$4238  &  324.865 & $-$42.642 & 358.237 & $-$48.332 &  20.1 &  2.01\,$\pm$\,0.08 & 8.0\,$\pm$\,1.2 & 13.1\,$\pm$\,3.0 & 7.7\,$\pm$\,1.3 & \nodata \\
0FGL  J2143.2+1741  &  325.807 & 17.688 & 72.016 & $-$26.051 &  14.5 &  2.57\,$\pm$\,0.12 & 14.1\,$\pm$\,2.2 & 30.7\,$\pm$\,6.2 & 12.0\,$\pm$\,1.7 & \nodata \\
0FGL  J2147.1+0931  &  326.777 & 9.519 & 65.805 & $-$32.236 &  19.9 &  2.53\,$\pm$\,0.10 & 16.6\,$\pm$\,2.1 & 34.7\,$\pm$\,6.1 & 16.6\,$\pm$\,1.6 & T \\
0FGL  J2157.5+3125  &  329.384 & 31.431 & 84.747 & $-$18.258 &  10.0 &  2.41\,$\pm$\,0.15 & 7.5\,$\pm$\,1.7 & 13.9\,$\pm$\,3.9 & 7.3\,$\pm$\,1.5 & \nodata \\
0FGL  J2158.8$$-$$3014  &  329.704 & $-$30.237 & 17.711 & $-$52.236 &  43.9 &  1.85\,$\pm$\,0.04 & 18.1\,$\pm$\,1.2 & 29.2\,$\pm$\,3.6 & 18.5\,$\pm$\,1.4 & T \\
0FGL  J2202.4+4217  &  330.622 & 42.299 & 92.569 & $-$10.398 &  12.3 &  2.24\,$\pm$\,0.12 & 8.5\,$\pm$\,1.8 & 12.8\,$\pm$\,4.3 & 8.0\,$\pm$\,2.0 & \nodata \\
0FGL  J2203.2+1731  &  330.815 & 17.532 & 75.715 & $-$29.529 &  12.7 &  2.25\,$\pm$\,0.13 & 6.9\,$\pm$\,1.4 & 17.0\,$\pm$\,3.5 & 8.3\,$\pm$\,1.6 & T \\
0FGL  J2207.0$$-$$5347  &  331.765 & $-$53.786 & 339.948 & $-$49.832 &  12.4 &  2.65\,$\pm$\,0.17 & 11.5\,$\pm$\,2.5 & 54.6\,$\pm$\,8.0 & 11.1\,$\pm$\,1.8 & T \\
0FGL  J2229.8$$-$$0829  &  337.452 & $-$8.495 & 55.326 & $-$51.701 &  16.8 &  2.67\,$\pm$\,0.12 & 15.9\,$\pm$\,2.4 & 27.7\,$\pm$\,5.7 & 12.0\,$\pm$\,0.4 & \nodata \\
0FGL  J2232.4+1141  &  338.117 & 11.690 & 77.372 & $-$38.592 &  15.2 &  2.61\,$\pm$\,0.12 & 14.0\,$\pm$\,2.3 & 24.6\,$\pm$\,6.2 & 11.2\,$\pm$\,1.3 & \nodata \\
0FGL  J2254.0+1609  &  343.502 & 16.151 & 86.125 & $-$38.187 &  149.1 &  2.41\,$\pm$\,0.02 & 246.1\,$\pm$\,5.2 & 385.8\,$\pm$\,20.5 & 221.6\,$\pm$\,4.3 & T \\
0FGL  J2325.3+3959  &  351.334 & 39.993 & 105.532 & $-$19.952 &  11.4 &  1.89\,$\pm$\,0.13 & 2.8\,$\pm$\,0.8 & 11.0\,$\pm$\,2.7 & 1.3\,$\pm$\,0.4 & T \\
0FGL  J2327.3+0947  &  351.833 & 9.794 & 91.159 & $-$47.821 &  17.1 &  2.73\,$\pm$\,0.12 & 18.3\,$\pm$\,2.6 & 51.0\,$\pm$\,8.4 & 15.8\,$\pm$\,1.6 & T \\
0FGL  J2345.5$$-$$1559  &  356.389 & $-$15.985 & 65.677 & $-$71.092 &  15.5 &  2.42\,$\pm$\,0.12 & 10.5\,$\pm$\,1.7 & 22.3\,$\pm$\,4.3 & 10.3\,$\pm$\,1.3 & T \\
\enddata
\tablenotetext{a}{Spectral index derived from a single power-law fit over the  0.2-100 GeV energy range}
\tablenotetext{b}{Flux ($E>100$ MeV, in  $10^{-8}$\pflux) derived from a single power-law fit over the 0.2-100 GeV energy range}
\tablenotetext{c}{Weekly averaged peak flux ($E>100$ MeV) in $10^{-8}$\pflux}
\tablenotetext{d}{Flux ($E>100$ MeV, in  $10^{-8}$\pflux) obtained by adding the fluxes estimated in the two energy ranges 0.1$-$\,1\,GeV and 1$-$\,100\,GeV}
\tablenotetext{\dagger}{Flux at $E>1$\,GeV in  $10^{-8}$\pflux. For these sources, only an upper limit is obtained for the 0.1$-$1 GeV flux  (see \cite{SourceList}). } 
\end{deluxetable}

\clearpage
\begin{deluxetable}{ccrrrrrrc}
\tablecolumns{8}
\tablewidth{0pc} 
\tabletypesize{\footnotesize}
%% Rotate to a landscape orientation
%\rotate
\tablecaption{Sources in both {\it Fermi}-LAT and {\it EGRET} samples}
\tablenum{4}
\tablehead{\colhead{LAT Name} &  \colhead{{\it EGRET} Name} &  \colhead{F$_{mean}^{F}$\tablenotemark{a}} &  \colhead{F$_{peak}^{F}$} &\colhead{index$^{F}$} &  \colhead{F$_{100}^{E}$\tablenotemark{a}} &  \colhead{F$_{peak}^{E}$} & \colhead{index$^{E}$}  &  \colhead{Type} \\  }
%% All data must appear between the \startdata and \enddata commands
\startdata
0FGL J0210.8$-$5100  & J0210$-$5055 & 24.3 & 76.2 & 2.28 & 85.5 & 134.0 & 1.99 & FSRQ \\
0FGL J0222.6+4302  & J0222+4253 & 25.8 & 49.6 & 1.96 & 18.7 & 25.3 & 2.01 & BLLac \\ 0FGL
J0238.6+1636  & J0237+1635 & 72.5 & 104. & 2.05 & 25.9 & 65.1 & 1.85 & BLLac \\ 0FGL
J0423.1$-$0112  & J0422$-$0102 & 8.0 & 13.3 & 2.37 & 16.3 & 81.7 & 2.44 &  FSRQ\\ 0FGL
J0457.1$-$2325  & J0456$-$2338 & 41.7 & 64.2 & 2.23 & 8.1 & 18.8 & 3.14 & FSRQ \\ 0FGL
J0516.2$-$6200  & J0512$-$6150 & 5.3 & 11.1 & 2.17 & 7.2 & 28.8 & 2.40 &  Unc \\ 0FGL
J0531.0+1331  & J0530+1323 & 24.3 & 39.4 & 2.54 & 93.5 & 351.0 & 2.46 & FSRQ \\ 0FGL
J0538.8$-$4403  & J0540$-$4402 & 37.6 & 49.6 & 2.18 & 25.3 & 91.1 & 2.41 & BLLac \\ 0FGL
J0722.0+7120  & J0721+7120 & 16.3 & 29.0 & 2.07 & 17.8 & 31.8 & 2.19 & BLLac \\ 0FGL
J0738.2+1738  & J0737+1721 & 4.6 & 7.46 & 2.10 & 16.4 & 37.5 & 2.60 & BLLac \\ 0FGL
J0855.4+2009  & J0853+1941 & 8.9 & 18.9 & 2.30 & 10.6 & 15.8 & 2.03 & BLLac \\ 0FGL
J0921.2+4437  & J0917+4427 & 8.6 & 15.6 & 2.34 & 13.8 & 40.8 & 2.19 & FSRQ \\ 0FGL
J0957.6+5522  & J0952+5501 & 8.7 & 12.8 & 2.01 & 9.1 & 47.2 & 2.12 & FSRQ \\ 0FGL
J1104.5+3811  & J1104+3809 & 15.3 & 20.9 & 1.76 & 13.9 & 27.1 & 1.57 & BLLac \\ 0FGL
J1159.2+2912  & J1200+2847 & 10.3 & 16.0 & 2.47 & 7.5 & 163.0 & 1.73 & FSRQ \\ 0FGL
J1229.1+0202  & J1229+0210 & 75.2 & 136. & 2.71 & 15.4 & 53.6 & 2.58 & FSRQ \\ 0FGL
J1256.1$-$0548  & J1255$-$0549 & 31.5 & 46.3 & 2.34 & 74.2 & 267.0 & 1.96 & FSRQ \\ 
0FGL J1325.4$-$4303  & J1324$-$4314 & 21.4 & 32.3 & 2.90 & 13.6 & 38.4 & 2.58 & RG (CenA) \\
0FGL J1333.3+5058 & J1337+5029 & 7.2 & 13.7 & 2.4 & 9.2  &  26.8  &  1.83  & FSRQ\\
0FGL J1457.6$-$3538  & J1500$-$3509 & 36.5 & 77.2 & 2.24 & 10.9 & 40.7 & 2.99 & FSRQ \\
0FGL J1512.7$-$0905  & J1512$-$0849 & 55.8 & 165. & 2.47 & 18.0 & 51.1 & 2.47 & FSRQ \\
0FGL J1517.9$-$2423  & J1517$-$2538 & 4.1 & 6.96 & 1.93 & 8.4 & 53.3 & 2.66 & BLLac \\
0FGL J1625.8$-$2527  & J1626$-$2519 & 16.0 & 28.4 & 2.39 & 21.3 & 90.2 & 2.21 & FSRQ \\
0FGL J1625.9$-$2423 & J1627$-$2419 & 19.9 & 32.1 & 2.45 &  15.8  &  55.2  & 2.21     & Unc\\
0FGL J1635.2+3809  & J1635+3813 & 22.0 & 49.7 & 2.43 & 58.4 & 108.0 & 2.15 & FSRQ \\ 
0FGL J1802.6$-$3939\tablenotemark{b}  & J1800$-$3955 & 25.4 & 64.0 & 2.23 & 9.8 & 189.0 & 3.10 & Unc \\ 
0FGL J1833.4$-$2106\tablenotemark{b}   & J1832$-$2110 & 42.0 & 56.8 & 2.61 & 26.6 & 99.3 & 2.59 & FSRQ \\
0FGL J1911.2$-$2011  & J1911$-$2000 & 22.4 & 52.2 & 2.42 & 17.5 & 47.6 & 2.39 &FSRQ \\ 
0FGL J1923.3$-$2101  & J1921$-$2015 & 13.0 & 41.6 & 2.31 & 4.6 & 31.0 & 2.10 & FSRQ \\ 
0FGL J2025.6$-$0736  & J2025$-$0744 & 47.9 & 73.5 & 2.30 & 21.2 & 74.5 & 2.38 & BLLac \\ 
0FGL J2056.1$-$4715  & J2055$-$4716 & 11.0 & 21.0 & 2.55 & 9.6 & 35.0 & 2.04 & FSRQ \\ 
0FGL J2158.8$-$3014  & J2158$-$3023 & 18.0 & 29.1 & 1.85 & 13.2 & 30.4 & 2.35 & BLLac \\ 
0FGL J2202.4+4217  & J2202+4217 & 8.4 & 12.8 & 2.23 & 11.1 & 39.9 & 2.60 & BLLac \\ 
0FGL J2232.4+1141  & J2232+1147 & 14.0 & 24.5 & 2.61 & 19.2 & 51.6 & 2.45 & FSRQ \\
0FGL J2254.0+1609  & J2254+1601 & 246. & 385. & 2.41 & 53.7 & 116.0 & 2.21 & FSRQ \\
\enddata
%% Include any \tablenotetext{key}{text}, \tablerefs{ref list},
%% or \tablecomments{text} between the \enddata and 
%% \end{deluxetable} commands
\tablenotetext{a}{$10^{-8}$\pflux}
\tablenotetext{b}{source located at $|b|<10^\circ$}
\end{deluxetable}

\clearpage
\begin{deluxetable}{llllll}
\setlength{\tabcolsep}{0.04in}
\tablewidth{0pc} 
\tabletypesize{\small}
\tablecaption{Correlation analysis for the radio and gamma-ray properties
\label{tab:radiogamma}}
\tablenum{5}
\tablehead{
%%%%%%%% column names
\colhead{Data} & \colhead{Method} & \colhead{Source type} & \colhead{\# objects } & \colhead{Correlation coeff.} & \colhead{Chance probability\tablenotemark{a}}   }
\startdata
$S_{8.4\,\mathrm{GHz}}-F_{>100\,\mathrm{MeV, peak}}$ & Spearman rank & All     & 106 & 0.42 & 0.0000045 \\
Log $L_r$-Log $L_\gamma$                       & Sp.\ r., partial & All     &  90 & 0.46 & 0.0000025 \\
\hline
$S_{8.4\,\mathrm{GHz}}-F_{>100\,\mathrm{MeV, peak}}$ & Spearman rank & FSRQ    &  57 & 0.19 & 0.080    \\
Log $L_r$-Log $L_\gamma$                       & Sp.\ r., partial & FSRQ    &  57 & 0.34 & 0.0047   \\
\hline
$S_{8.4\,\mathrm{GHz}}-F_{>100\,\mathrm{MeV, peak}}$ & Spearman rank & BL Lacs &  42 & 0.49 & 0.00055   \\ 
Log $L_r$-Log $L_\gamma$                       & Sp.\ r., partial & BL Lacs &  30 & 0.60 & 0.00023   \\
\enddata
\tablenotetext{a}{if the samples were unbiased}
\end{deluxetable}

\clearpage
\begin{deluxetable}{lc}
\tablewidth{0pt}
\tablecaption{Composition of the $|$b$|\geq$10$^{\circ}$ sample (TS$\geq$100). 
\label{tab:sampleb10}}\tablenum{6}
\tablehead{
%%%%%%%% column names
\colhead{CLASS} & \colhead{\# objects }}
\startdata
Total                         & 132 \\
FSRQs                         & 57\tablenotemark{a}  \\
BL Lacs                       & 42\tablenotemark{a} \\
Uncertain\tablenotemark{b}    & 5\tablenotemark{a}  \\
Radio Galaxies                & 2\tablenotemark{a}  \\
Pulsars                       & 7\tablenotemark{c} \\
Anti-associations             & 4 \\

% ----- The ones below are the UNID
Low confidence associations   & 10 \\
Unassociated  sources         & 5 \\

\enddata
\tablenotetext{a}{Part of the high confidence sample.}
\tablenotetext{b}{Blazars with uncertain classification.}
\tablenotetext{c}{ Five LAT detected pulsars plus 0FGL J0025.1-7202 (47 Tuc) and 0FGL  J0538.4-6856 (associated with the Large Magellan Cloud, see \cite{SourceList}).}
\end{deluxetable}

\begin{deluxetable}{lcccc}
\tablewidth{0pt}
\tablecaption{Results of the best power-law fits to the source counts
distributions. Errors within brackets are systematic uncertainties due
to the incompleteness of the sample. The lower part of the table shows
the results for the {\it flux-limited} portion of the sample.
\label{tab:logn}}
\tablenum{7}
\tablehead{
%%%%%%%% column names
\colhead{SAMPLE} & \colhead{\# Objects }     &
\colhead{$\alpha$} & \colhead{A\tablenotemark{a}} & 
\colhead{EDB fraction\tablenotemark{b}}
}
\startdata
All\tablenotemark{c}  & 121 & 2.59$\pm0.12$ & 2.62$\pm0.24$  & 7.2\,\% \\
Blazars & 106 & 2.50$\pm0.12$ & 2.24$\pm0.22$($\pm$0.24)   & 6.1\,\% \\
FSRQs   & 57  & 2.60$\pm0.14$ & 2.15$\pm0.28$($\pm$0.32)   & 3.1\,\% \tablenotemark{d}  \\
BL Lacs & 42  & 2.34$\pm0.15$ & 0.41$\pm0.06$($\pm$0.06)   & 1.0\,\% \\
\hline
FSRQs   & 29  & 2.52$\pm0.20$ & 1.93$\pm0.35$($\pm$0.09)   & 2.6\,\%\tablenotemark{d} \\  
BL Lacs & 9   & 2.50$\pm0.37$ & 0.48$\pm0.16$($\pm$0.02)   & 1.3\,\%\\
\enddata
\tablenotetext{a}{In units of 10$^{4}$\,cm$^{2}$ s deg$^{-2}$.}
\tablenotetext{b}{Fraction of the EGRET diffuse extragalactic background
\citep{sreekumar98} resolved into sources 
by LAT  for  $4\times 10^{-8}<$F$_{100}<10^{-7}$\,ph cm$^{-2}$ s$^{-1}$.}

\tablenotetext{c}{Includes all sources except 7 pulsars  and 4 
anti-associated objects.}

\tablenotetext{d}{The lower limit of integration in Eq.~\ref{eq:diffuse}
has been set to $6\times 10^{-8}$\,ph cm$^{-2}$ s$^{-1}$.}

\end{deluxetable}

\begin{deluxetable}{lcc}
\tablewidth{0pt}
\tablecaption{Results of the V/V$_{MAX}$ test.
\label{tab:vvmax}}
\tablenum{8}
\tablehead{
%%%%%%%% column names
\colhead{SAMPLE} & \colhead{\# Objects } & \colhead{$<$V/V$_{MAX}$$>$}
}
\startdata
FSRQs                    & 57  & 0.645 $\pm0.043$ \\
BL Lacs                  & 31  & 0.430 $\pm0.055$ \\
BL Lacs\tablenotemark{a} & 42  & 0.472 $\pm0.046$\\ 
BL Lacs\tablenotemark{b} & 42  & 0.473 $\pm0.046$\\ 
All with z$>$0           & 92  & 0.512 $\pm0.031$ \\
\hline
FSRQs   \tablenotemark{c}  & 29  & 0.654 $\pm0.061$ \\
BL Lacs \tablenotemark{c}  & 8   & 0.542 $\pm0.103$ \\
\enddata
\tablenotetext{a}{For all the 11 BL Lacs without redshift,  we have assumed
z = $<$z$>$ where $<$z$>$=0.38. }
\tablenotetext{b}{For all the 11 BL Lacs without redshift,  we have drawn
a redshift measurement from a Gaussian distribution with mean 0.38 and 
dispersion 0.34.} 
\tablenotetext{c}{Flux-limited sample.}

\end{deluxetable}

\begin{deluxetable}{lccr}
\tablewidth{0pt}
\tablecaption{Results of best-fit power-law  models to the 
GLFs of FSRQs  in different redshift bins. 
The lower part of the table shows the results for the 
{\it flux-limited} portion of the sample. 
Errors represent the 68\,\% confidence level. Uncertainties within
bracketes are systematic errors due to the incompleteness of the sample. 
\label{tab:fsrq_fits}}
\tablenum{9}
\tablehead{
%%%%%%%% column names
\colhead{Redshift bin} & \colhead{\# Objects } 
& \colhead{$\beta$} & \colhead{Normalization\tablenotemark{a}}
}
\startdata
z = 0.0 -- 0.9     & 20 & 1.57$\pm0.10$ & 2.43$\pm0.52$($\pm0.36$) \\
z = 0.9 -- 1.4     & 17 & 2.56$\pm0.29$ & 5.59$\pm1.33$($\pm0.83$)  \\
z $>$ 1.4          & 20 & 2.58$\pm0.19$ &13.07$\pm2.92$($\pm1.96$) \\
\hline
z = 0.0 -- 0.9     & 10 & 1.46$\pm0.13$ & 1.73$\pm0.54$($\pm0.08$) \\
z = 0.9 -- 1.4     & 8  & 2.65$\pm0.48$ & 6.62$\pm2.34$($\pm0.33$) \\
z $>$ 1.4          & 11 & 2.63$\pm0.30$ &17.76$\pm5.35$($\pm0.88$) \\

\enddata
\tablenotetext{a}{Normalization of the GLF model at 10$^{48}$\,erg s$^{-1}$
expressed in units of $10^{-11}$\,erg$^{-1}$\,s\,Mpc$^{-3}$.}
\end{deluxetable}

\begin{deluxetable}{lccr}
\tablewidth{0pt}
\tablecaption{Results of best-fit power-law models to GLFs of BL Lacs
 in different redshift bins.
The lower part of the table shows the results for the 
{\it flux-limited} portion of the sample.
 Errors represent the 68\,\% confidence level.
Uncertainties within
bracketes are systematic errors due to the incompleteness of the sample. 
\label{tab:bllac_fits}}
\tablenum{10}
\tablehead{
%%%%%%%% column names
\colhead{Redshift bin} & \colhead{\# Objects } 
& \colhead{$\beta$} & \colhead{Normalization\tablenotemark{a}}
}
\startdata
z = 0.0 -- 0.3     & 15 & 2.08$\pm0.16$ & 7.67$\pm1.98$($\pm3.06$) \\
z $>$ 0.3          & 16 & 2.10$\pm0.11$ & 4.75$\pm1.18$($\pm1.90$) \\
z $>$ 0.0          & 31 & 2.17$\pm0.05$ & 4.19$\pm0.75$($\pm1.67$) \\
\hline
z $>$ 0.0          & 8 & 1.90$\pm0.11$ & 6.70$\pm2.36$($\pm0.33$) \\
\enddata
\tablenotetext{a}{Normalization of the GLF model at 10$^{48}$\,erg s$^{-1}$
expressed in units of $10^{-12}$\,erg$^{-1}$\,s\,Mpc$^{-3}$.}
\end{deluxetable}

\end{document}